\documentstyle[12pt]{article}
\makeatletter
\def\setb@se#1{\baselineskip=#1 \normalbaselineskip=#1}
\setb@se{12pt}
\long\def\title#1{\vspace*{11.5pc}{\pretolerance=10000\raggedright
  \setb@se{12pt}\bf #1\par}\nobreak\ignorespaces}
\long\def\author#1{\vspace{4pc}\begin{list}{\hfill}%
{\topsep=0pt\parskip=0pt\parsep=0pt\partopsep=0pt\listparindent=0pt%
\itemsep=0pt\rightmargin=0pt\labelsep=0pt\labelwidth=5pc\leftmargin=5pc}%
\item\normalsize{#1}\end{list}\vspace{14pt}}
\long\def\affil#1{\begin{list}{\hfill}%
{\topsep=0pt\parskip=0pt\parsep=0pt\partopsep=0pt\listparindent=0pt%
\itemsep=0pt\rightmargin=0pt\labelsep=0pt\labelwidth=5pc\leftmargin=5pc}%
  \item\normalsize{\rm #1}\end{list}\vspace{7pt}}
\long\def\beginabstract{\vspace{21pt plus 7pt minus 7pt}\begin{list}{\hfill}%
{\topsep=0pt\parskip=0pt\parsep=0pt\partopsep=0pt\listparindent=0pt%
\itemsep=0pt\rightmargin=0pt\labelsep=0pt\labelwidth=5pc\leftmargin=5pc}%
\item\normalsize{\bf Abstract. }}
\long\def\endabstract{\end{list}\vspace{28pt plus14pt minus 14pt}%
\normalsize\noindent}
\def\@sect#1#2#3#4#5#6[#7]#8{\ifnum #2>\c@secnumdepth
  \def\@svsec{}\else
  \refstepcounter{#1}\edef\@svsec{\csname the#1\endcsname.\hskip 0.5em}\fi
  \@tempskipa #5\relax
   \ifdim \@tempskipa>\z@
     \begingroup #6\relax
       \@hangfrom{\hskip #3\relax\@svsec}{\interlinepenalty \@M #8\par}%
     \endgroup
    \csname #1mark\endcsname{#7}\addcontentsline
      {toc}{#1}{\ifnum #2>\c@secnumdepth \else
                   \protect\numberline{\csname the#1\endcsname}\fi
                 #7}\else
     \def\@svsechd{#6\hskip #3\@svsec #8\csname #1mark\endcsname
                   {#7}\addcontentsline
                        {toc}{#1}{\ifnum #2>\c@secnumdepth \else
                          \protect\numberline{\csname the#1\endcsname}\fi
                    #7}}\fi
  \@xsect{#5}}

\def\section{\@startsection {section}{1}{\z@}{-28pt plus -14pt minus
-14pt}{2.3ex plus .2ex}{\normalsize\bf}}
\def\subsection{\@startsection{subsection}{2}{\z@}{-14pt plus -8pt
minus -4pt}{1.5ex plus .2ex}{\normalsize\bf}}
\def\subsubsection{%
    \@startsection{subsubsection}{3}{\z@}{-14pt plus
    -8pt minus -4pt}{-1.5ex plus -.2ex}{\normalsize\bf}}
\def\paragraph{\@startsection
     {paragraph}{4}{\z@}{3.25ex plus 1ex minus .2ex}{-1em}{\normalsize\bf}}
\def\subparagraph{\@startsection
     {subparagraph}{4}{\parindent}{3.25ex plus 1ex minus
     .2ex}{-1em}{\normalsize\bf}}

\def\caption{\refstepcounter\@captype \@dblarg{\@caption\@captype}}
\long\def\@caption#1[#2]#3{\par\addcontentsline{\csname
 ext@#1\endcsname}{#1}{\protect\numberline{\csname
 the#1\endcsname}{\ignorespaces #2}}\begingroup
   \@parboxrestore
   \hspace*{28pt}
   \parbox{394pt}{\@makecaption{{\bf\csname
            fnum@#1\endcsname}}{\ignorespaces #3}}\par
\vspace{7pt}\endgroup}

\long\def\@makecaption#1#2{
   \vskip 14pt
   \setbox\@tempboxa\hbox{#1{\bf.} #2}
   \ifdim \wd\@tempboxa >\hsize   
       #1{\bf.} #2\par            
     \else                        
       \hbox to\hsize{\box\@tempboxa\hfil}
   \fi}

\newcommand{\boldarrayrulewidth}{1pt} 

\def\bhline{\noalign{\ifnum0=`}\fi\hrule \@height
                    \boldarrayrulewidth \futurelet
                    \@tempa\@xhline}

\def\@xhline{\ifx\@tempa\hline\vskip \doublerulesep\fi
      \ifnum0=`{\fi}}

\newcommand{\topline}{\bhline\bs}
\newcommand{\midline}{\bs\hline\bs}
\newcommand{\bottomline}{\bs\bhline}

\def\thebibliography#1{\section*{REFERENCES\@mkboth
  {REFERENCES}{REFERENCES}}\list
  {\hfil[\arabic{enumi}]}{\itemsep=0pt\labelsep=7pt\itemindent=-14pt
    \settowidth\labelwidth{[#1]}
    \leftmargin\labelwidth
    \advance\leftmargin\labelsep
    \advance\leftmargin -\itemindent
    \usecounter{enumi}}\setb@se{12pt}\small
    \def\newblock{\hskip .11em plus .33em minus .07em}
    \sloppy\clubpenalty4000\widowpenalty4000
    \sfcode`\.=1000\relax}
\def\references{\section*{REFERENCES\@mkboth
{REFERENCES}{REFERENCES}}\list{}{\itemsep=0pt\labelsep=0pt\itemindent=-28pt
\labelwidth=0pt\leftmargin=28pt}\setb@se{12pt}\small
\def\newblock{\hskip .11em plus .33em minus .07em}
\sloppy\clubpenalty4000\widowpenalty4000
\sfcode`\.=1000\relax}
\newcommand{\etal}{{\em et al\/}\ }
\newcommand{\dash}{------}

\newcommand{\bs}{\noalign{\vspace{7pt plus2pt minus2pt}}}

\makeatother

\newcommand{\beq}{\begin{equation}}
\newcommand{\eeq}{\end{equation}}
\newcommand{\bq}{\begin{quotation}}
\newcommand{\eq}{\end{quotation}}
\newcommand{\bc}{\begin{center}}
\newcommand{\ec}{\end{center}}

\newcommand{\BFACE}[1] {\mbox{\boldmath $#1$} }

\begin{document}

\title{CHAOS IN THE EINSTEIN EQUATIONS - CHARACTERIZATION AND
IMPORTANCE?\footnote{To appear in NATO ARW on ``Deterministic
                     Chaos in General Relativity'',
                     D. Hobill (ed.). Plenum Press., N.Y., 1994}}

\author{Svend E. Rugh \footnote{e-mail: rugh@nbivax.nbi.dk}}.

\affil{The Niels Bohr Institute, University of Copenhagen, \\
Blegdamsvej 17, 2100 K\o benhavn \O, DENMARK}

\beginabstract
Is it possible to define what we could mean by chaos in a
space-time metric (even in the simplest toy-model
studies)? Is it of importance for phenomena we may
search for in Nature?
\endabstract

\section{INTRODUCTION}

Theoretical physics would not die out
even if we had {\em already}
found the ``master plan'', the ``master law'' (T.O.E.)
for how the Universe in which we live is
constructed. Namely, given the knowledge of the
basic laws of physics (on each level on the Quantum Staircase, say)
it is still a major project to try
to deduce their {\em complex} consequences, i.e. to
find out the
complex ways in which matter (and their interactions)
organizes itself into living and nonliving forms.

The word ``chaos''
is somewhat of an unlucky choice since
we may too easily associate it with something which is structureless.
Chaotic systems do not lack structure.
On the contrary, chaotic systems exhibit far more interesting and richer
structure in their dynamical behavior than integrable systems.

However, chaotic systems have an aspect of  unpredictability
and ``simulated'' disappearance of information:
Information walks down to the small scales and is replaced by noise walking up
from the small scales. I.e. chaos pumps {\em information} up and down
the chain of decimals in the phase space coordinates!
(due to the exponential amplification of uncertainties in the
specification of the initial state).

Many physical systems, governed by {\em non-linear}
equations of motion, will exhibit chaos.
By now, examples are known from all disciplines in physics.

Due to the highly non-linear {\em self-interaction} of
the gravitational and non-Abelian
gauge fields, the time evolution of (generic) configurations of
``gauge fields'' and ``gravitational fields'' will be non-integrable
even without any coupling to
material bodies.
For example, as evidenced by the simplest toy-model
studies, we expect that the Einstein equations
- in scenarios involving {\em strong} field strength, where the
non-linearity of the equations is important (e.g. probing the
Einstein equations near space-time singularities) -
will exhibit {\em chaotic solutions} (rather than integrable ones)
if not too much symmetry is imposed on the field configuration.
The same is true for non-Abelian gauge fields. While on one
hand this is {\em not very surprising} (the Einstein and
Yang-Mills
equations are highly {\em non-linear} theories and one of
the lessons from chaos theory has been that
{\em even the simplest} non-linear equations usually exhibit ``chaos''!)
there are, on the other hand, several issues of interest:

\begin{quotation}

\underline{\bf 1}. {\em Can ``standard indicators of chaos'' be used} to
characterize the ``metrical chaos''?
In other words: Are there
some {\em deeper problems} in
characterization of chaos in this context compared to other chaotic
physical models?

\underline{\bf 2}. What is the
{\em structure} of this (non-dissipative)
chaos? How does
chaos ``look like'' in simple toy-models of
the classical Yang-Mills equations and the classical
Einstein equations?

\underline{\bf 3}. Is it of any {\em physical significance} that
field configurations of the fundamental forces at the
classical (or semi-classical) level may exhibit
chaotic, irregular non-integrable solutions?

\end{quotation}

Let us turn our attention to the gravitational field:
It is well known that individual orbits of test particles (bodies)
in a {\em given}
gravitational field can exhibit ``chaos". This is the case for the
motion of test particles in Newton's theory of gravitation
(e.g. ``chaotic" orbits of individual stars
in the potential generated by the other stars in a galaxy)
and also in general relativity: Cf., e.g.,
the study by
G. Contopoulos of periodic orbits and chaos
around two black holes, G. Contopoulos (1990),
the study of chaos around a single black hole,
L. Bombelli and E. Calzetta (1992) and the study of chaotic motion
of test particles around a black hole immersed in a magnetic
field, V. Karas \etal (1992).
Also an extended object like a (cosmic) string may
jump chaotically around the equatorial plane of a
black hole, cf.\ A.L. Larsen (1993).

Because of their nonlinearity, the Einstein field
equations however permit spacetime
to be curved (``gravity generates gravity") - even in the absence of any
nongravitational energy - and the dynamical evolution of the spacetime
metric ``$ g_{\mu \nu} (x) $" itself is governed by
{\em highly nonlinear} equations
and may allow
solutions of the chaotic type for the ``metric" field itself.

If we explicitly write down (cf.\ Kip S. Thorne (1985))
the vacuum Einstein equations $G_{\mu \nu} = 0$
as differential equations for the ``metric density''
${\tilde{g}}^{\mu \nu} \equiv \sqrt{-g} g^{\mu \nu}$
we have a very complicated set of partial differential equations
for $\tilde{g}^{\mu \nu}$,\footnote{Note, that $\tilde{g}_{\mu \nu}$
(the inverse of $\tilde{g}^{\mu \nu}$) is
a highly nonlinear algebraic function of $\tilde{g}^{\mu \nu}$.
Repeated indices are to be summed, commas denote partial
derivatives, e.g. $\tilde{g}^{\alpha \beta}_{, \mu} \equiv
\partial \tilde{g}^{\alpha \beta}/ \partial x^{\mu} $.
The coordinate system has been specialized so the metric is in
``deDonder gauge'', $\tilde{g}^{\alpha \beta}_{, \beta} = 0$ (see
K.S. Thorne (1985)).}

\newpage

\begin{eqnarray}   \label{horrendous}
\tilde{g}^{\mu \nu}
\frac{\partial^2 \tilde{g}^{\alpha \beta}}{\partial
x^{\mu} \partial x^{\nu}} & = &
\tilde{g}^{\alpha \nu}_{, \mu} \tilde{g}^{\beta \mu}_{, \nu} +
\frac{1}{2} \tilde{g}^{\alpha \beta} \tilde{g}_{\lambda \mu}
\tilde{g}^{\lambda \nu}_{, \rho} \tilde{g}^{\rho \mu}_{, \nu} \nonumber \\
& + &
\tilde{g}_{\lambda \mu} \tilde{g}^{\nu \rho}
\tilde{g}^{\alpha \lambda}_{, \nu} \tilde{g}^{\beta \mu}_{, \rho} -
\tilde{g}^{\alpha \lambda} \tilde{g}_{\mu \nu}
\tilde{g}^{\beta \nu}_{, \rho} \tilde{g}^{\mu \rho}_{, \lambda} -
\tilde{g}^{\beta \lambda} \tilde{g}_{\mu \nu}
\tilde{g}^{\alpha \nu}_{, \rho} \tilde{g}^{\mu \rho}_{, \lambda} \nonumber \\
& + & \frac{1}{8}
(2 \tilde{g}^{\alpha \lambda} \tilde{g}^{\beta \mu} -
\tilde{g}^{\alpha \beta} \tilde{g}^{\lambda \mu})
(2 \tilde{g}_{\nu \rho} \tilde{g}_{\sigma \tau} -
\tilde{g}_{\rho \sigma} \tilde{g}_{\nu \tau})
\tilde{g}^{\nu \tau}_{, \lambda} \tilde{g}^{\rho \sigma}_{, \mu} \; \; .
\end{eqnarray}
The left hand side of (\ref{horrendous}) is a kind of curved spacetime
wave operator ``$\Box$'' acting on $\tilde{g}^{\alpha \beta}$
(giving a propagation effect of the gravitational degrees of freedom)
whereas the right hand side is a sort of ``stress-energy pseudotensor''
for the gravitational field which is quadratic in the first derivatives
of $\tilde{g}^{\alpha \beta}$ and acts as the source for
``$\Box$'' $\tilde{g}^{\alpha \beta}$.

At first, general relativity, i.e. Einsteins theory of gravitation,
is not even a dynamical theory in the usual sense.
It does not, from the very beginning, provide us
with a set of parameters (describing the gravitational degrees of freedom)
evolving in  ``time''.
``Time'' loses here its absolute meaning as opposed to the classical dynamical
theories where the ``Newtonian time'' is taken for granted.
The division between space and time
in general relativity comes through foliating the space-time manifold $\cal M$
into spacelike hypersurfaces $\Sigma_t$. The metric $g_{\mu \nu}$ on $\cal M$
induces a metric $g_{ij}$ on $\Sigma_t$
and can be parametrized in the form
$$
g_{\mu \nu} = \left(
\begin{array}{cc}
N_i N^i - N^2 & N_j\\
N_i & g_{ij}
\end{array}
\right)
$$
bringing the metric on the $3+1$ form\footnote{Albert Einstein
taught us to treat space and time on an equal footing in a
four-dimensional spacetime manifold and build up a
space-time covariant formulation of the theory,
yet to deal with dynamics we manifestly have to break the space-time
covariance of the formulation.}
\begin{equation}    \label{lapseshift}
ds^2 = - N^2 dt^2 + g_{ij} (dx^i + N^i dt)(dx^j + N^j dt)
\end{equation}
where $N$ and $N_i$ are called lapse function and shift
vector respectively. (Cf, e.g., MTW \S 21.4)

\begin{quotation}
\noindent
{\em Only after splitting the space-time into
space and time (the 3+1 ADM splitting)
we yield the possibility to treat the evolution of the
metric under the governing Einstein equations as a dynamical
system on somewhat equal footing as other dynamical systems
which have some (physical) degrees of
freedom evolving in ``time''. }\footnote{Compare, e.g., with the
characterization of turbulence in
connection with fields like Navier-Stokes flows or $SU(2)$ Yang Mills
fields.
In these cases there are no such fundamental problems or ambiguities as
concerns the fields being
in certain well defined space points and evolving in a well defined
``external'' time $t$. Such fields evolve in a flat Minkowskij
metric (if they are not coupled to gravitational fields)
and no ``mixing'' between space and
time concepts (up to ``stiff'' Lorentz transformations)
occurs. }
\end{quotation}

It is of interest to our discussion on the dynamics of the Einstein
equations (chaotic or not) to know
that one may show that the
Einstein equations indeed admit a well posed {\em initial value}
formulation, so the Einstein equations do
{\em determine} the evolution of the metric
(up to gauge transformations) {\em uniquely}
from given initial conditions and the solutions
admit a {\em Cauchy stability criteria} which
establishes that the solutions depend
{\em continuously} on the initial data. Cf., e.g.,
discussion in MTW \S 21.9, S.W. Hawking and G.F.R. Ellis (1973), sec.7,
and R.M. Wald (1984), sec.10.

Given the complexity of the Einstein equations (\ref{horrendous})
by comparison, e.g., with the Navier-Stokes equations, it is
not surprising
that we at present have only little
understanding of the dynamics of
the Einstein equations involving strong gravitational fields.

{}From the observational side
not many observable phenomena which involve the
gravitational field need the full non-linear Einstein equations and
in ``daily life'' gravity often the {\em linearized}
equations suffice.\footnote{Note, however, that
{\em nonlinear effects} of the Einstein equations may
show up to be
important even in regions where one would imagine the linearized
equations to be sufficient. For example,
one has a {\em nonlinear} effect in the form of a permanent
displacement of test masses (of the same order of magnitude
as the {\em linear} effects) after the passage of a gravitational
wave train - even when the test masses are
placed at arbitrary distances from the gravitational
wave source. See D. Christodoulou (1991).}
Among the three
{\em classic} tests of
Einstein's theory only the precession of the
perihelia of the orbits of the inner planets
tests a non-linear aspect of the Einstein equations.

To find scenarios where the full non-linearities of the Einstein
equations are important
one has to search among astrophysical and cosmological
phenomena far removed from ``daily life'' gravity.

Among non-linear phenomena in general relativity (geons, white and
black holes, wormholes, cf.\ Kip S. Thorne (1985), solitons,
cf.\ G.W. Gibbons (1985)) the formation of spacetime singularities
is one of the most remarkable phenomena
which appears in nonlinear solutions of
the {\em classical} Einstein equations
under a variety of circumstances, cf.\ the singularity theorems
by S.W. Hawking and R. Penrose (1970).

The singularity theorems of Hawking and Penrose tell us, however,
practically nothing
about the structure of such spacetime singularities.
What do they look like?

\subsection{Higly symmetric gravitational collapses as
a laboratory
for testing ideas about how to characterize ``chaos'' in a general
relativistic context.}

To capture a nonlinear aspect of the Einstein equations it is natural
to probe them in scenarios involving strong field strength,
e.g. near curvature singularities.

Without some restrictions of symmetry imposed on the spacetime metric
$g_{\mu \nu} (x)$ the Einstein equations are intractable (though
considerable progress has been made in numerical relativity
of solving cases with little symmetry).

We shall consider a simple example and use it to test ideas and concepts
about chaos. (If we are not able to agree upon
how chaos should be defined
in this simple example we may very well give up all hope to develop
indicators of chaotic behavior
in more complicated examples of spacetime metrics).

Thus, we restrict, for simplicity, attention to
toy-model metrics with
{\em spatially homogeneous} three-dimensional
space-like slices (hypersurfaces): Then
the gravitational fields are the same at every point
on each of the surfaces of homogeneity and one may thus represent
these fields via {\em functions of time} only!

More explicitly, spatially homogeneous 3-geometries are 3-manifolds
on which a three-dimensional Lie group acts transitively.
On the 3-manifold this symmetry is encoded in the
existence of
three linearly independent spacelike
Killing vectors $\xi_i, \; i=1,2,3$,
satisfying the Lie algebra $[\xi_i, \xi_j] = {\cal C}^k_{ij} \xi_k$
where ${\cal C}^k_{ij}$ are the structure constants of the Lie
algebra.\footnote{The classification of three dimensional
Lie-algebras dates back to L. Bianchi (1897) and the
spatially homogeneous metrics are therefore often referred to
as Bianchi metrics. See e.g. M.A.H. MacCallum (1979, 1983).}

The particular collapse (``big crunch''), the
mixmaster collapse, we shall consider has
the same non-Abelian isometry group $SU(2)$ on the
three-space (i.e. same topology $\sim R \times S^3$) as the compact
FRW-collapse but it contains three
scale-factors $a,b,c$ instead of just one. A {\em `freely falling
astronomer'} who falls into the spacetime singularity of the `big crunch'
will experience a growing tidal field,
in which he is {\em compressed} along two directions and
{\em stretched} (expanded) in
one direction, the directions being permuted infinitely many times in
a {\em not-predictable} way.

The possibility of chaos has been
investigated only for very few toy model studies of spacetime metrics!
Whether one should
expect the Einstein equations to generate ``chaos" in generic cases
(for strong gravitational fields, i.e., high curvatures) the answer
is absolutely: Nobody knows!

\vspace{0.5 cm}

The paper is organized as follows:
In sec.2 we describe various aspects of the
mixmaster gravitational collapse.
Not surprisingly, a collapse to a spacetime singularity
is prevented if one
includes matter with negative energy and pressure.
In that case, however, the behavior of the spacetime metric is
very interesting, very irregular and highly unpredictable,
and oscillations of the three-volume occur.
(Due to the negative pressure and energy density the
attraction of matter turns into an unphysical repulsion
preventing the ``universe'' from collapsing).
If the metric is evolved according to the vacuum Einstein equations,
the dynamics has, after some transient, a monotonically
declining three-volume and the degrees of freedom
of the spacetime metric is fast attracted into an interesting
self-similar, never ending oscillatory behavior on approach
to the big crunch singularity.
This may be understood as a never ending sequence of short bounces
against a potential boundary generated by the three-curvature
scalar ${}^{(3)} R$ on the three-space.
This scattering potential becomes, to a very good approximation,
infinitely hard when the metric approaches the singularity, and the
collapse dynamics may, in that limit, be captured by a set
of simple algebraic transition rules (maps),
for example the so-called ``Farey map'', which is a
strongly intermittent map (this map has, as sub-map, the Gauss
map which is well known in chaos theory and which
has positive Kolmogorov entropy). \footnote{Results
obtained in the first part of
sec.2 were also arrived at by D. Hobill \etal in completely
independent investigations.}

In sec.3 we describe the problem - inherent to general
relativity - of transferring standard
indicators of chaos, in particular the spectrum of Lyapunov exponents,
to the general relativistic context, since they are highly
gauge dependent objects.
This fact was pointed to and emphasized in
S.E. Rugh (1990 a,b). I can only moderately agree that
this observation was arrived at independently by
J. Pullin (1990).
By referring to a specific gauge (the Poincar\'{e} disc) Pullin misses
the point (in my opinion). No ``gauge'' is better than others.
One should try to develop indicators which capture
chaotic properties of the gravitational field (``metric chaos'')
in a way which is invariant under
spacetime diffeomorphisms - or prove that this can not
be done! (H.B. Nielsen and S.E. Rugh).
This program of research is still in its infancy.

In sec.4 and sec.5 some (even) more wild speculations are
offered concerning the generality and applicability of the concept of
``metrical chaos'' etc. One would like to argue - but it is not easy -
that {\em non-integrability} of the Einstein equations, is a generic
phenomenon when considering scenarios involving really strong
gravitational fields, e.g. near Planck scales where the gravitational
field should be treated quantum mechanically. Whether there are
implications of ``metrical chaos'' on the quantum level is a question
which not possible to address since no good candidate for a
theory of quantum gravity is known.
In the context of the Wheeler-DeWitt equation
(which however involves arbitrariness, cf.\ e.g. the factor
ordering problem)
one may address this question for the mixmaster
gravitational collapse. This is beautifully illustrated
in the ``Poincar\'{e} disc gauge'' (which I describe shortly)
and was already considered by Charles W. Misner
twenty years ago. The mixmaster collapse dynamics is however so
special (the scattering domain of the Poincar\'{e} disc tiles
the disc under the action of an ``arithmetic group'')
that its quantization exhibits ungeneric
features, relative to more generic Hamiltonian models
(of similar low degree of dimensionality) studied in
the discipline of ``Quantum Chaos''\footnote{``Quantum Chaos'', or
what Michael Berry has named ``Quantum Chaology'',
cf.\ e.g. M. Berry (1987), investigates
the semi-classical or quantum behavior characteristic of
Hamiltonian systems whose classical
motion exhibits chaos.}. This illustrates, once again, that
the mixmaster gravitational collapse is a very beautiful,
yet very special, example of chaos (algebraic chaos).
However, for our purpose, to use it as a toy-model
to investigate the applicability of indicators of chaos in the
general relativistic context,
it serves as a good starting point.

It is interesting whether ``metrical chaos''
(not yet defined) has potential applications for phenomena occurring
in Nature. Certainly, non-integrability (i.e. lack of first integrals
relative to the number of degrees of freedom) and non-linear
effects may show up even in scenarios involving rather weak fields
(cf.\ e.g. D. Christodoulou (1991)).

Considering the possibility of the early Universe to be described
by the mixmaster metric, we note in sec.6
that, according to the Weyl curvature hypothesis
(of R. Penrose), which suggests that the Weyl curvature tensor should
vanish at the initial singularity (at ``big bang''), the mixmaster
metric has too big Weyl curvature to be implemented in our
actual Universe at Planck scales, say. The Guth/Linde inflationary
phase may modify this viewpoint.

In sec.6 also some more general reflections are put forward concerning
the ``chaotic cosmology'' concept by Charles W. Misner \etal which
attempt at arriving at our present Universe from (almost) arbitrary
initial conditions.

\newpage

\section{THE MIXMASTER GRAVITATIONAL COLLAPSE GIVES
A ``HINT'' OF THE SORT OF COMPLEXITY (``METRICAL CHAOS'') ONE
MAY HAVE FOR SOLUTIONS TO EINSTEIN EQUATIONS.}

The mixmaster gravitational collapse is a very
famous\footnote{Cf., e.g.,
Ya.B. Zel'dovich and I.D. Novikov (1983), esp. \S 22;
C.W. Misner, K.S. Thorne and J.A. Wheeler
(1973), esp. \S 30 or
L.D. Landau and E.M. Lifshitz (1975),
esp. \S 116-119. See, also, J.D. Barrow (1982).}
gravitational collapse (a ``big crunch'')
which generalizes the
compact FRW collapse and
which gives us a ``hint'' of the sort of complexity (``metrical chaos'')
one should expect for gravitational
collapses which have more degrees of freedom
than the simple (integrable) FRW-collapse.

We imagine that a ``3+1'' split has been performed,
splitting the spacetime manifold
into the topological
product of a line (the ``time" axis) and the
three-dimensional spacelike hypersurfaces $\Sigma_t$
(the dynamical degrees of freedom
are the {\em spatial} components of the metric,
the induced metric $g_{ij}$ on $\Sigma_t$,
which evolves in the ``time'' parameter ``$t$'').
In fact, we shall operate in a
``synchronous reference frame''
(Landau and Lifshitz \S 97 and
MTW, \S 27.4)
which brings the spacetime metric (\ref{lapseshift}) on the
very simple form
$ ds^2 = -dt^2 + g_{ij} dx^i dx^j $.

\subsection{What are the degrees of freedom in this toy model?}

\indent The symmetry-ansatz for the metric\footnote{If the
metric is coupled to matter, e.g. perfect fluid matter, the
assumption of diagonality of $\gamma_{ij}(t)$ is a
{\em simplifying} ansatz. If no non-gravitational matter is
present, the vacuum Einstein equations will automatically
make the off-diagonal components of $\gamma_{ij}$ vanish for a
space with invariance group $G = SU(2)$, see e.g.
Bogoyavlensky (1985), p. 34.} is:
\begin{equation}  \label{metric}
ds^2 = -dt^2 + \gamma_{ij}(t)
\BFACE{\omega}^i (x) \BFACE{\omega}^j (x) \; \; , \; \;
\gamma_{ij} (t) = diag(a^2(t), b^2(t), c^2(t))
\end{equation}
The spacetime has the topology {\bf{R}} $ \times S^3 $
(product of a time axis and the compact three-sphere). The
three-space is invariant under the G=SU(2) group, as
expressed by the
SU(2) invariant one-forms $ \BFACE{\omega}^i(x) \; , \; i = 1,2,3 \; $
which satisfy $d \BFACE{\omega}^i = \epsilon^i_{jk}
\BFACE{\omega}^j \wedge \BFACE{\omega}^k $ where
$\epsilon^i_{jk} $ is the completely antisymmetric tensor of
rank $3$.
In terms of Euler angle coordinates $(\psi, \; \theta, \; \phi)$ on
$SU(2)$ which take values in the range $
0 \leq \psi \leq 4 \pi \; , \; 0 \leq \theta \leq  \pi \; , \;
0 \leq \phi \leq 2 \pi $ (see also MTW (1973), p.808)
we have
\begin{eqnarray} \label{oneformsothree}
\BFACE{\omega}^{1} &=& \cos \psi \; d \theta +\sin \psi \;
\sin \theta \;d \phi  \nonumber   \\
\BFACE{\omega}^{2} &=& \sin \psi \; d \theta -\cos \psi
\; \sin \theta \;d \phi    \\
\BFACE{\omega}^{3} &=& d \psi + \cos \theta \;d \phi    \nonumber
\end{eqnarray}
Written out in terms of the coordinate differentials
$d\psi, \; d\theta, \; d\phi$
we get for the line element of the spacetime metric (\ref{metric})
\begin{eqnarray}
ds^2 &=& -dt^2 +  c(t)^2 \; d\psi^2 +
(a(t)^2 \cos^2 \psi + b(t)^2 \sin^2 \psi) d\theta^2    \nonumber \\
&+&  \left\{ \sin^2 \theta (a(t)^2 \sin^2 \psi + b(t)^2 \cos^2 \psi) +
c(t)^2 \; \cos^2 \theta  \right\} \; d\phi^2         \\
&+& (a(t)^2 - b(t)^2) \sin 2\psi \; \sin \theta \; d\theta \; d\phi +
2 c(t)^2 \; \cos \theta \; d\psi \; d\phi   \nonumber
\end{eqnarray}
This is the toy-model spacetime metric (with $a(t), b(t), c(t)$,
the three scale-functions, as degrees of freedom)
which we want to evolve
on approach to the ``big crunch'' space-time singularity
where the three-volume of the metric collapses to zero.

The space is closed and the three-volume (Landau and Lifshitz p.390)
of the compact space is given by
\begin{equation} \label{volume}
V = \int \sqrt{-g} \; d \psi \; d\theta \; d\phi =
\int\sqrt{\gamma} \; \BFACE{\omega}_1
\wedge\BFACE{\omega}_2\wedge \BFACE{\omega}_3  = 16 \pi^2 \; abc
\end{equation}
When $a=b=c=R/2 \;$ the space reduces to a
space of constant positive curvature with radius of curvature $R=2a$,
which is the metric of highest symmetry
on the group space $SU(2)$. The volume (\ref{volume}) then reduces
to the three-volume
$V = 2 \pi^2 R^3(t)$ of the compact (isotropic)
{\em Robertson-Walker} space.
If we couple the gravitational field to perfect fluid matter,
the cosmological model with the ansatz (\ref{metric}) for the metric is
an anisotropic generalization of the well known compact FRW model: It
has different ``Hubble-constants" along different directions in the
three-space. One may also interpret
the metric (\ref{metric}) as a closed FRW universe on which is superposed
circularly polarized gravitational waves with the longest wavelength that
will fit into a closed universe (cf.\ D.H. King (1991) and references
therein).

Since the metric (\ref{metric}) is spatially
homogeneous, the
full non-linear Einstein equations
for the metric are a set of ordinary (non-linear)
differential equations.
To see this more explicitly,
introduce in place of the quantities $ a,b,c $ , their logarithms
$
\alpha = \ln a \; , \; \beta = \ln b \; , \; \gamma = \ln c
$
and a new time variable $ \tau = \int dt/abc $
in place of the proper (synchronous)
time $ t $ appearing in the metric (\ref{metric}),
cf.\ Landau and Lifshitz (1975), \S 116-119.
With the inclusion of a perfect fluid matter source,
the space-space components of Einstein's equations then reads
\begin{eqnarray} \label{numspacespace}
2 \alpha_{\tau \tau} & = & \frac{d^2}{d \tau^2} (\ln a^2) =
(b^2 - c^2)^2 - a^4
+ 8 \pi G (\rho - p) \; a^2 b^2 c^2       \nonumber \\
2 \beta_{\tau \tau} & = & \frac{d^2}{d \tau^2} (\ln b^2) =
(c^2 - a^2 )^2 - b^4
+ 8 \pi G (\rho - p) \; a^2 b^2 c^2        \\
2 \gamma_{\tau \tau} & = & \frac{d^2}{d \tau^2} (\ln c^2) =
(a^2 - b^2)^2 - c^4
+ 8 \pi G (\rho - p) \; a^2 b^2 c^2       \nonumber
\end{eqnarray}
and the time-time component reads
\begin{equation}  \label{numtimetime}
(\alpha + \beta + \gamma)_{\tau \tau}
- 2(  \alpha_{\tau} \beta_{\tau} + \alpha_{\tau} \gamma_{\tau}
+ \beta_{\tau} \gamma_{\tau}) =
- 4 \pi G (\rho + 3p) \; a^2 b^2 c^2
\end{equation}
The quantities $p$ and $\rho$ denote the pressure and the energy
density of the fluid. \\
Adopting the standard ansatz that a barotropic equation
\begin{equation} \label{barotropic}
p = (\gamma - 1) \rho
\end{equation}
($\gamma$ constant)
relates the two quantities one may easily show that the equations of motion
(\ref{numspacespace}),(\ref{numtimetime}) have a first integral
\begin{equation} \label{numfirstintegral}
\tilde{I} = I - 8 \pi G \rho \; a^2 b^2 c^2 = 0
\end{equation}
where
\begin{equation} \label{Ifirstintegral}
I =
\alpha_{\tau} \beta_{\tau} + \alpha_{\tau} \gamma_{\tau}
+ \beta_{\tau} \gamma_{\tau} -
\frac{1}{4} ( \; a^4 + b^4 + c^4 \;)
+ \frac{1}{2} (\; a^2 b^2
+ a^2 c^2 + b^2 c^2 \; )
\end{equation}
To be in accordance with the full set of Einstein equations the solution
should have $\tilde{I} = 0$.
The dynamical equations for the
compact FRW cosmology is recovered in the case of $a=b=c=R/2$.

Important astrophysical examples
of the barotropic equation (\ref{barotropic})
are $\gamma = 1, \; 4/3 , \; 2$ corresponding to the cases of
``dustlike'' matter,
``radiation'' matter and ``stiff matter'' respectively.
The energy density scales
with the volume of the space as $\rho \sim V^{-\gamma}$,
see also e.g. Landau and Lifshitz (1975)
or Kolb and Turner (1990).

One may show
(cf.\ also Landau and Lifshitz (1975), p.390)
that sufficiently near the singularity the
perfect fluid matter
terms (appearing on the right hand
side of the equations
(\ref{numspacespace}) and (\ref{numtimetime}))
may be neglected if the equation of state
$ p \leq 2/3 \rho $.
Thus, it is sufficient to investigate the
``empty space equation'' (the vacuum Einstein equations)
$ \; R_{\mu \nu} = 0 \;$
even if ``dust" ($\gamma = 1$ in equation (\ref{barotropic}))
and ``radiation" fluids
($\gamma = 4/3$)
are included in the
mixmaster big crunch collapse: One
says that the mixmaster cosmology is ``curvature dominated'' in the region
sufficiently near the space-time singularity! In physical terms this means
that sufficiently near the space-time
singularity the self-interaction of the
gravitational field completely dominates the dynamical evolution and
contributions from (non-gravitational) matter may be
neglected in the study.
This conclusion
clearly does not apply to the case of
a ``stiff matter" fluid (where $p = \rho \sim (abc)^{-2}$) and -
of course - does not a priori apply to sources of other physical origin.
Such other material sources (non-Abelian Yang Mills fields, etc.)
might very well couple to and significantly alter the dynamical
structure of the gravitational collapse.

Note, that in the reversed time-direction,
when the mixmaster spacetime metric (\ref{metric})
is evolved away from its singularity
(i.e. as the volume $V = 16\pi^2 abc$ of the
space increases)
the matter terms gradually become more important
and eventually dominate the dynamical evolution of
the mixmaster metric (\ref{metric}) and the matter terms may
lead to isotropization (though, not fast enough to explain the
observed isotropy today, cf.\ e.g.
Doroshkevich, Lukash and Novikov (1973) and Lukash (1983)).

Despite the fact that this metric is widely known,
it is  remarkable that only quite recently
(cf.\ X. Lin and R.M. Wald (1990))
it has been rigorously proven to recollapse (this is also true
in the ``vacuum" case, i.e. when
no perfect fluid, or any other matter source, is included in the model).
So, we know that the mixmaster
spacetime metric
has {\em two} spacetime singularities (like the
compact FRW cosmology):
A ``big bang" and a ``big crunch"!

\subsection{The three-volume of the mixmaster space-time metric
cannot oscillate if evolved according to the vacuum Einstein equations}

\indent The three-volume $V = 16 \pi^2 abc$
of the model-universe cannot oscillate. This
can be shown in several (not truly independent) ways:

One may derive this fact directly from
the  governing set of differential equations
(see S.E. Rugh (1990a) and
S.E. Rugh and B.J.T. Jones (1990)). We note, that statements
on monotonicity of the three-volume are equivalent whether given
in $t$ or in $\tau$ time, since
$ dt = abc \; d\tau \; $  and $ \;  abc > 0. $
The property of $\; \ln V \;$ being a
{\em concave} function (negative second derivative) does
not translate from
$t$ to $\tau$ time. Below we show that $\; \ln V \;$ is a
concave function in the
$t$ time variable (but not in $\tau$ time).

Neglecting, for notational convenience, the factor $16 \pi^2$ in the
expression for the three-volume, we have
$\ln V \equiv \ln a + \ln b + \ln c \equiv \alpha +
\beta + \gamma$,
and the $R_{00}$ equation for the
mixmaster metric reads
\begin{equation}
\frac{1}{2} (\alpha + \beta + \gamma )_{\tau \tau} \equiv
\frac{1}{2} ( \ln V )_{\tau \tau}
\equiv \alpha_{\tau} \beta_{\tau} + \alpha_{\tau} \gamma_{\tau} +
\beta_{\tau} \gamma_{\tau}
\end{equation}
{}From the definition $\partial_{t} = (abc)^{-1} \partial_{\tau}
= V^{-1} \partial_{\tau} $ one arrives at
$
\partial^2_{t} = V^{-2} \left\{ \partial^2_{\tau}  -
(\ln V)' \partial_{\tau} \right\}
$
and hence
\begin{eqnarray}  \label{Vdotdottau}
V^2 \;  \partial_{t}^{2}(\ln V) & = &
V \; \partial_{t}^{2}\;  V - (\partial_{t} \; V )^2 =
(\ln V)'' - ((\ln V)')^2  \nonumber \\
& = & 2 (\alpha' \beta' + \alpha' \gamma' + \beta' \gamma' ) -
( \alpha' + \beta' + \gamma' )^2  \nonumber  \\
& = &  - \alpha'^2 - \beta'^2 - \gamma'^2  \leq 0
\end{eqnarray}
It follows that ($\ln V$) -  and therefore
the volume itself, $V$, can have {\em no local minimum}
(where we should have $\dot{V}=0, \ddot{V} > 0 $).
As a corollary it follows that volume oscillations are
not possible: After a transient
(eventually passing a maximum in the three-volume), the volume of
the three-space should decrease
{\em monotonically} as the mixmaster gravitational collapse evolves
towards the ``big crunch" singularity.

We may, also, consider the Raychaudhuri equation,
A. Raychaudhuri (1955),
which governs an equation for the relative
volume expansion $ \; \Theta = \partial_t (log \left\{
Volume \right\} ) \; $ with respect to the coordinate $ t $ time.
The equation was independently discovered by Lev Landau and
A. Raychaudhuri (see S.W. Hawking \& G.F.R. Ellis (1973), p.84) and
is derived from the Einstein equations for a spacetime metric in
a synchronous reference frame coupled to
co-moving perfect fluid matter.
The general form of Raychaudhuri's equation is:
{\small
$$
\left(
\begin{array}{c}
expansion \\
derivative
\end{array}
\right) = -
\left(
\begin{array}{c}
energy \; \; density          \\
term
\end{array}
\right) \; -  \;
\left(
\begin{array}{c}
shear          \\
term
\end{array}
\right) \; - \;
\left(
\begin{array}{c}
expansion          \\
term
\end{array}
\right) \; + \;
\left(
\begin{array}{c}
vorticity         \\
term
\end{array}
\right)
$$
}
or
\begin{equation}
\dot{\Theta} = - R_{\mu \nu} u^{\mu} u^{\nu} - \sigma^2
- \frac{1}{3} \Theta^2 +  2 \Omega^2 \; \; .
\end{equation}
A dot denotes derivation with respect to the time $ t $.
The term $ R_{\mu \nu} u^{\mu} u^{\nu}
= 4 \pi G (\rho + 3 p ) $  refers to the {\em co-moving}
perfect fluid  source (the nongravitational matter)
included in the cosmological model, whereas
$\sigma^2$ and $\Omega^2$ denote the ``shear" and ``vorticity" scalars
contracted from the shear and vorticity tensors of the metric field
(see, also, S.W. Hawking and G.F.R. Ellis (1973)).
One may show that
the metric (\ref{metric}) has no vorticity $ \Omega^2 = 0 $.
For the quantity
$$ {\cal{G}} \; \equiv \; \sqrt[3]{abc} \;
\propto \; \sqrt[3]{Volume} \; \; , $$
which is related to the relative volume expansion $\Theta$
as
$$ \Theta = \frac{d/dt \left\{ Volume \right\} }{Volume} =
3 \frac{{\dot{\cal G}}}{{\cal G}} \; \; , \; \;
{\dot{\Theta}} + \frac{1}{3} \Theta^2 =
3 \frac{ {\ddot{\cal G}} }{{\cal G}} \; \; , $$
one obtains, after a little algebra, the equation
\begin{equation}
\frac{ {\ddot{\cal{G}}} }{ {\cal{G}} } =
\frac{ d^2/dt^2 (\sqrt[3]{abc}) }{\sqrt[3]{abc} } =
-\frac{4 \pi G}{3} (\rho + 3 p)
- \frac{2}{3} \sigma^2 \leq 0   \; \; .
\end{equation}
Thus, provided $\rho$ and $p$ are not negative, ${\ddot{\cal{G}}}$ is always
non-positive implying that ${\cal{G}}$ cannot have a local minimum
(where one should have ${\dot{\cal{G}}} = 0$ and ${\ddot{\cal{G}}} > 0 $).
I.e., according to Raychaudhuri's equation, the three-volume
for any cosmological model without vorticity $ \Omega = 0 $ (our diagonal
mixmaster toy model collapse (\ref{metric}) belongs to this
class) cannot be oscillatory, but can
only have one maximum like the FRW cosmology. In the empty case,
$ p = \rho = 0 $, the conclusion applies equally well.

\subsection{Volume oscillations as a ``probe" on numerical solutions}

\indent The fact that solutions to the vacuum Einstein equations
should not have
oscillating three volumes is no surprise. The idea,
however, is to use the property of ``no oscillations"
as a ``probe" to examine the
validity of some previous investigations which
were done on this model.
In the references
Zardecki (1983), Francisco and Matsas (1988), and in fact in
Barrow (1982, 1984) and Barrow and Silk (1984),
the volume behavior of the depicted evolutions
is not in accordance with the conclusion arrived
at here.
They have such oscillations and can therefore not be proper
solutions in agreement with the Einstein equations
for positive or zero non-gravitational energy densities.
In fact, some of these models effectively
included {\em stiff} matter with {\em negative} mass densities.
(cf.\ S.E. Rugh (1990a) and
S.E. Rugh and B.J.T. Jones (1990)).
That is,
one may easily show\footnote{In the case of ``stiff matter'',
$p = \rho$, i.e. $\gamma = 2$ in (\ref{barotropic}), and we
have $\rho \sim V^{-2}$. The 11-,22-,33-components of the
Einstein field equations (\ref{numspacespace}) attain
the same form as the vacuum equations. The exception is
the 00-component of the Einstein equations
(\ref{numtimetime}) and thereby the first
integral constraint (\ref{numfirstintegral}).
Since the energy density scales with the three volume as
$\rho \sim V^{-2}$ we put $\rho = \rho_0 (abc)^{-2}$
which gives $I = 8 \pi G \rho_0$. For the
vacuum Einstein equations we ought to have $I \equiv 0$.
However, if initial data fail to satisfy the
zero density constraint, the corresponding solutions act
as if we had included ``stiff matter'' with
$p_0 = \rho_0 = (8 \pi G)^{-1} I \neq 0$.} that vacuum solutions
which fail to satisfy the first integral constraint $I = 0$,
with $I$ given in (\ref{Ifirstintegral}),
correspond to the inclusion of
``stiff matter'' (i.e. a perfect fluid with equation of state
``$p = \rho$'') with energy density
$$ p = \rho = (8 \pi G)^{-1} \frac{I}{a^2b^2c^2} \; \; .  $$
The character of the solutions depends on the sign
of $I$. \\
For $I < 0$ the negative mass densities
create a {\em repulsion effect}, which causes
oscillations in the three volume $ \; V = 16 \pi^2 abc \;$.
In such solutions a typical evolution of the parameters
$ \; a(t),b(t),c(t) \;$ of the metric (\ref{metric}) will
be like in fig.1 where the scale functions
- or rather their logarithm's
$ \alpha = \log a \; , \; \beta = \log b \;, \gamma = \log c $ -
are plotted against the standard (Landau and Lifshitz, Vol. II, \S 118)
time variable $ \tau = \int dt/abc $.

\begin{figure}
\vspace{8.5 cm}
\caption[xxx]{{\small A typical evolution of the logarithmic scale functions
$\alpha = \log a, \; \beta = \log b, \; \gamma = \log c$ (the thick curve
is $\alpha = \log a$)
as a function of $\tau$-time, $\tau = \int dt/abc$, is
rather interesting if we select initial conditions which
have $I < 0$ and thus effectively introduces stiff matter with
negative energy density. Apparently the evolution of the
scale functions of the metric (\ref{metric})
is highly irregular. }}
\end{figure}

\begin{figure}
\vspace{8.5 cm}
\caption[xxx]{{\small
Behavior of the logarithm of the three-volume
$V = 16 \pi^2 abc$ as a function of the
$\tau$-time $\tau = \int dt/abc$
corresponding to the behavior of the scale functions
in fig.1. Due to the effective inclusion of
negative energy densities
(see text) the solutions do not satisfy the reasonable conditions on the
energy momentum tensor required for the
Hawking-Penrose singularity theorems
to apply. The metric does not evolve towards a singularity: The
negative mass density creates a repulsion effect
and, as a result, the three-volume
oscillates in time.  } }
\end{figure}

\noindent
One may also display the ``degree of anisotropy" of
the spacetime metric in the anisotropy
variables (ADM variables, see later)
\begin{equation} \label{ADMbeta}
\vec{\BFACE{\beta}} = (\beta_{+}, \beta_{-}) =
(\; \frac{1}{6} \log (\frac{ab}{c^2}) \; ,
\; \frac{1}{2 \sqrt{3}} \log (\frac{a}{b}) \; ).
\end{equation}
${\vec{\BFACE{\beta}}} = {\vec{0}} $
means no anisotropy, while a huge $ | {\vec{\BFACE{\beta}}} | $
means that our model (this is the case for proper solutions
near the singularity, cf.\ fig.5) is very
anisotropic. In terms of this parameterization the
metric (\ref{metric}) reads, cf., e.g.,
MTW (1973),
\begin{equation} \label{metricADMvariables}
ds^2 = -dt^2 + e^{-2 \Omega} (e^{2 \beta})_{ij} \BFACE{\omega}^i (x)
\BFACE{\omega}^j (x)
\end{equation}
where $\Omega = - \frac{1}{3} \log (abc)$ and
$\beta_{ij}$ is the traceless, diagonal matrix with
the diagonal elements
$\beta_+ + \sqrt{3} \beta_-$,
$\beta_+ - \sqrt{3} \beta_-$ and
$- \beta_+$.
The evolution of the metric is decomposed in {\em expansion}
(volume change parametrized by $\Omega$) and {\em anisotropy}
(shape change parametrized by
$ \vec{\BFACE{\beta}} = (\beta_{+}, \beta_{-}) $).
The trajectories of the ``anisotropy'' of the
(erratic volume oscillating) mixmaster model,
which correspond to the sketched solutions above, fig.1,
is displayed in fig.3 (the $C_{3v}$ symmetry is apparent and
is of course expected from the symmetry under the interchange
$a \leftrightarrow b \leftrightarrow c$ of the scale factors in the
metric (\ref{metric})).

\begin{figure}
\vspace{14.0 cm}
\caption[xxx]{{\small Solutions corresponding to fig.1 but
mapped out in the anisotropy variables
$\vec{\BFACE{\beta}} = (\beta_+, \beta_-)$. Not surprisingly,
this behavior is reflected in positive values of the maximal
characteristic Lyapunov exponent.
}}
\end{figure}

\vspace{0.5 cm}

\noindent
{\bf The mixmaster equations pass the Painlev\'{e} test} \\
It is interesting that Contopoulos \etal (1993) recently have
performed a Painlev\'{e} analysis on the set of
mixmaster equations (\ref{numspacespace}), for $\rho = p = 0$,
and find that the set of
mixmaster equations pass the Painlev\'{e} test.
Apparently, this analysis
does not utilize the additional information from the
first integral constraint $I = 0$. Thus the equations of
motion for the mixmaster space-time metric
also pass
the Painlev\'{e} test for the case $I < 0$ which, according to
Contopoulos \etal, is a strong indication that
the trajectories corresponding to fig.1 and fig.3
(as well as fig.4 and fig.5 for $I=0$) are
integrable, i.e. that two additional
{\em constants of motion (symmetries)} besides the Hamiltonian
can be found. If an integrable system could produce an orbit
like the one in fig.3 it would be surprising. We have previously
searched for such additional integrals.\footnote{F.Christiansen,
H.H.Rugh and S.E.Rugh, unpublished investigations}
Our results, so far, indicate the lack of existence of
such additional integrals in the equations of motion
for the case $I < 0$.

\subsection{The mixmaster gravitational collapse
evolved towards the spacetime singularity
under the governing vacuum Einstein equations}

The dynamics of the spacetime metric (\ref{metric})
will be very complicated
(though not as complicated as in figures 1,2 and 3 above)
when evolved according to
the vacuum Einstein equations $ R_{\mu \nu} = 0 $.
The deterministic evolution of the scale factors
$ a(t), \; b(t),\; c(t)$ and their first derivatives is described by the
set of six coupled, first order, differential equations
(\ref{numspacespace})
(we put $p = \rho = 0$) constrained by the first integral
(\ref{numfirstintegral}). Due to the scale invariance
of the Einstein equations there are four degrees of freedom
in the problem.
We may distinguish between
solutions which are axisymmetric (an integrable case)
and solutions without any axisymmetry.

With rotational invariance about one axis\footnote{Note that the FRW solution,
being invariant under rotations about any axis in
the SU(2) homogeneous 3-space (i.e. isotropic) is not obtained from the
Taub solution by putting a=b=c. The Taub solution is a
{\em vacuum} solution whereas
the FRW solution is a solution with matter (a perfect fluid) in the model.
}   (a=b, say) in the SU(2) homogeneous
3-space, we yield
for the metric (\ref{metric})
the special case of the Taub spacetime metric
of the form
$$ ds^2 = -dt^2 +
a^2(t) (  (\BFACE{\omega}^{1})^2 + (\BFACE{\omega}^{2})^2  ) +
c^2(t) (\BFACE{\omega}^{3})^2 \; \; \; ,$$
and in this axisymmetric case the vacuum equations admit
the {\em exact} solution
(given in A.H.Taub (1951))
\begin{equation} \label{Taub}
a^2=b^2=\frac{p}{2} \frac{\cosh(2p \tau +
\delta_1)}{\cosh^2(p \tau + \delta_2)}
\; ,
\; \; c^2= \frac{2p}{\cosh(2p \tau + \delta_1)} \; ,
\end{equation}
where $ p, \delta_1, \delta_2 $ are constants.
However, this axisymmetric solution is {\em unstable}
with respect to small perturbations in the
parameter space of scale functions $( a,b,c,... )$,
see also C.W.Misner (1969).
For a numerical investigation of the Taub solutions, see
C. Behr (1962).

The dynamical behavior for a typical mixmaster gravitational
collapse (without additional symmetries)
is displayed in fig.4 and fig.5 where
the three axes $(a,b,c)$
are followed as a function of the
standard Landau and Lifshitz time coordinate
$ \tau = \int dt/abc \rightarrow \infty $.
Near the singularity where the scale functions collapses to
zero $a,b,c \rightarrow 0$
we prefer to display the logarithm
$ \; \alpha = \log a, \; \beta = \log b, \;
\gamma = \log c $ of the scale functions.
The ``anisotropy" of the toy-model grows without limit
on approach to the big crunch singularity\footnote{If the
anisotropy-variables $\vec{\BFACE{\beta}}$, calculated
according to (\ref{ADMbeta}),
are mapped out corresponding to the
long $\tau$ time behavior of the
$\alpha , \beta , \gamma$ variables depicted in fig.5,
they will be pieces of
straight line segments corresponding to bounces from outward
expanding (almost infinitely steep) potential walls.
It is like a game of
billiards played on a triangular shaped table with outward
expanding table boundaries, with a ball which moves faster than the
outward expanding boundaries (cf., also,
displayed figures of this behavior in S.E. Rugh (1990a) and
B.K. Berger (1990)).}

We have selected a set of reference initial conditions as in
A. Zardecki (1983),
but have adjusted the value of $c'$ to make the first integral vanish
to {\em machine precision}.
(Such an adjustment is indeed  necessary. Cf. discussions in
S.E. Rugh (1990a) and D. Hobill \etal (1991)).
This yields the starting conditions
\begin{eqnarray} \label{startingconditions}
&a& = 1.85400.. \; \, \; b = 0.438500.. \; , \; c = 0.085400..  \nonumber \\
&a'&= -0.429200..  \; , \; b' = 0.135500.. \; , \;
c' = 2.964843279......
\end{eqnarray}
All integrations have been performed by a
standard {\em fourth order Runge Kutta} algorithm,
and each calculation takes less than one CPU-minute.

We distinguish between two qualitatively different
phases of the evolution of the spacetime metric (\ref{metric}) towards
the ``big crunch'' singularity:

\begin{quotation}
\noindent
{\em Transient behavior:}
The model cosmology
trajectory ``catches up" its initial
conditions, and eventually passes a maximum in three-volume. The volume then
begins to decrease monotonically. (Fig.4) \footnote{If we want a more
realistic model cosmology at this stage, matter terms should be included
here, since the omission of matter contributions for the
mixmaster toy-model cosmology is only justified
when we are sufficiently near the ``big crunch" singularity.} \\

\noindent
{\em Asymptotic behavior near the singularity}:
The evolution towards the ``big crunch" singularity
is fast attracted\footnote{Note that the sequence of
oscillations of the scale functions, as shown in fig.5,
is an attractor for almost any initial condition,
but not initial conditions with axisymmetry.}
into an infinite
sequence of oscillations of the three scale functions (fig.5).
We may identify the Kasner segments (Kasner epochs)
between each ``bounce'', and the combinatorial model
by Belinskii \etal describes well the
transitions from one Kasner segment to the next. (Table 1).
\end{quotation}

\begin{figure}
\vspace{15.0 cm}
\caption[xxx]{
{\small Transient evolution of the three scale functions of the
mixmaster spacetime metric governed by
the {\em vacuum} Einstein equations with starting conditions
(\ref{startingconditions}). We have displayed  the ``{\em transient}"
$\tau$ time interval $0 \leq \tau = \int dt/abc  \leq 1$.
(From S.E. Rugh (1990 a)).
The three-volume has a maximum at $\tau \approx 1/5$ and
then decreases monotonically
(turning into an oscillatory behavior of
the scale functions as shown in fig.5)
towards the big crunch singularity. (It is interesting, and
surprising, if this displayed behavior should turn out to be
integrable, as suggested by Contopoulos \etal (1993)).
}}
\end{figure}

\begin{figure}
\vspace{18.0 cm}
\caption[xxx]{{\small Asymptotic evolution of the
three scale functions (typical figures from a numerical experiment
described in S.E. Rugh (1990a)):
The axes $\alpha = \ln a, \; \beta = \ln b, \; \gamma = \ln c $ of the
toy-model gravitational collapse, given by the
metric (\ref{metric}), is followed in $\tau$-time
($\tau = \int dt/abc$)
towards the ``big crunch" singularity
under the governing {\em vacuum} Einstein equations.
The spacetime singularity is reached in
{\em finite} $t$ time,
but in $\tau$ time the dynamical evolution is stretched out
to infinity $\tau \rightarrow \infty$.
The three-volume of the space shrinks monotonically to zero,
the anisotropy of the model grows without limit and the Weyl curvature
$C^2 = C_{\alpha \beta \gamma \delta} C^{\alpha \beta \gamma \delta}$
also diverges on approach to the spacetime singularity.
}  }
\end{figure}

\noindent
{\bf Extracting lower dimensional signals from the
gravitational collapse} \\
In an ever expanding phase space (cf., also, discussions later)
it is natural to try to {\em ``project out''}
some lower dimensional (compact) signal\footnote{It
is not uncommon to ``probe''
chaos in dynamical flows which take
place in higher dimensional phase spaces by extracting
lower dimensional time-signals from the flow. (E.g.
probing turbulence and complexity in Navier-Stokes
hydrodynamical flows by measuring time signals
of a temperature probe, placed at some given space point in the fluid).
Extraction of a single physical variable $\xi(t)$ with ``chaotic'' behavior
(a ``time'' signal or a discrete map $\xi_n$)
mirrors ``chaos'' in the full phase
flow.}, say, which captures some of the
{\em recurrent} ``chaotic'' properties of the model.
The dynamical behavior of the toy-model gravitational
collapse is indeed in certain aspects chaotic, as captured
by the parameter
\begin{equation}
u = \frac{min \left\{ \alpha', \beta', \gamma' \right\} }{
\alpha' + \beta' + \gamma' - (min \left\{ \alpha', \beta', \gamma' \right\} +
max \left\{ \alpha', \beta', \gamma' \right\} ) }
\end{equation}
Extracting this parameter ``u'' (the so-called ``Lifshitz-Khalatnikov"
parameter ``$u \in$ {\bf R}",
see also Landau and Lifshitz, \S 116-119) from the
displayed trajectories we get table 1.
We find that the gravitational
collapse is {\em extraordinarily well}
described by
the {\em BKL}-combinatorial model which is summarized
below, following O.I.Bogoyavlensky and S.P.Novikov (1973).
``{\em BKL}'' refers to the originators of this combinatorial
description of the behavior of the scale functions:
V.A. Belinskii, I.M. Khalatnikov and E.M. Lifshitz.
See also I.M. Khalatnikov et al. (1985)
and references therein.\footnote{The accordance with
the {\em BKL}-combinatorial model
for the dynamical evolution of the mixmaster collapse
has been further investigated in the
work of B.K. Berger, cf., e.g., Berger (1993).
For a description of the more complete 4-parameter map, derived
by Belinskii \etal , see also D. Chernoff and J.D. Barrow (1983). }

\begin{table}[ht]
\begin{center}
\begin{tabular}{lll}
\topline
$\tau = \int dt/abc$ & $u$ & $1/u$  \\
\midline
1 & 5.564816 &   \\
6 & 4.564816 &   \\
16 & 3.564815 &  \\
27 & 2.564816 &  \\
48 & 1.564816 &  \\
96 & 1.770488 & 0.5648160 \\
305 & 1.297878 & 0.7704884 \\
4700 & 3.357077 & 0.2978782  \\
50000 & 2.357077 &  \\
180000 & 1.357077 &  \\
500000 & 2.800516  & 0.3570771  \\
3000000 & 1.800510  &  \\
19000000 & 1.249229 & 0.8004938  \\
75000000 & 4.013535 & 0.2491569  \\
\bottomline
\end{tabular}
\end{center}
\caption[xxx]{
{\small This table summarizes the evolution of numerically
extracted values of
the Lifshitz-Khalatnikov parameter ``u'' corresponding to
solutions of the vacuum Einstein equations as those depicted in
fig.5. The description of the evolution of the ``u'' parameter
in terms of the ``Farey map'' or ``Gauss map'' is
in complete agreement with this table.
{}From S.E. Rugh (1990a), p.98. (The $\tau$-time values
offered in the table correspond to $\tau$-times when the collapse
orbit is well beyond a bounce and has reached a new
straight line behavior (``Kasner epoch'') in fig.5). }  }
\end{table}

With the parametrization
$$ (p_1(u),p_2(u),p_3(u)) = ( -u, 1+u , u(1+u) ) / (1 + u + u^2) $$
the {\em BKL}-piecewise approximation
of the scale functions $a,b,c$ by the power law
functions (Kasner epochs)
$t^{p_1},\; t^{p_2}, \; t^{p_3}, \; p_{i} = p_i(u),$
is (on approach to the spacetime singularity)
described by the sequence of states
\begin{equation}
(u_0,\sigma_0) \rightarrow (u_1,\sigma_1) \rightarrow
(u_2,\sigma_2) \rightarrow ....
\end{equation}
where the {\em BKL}-``Kasner state" transformation
(``alternation of Kasner epochs") is given by
\begin{equation}  \label{BKLuBN}
\left\{ \begin{array}{ll}
(u,\sigma) \rightarrow (u-1, \; \sigma \sigma_{12}) & (2 \leq u < \infty) \\
   & \\
(u,\sigma) \rightarrow (1/(u-1), \; \sigma \sigma_{12}
\sigma_{23}) & (1 \leq u \leq 2)
\end{array}
\right.
\end{equation}
the ``Kasner state" being described by the pair
\begin{equation}
(u, \sigma) \; \; \;;
\sigma = \left(
\begin{array}{ccc}
1 & 2 & 3 \\
i & j & k
\end{array}
\right) \; , \;
\sigma_{12} = \left(
\begin{array}{ccc}
1 & 2 & 3 \\
2 & 1 & 3
\end{array}
\right) \; , \;
\sigma_{23} = \left(
\begin{array}{ccc}
1 & 2 & 3 \\
1 & 3 & 2
\end{array}
\right)  \; ,
\end{equation}
and $\sigma$ denoting the permutation of the three Kasner axes.

\subsection{The (compressed and stretched)
astronomer who falls into the spacetime singularity}

A {\em `freely falling astronomer'}
who falls into the mixmaster spacetime singularity
will experience a tidal field, in which
he is {\em compressed} along two directions and
{\em stretched} (expanded) in
one. The directions of these gravitationally induced stresses are
permuted infinitely many times (in a not-predictable way!)
on approach to the space time singularity.
A picture in Kip S. Thorne (1985) sketches
the tidal gravitational forces felt by an observer (an astronomer)
who falls into the
singularity. Such tidal forces are produced by spacetime
curvature.  The astronomer, who in this
example makes up the tidal field instrument
(see, e.g., MTW, p.400-404),
feels, in his local inertial frame, tidal accelerations
given by the equation of geodesic deviation,
$ d^2 \xi_j / dt^2 = R_{joko} \; \xi^{k}\; $, where
$\xi$ is the separation vector between two freely falling test particles
(two reference points in the body of an astronomer, falling freely along
geodesics - if we neglect
internal elastic forces in the body of the observer
(justified, if the  spacetime curvature is big?).
There is in the local inertial frame of the infalling observer a preferred
choice of coordinate axes ($i=1,2,3$) which diagonalizes the tidal field.
In terms of these ``principal axes" the component $R_{ioio}$ produces,
according
to the equation of geodesic deviation,
$ \ddot{\xi}_i / {\xi}_i = - R_{ioio} $,
a tidal {\em compression} or {\em stretching} along
direction ``$i$", depending on whether
$R_{ioio}$ is positive or negative.

One may swindle a bit and write down the
Riemann curvature components of the Kasner-segments
as if they were given by the Riemann curvature
of the Kasner metric (instead of the Riemann tensor components of the
spacetime metric (\ref{metric})).
The Riemann tensor components of the Kasner metric
reads
$R_{ioio} = -p_i \;(p_i - 1)\; t^{-2} \;, i=1,2,3$,
and we get, in terms of the Lifshitz-Khalatnikov parameter
$u \in \BFACE{R}$,
the following expressions for the tidal stresses:\\
Along the two axes of compression
$$ \frac{\ddot{\xi}}{\xi} = -\frac{u(u+1)}{(1+u+u^2)^2} \;
t^{-2} \; \; \;
\mbox{resp.} \; \; \;
\frac{\ddot{\xi}}{\xi} = -\frac{u^2 (u+1)}{(1+u+u^2)^2}
\; t^{-2}  \; \; . $$
Along the axis of expansion
$$ \frac{\ddot{\xi}}{\xi} = +\frac{u(u+1)^2}{(1+u+u^2)^2} \;
t^{-2} \; > \; 0 \; \; . $$
The tidal stresses grow up like $\sim t^{-2}$ where
$t$ denotes the {\em finite} time distance (as measured by the
infalling observer) until the spacetime singularity is reached.
The successive shifts
$ \; u \rightarrow u - 1 \; \& \; C \; $
in the parameter ``u",
are governed by the combinatorial model (\ref{BKLuBN}).
During each ``era'' (comprised of cycles
$u \rightarrow u - 1$ until $u$ reaches the value
$1 \leq u \leq 2$) one of the principal axes experiences
a continual tidal compression while the other two principal
axes oscillate between compression and stretch. At a
change of ``era'' (i.e. when $u \rightarrow 1/(u-1)$
for $1 \leq u \leq 2$) there is a change in the
role of the axes (another of the three principal
axes experiences a continual tidal compression while
the other two axes oscillate). Note, that
``real'' {\em physical} quantities, measurable by
{\em tidal field instruments}
exhibit chaotic oscillations which condense
infinitely on approach to the spacetime
singularity at $t \rightarrow 0$.

The astronomer may feel a little ``worried'', being compressed like this
in two directions and stretched (expanded) in one. The directions of
these gravitationally induced stresses are even permuted
infinitely many times on approach to the spacetime singularity.

\subsection{The ``Farey map'' with ``strong intermittency''
encodes the ``per bounce'' dynamics of the mixmaster
gravitational collapse}

The {\em unpredictability} of the mixmaster gravitational collapse
does not originate from the oscillations of the scale functions
(as described by $u \rightarrow u-1$) {\em within}
a given major cycle (an ``era'' of oscillations), but
rather from the {\em shifts} between major cycles (described by
$u \rightarrow 1/(u-1), \; 1 \leq u \leq 2$).
These shifts give rise to the
(highly chaotic) Gauss map, cf.\ e.g. J.D. Barrow (1982).
The Gauss map is  well known from ``chaos theory"
(cf., e.g., R.M. Corless \etal (1990)
and references therein)
and acts as a {\em left shift} on the continued fraction representation
of numbers on the unit interval.
The Gauss map has positive
Kolmogorov entropy, $ \; h = \pi^2 / 6 (\ln 2) \gg 0 \; $ and
it has the Bernoullian property\footnote{The Gauss map is Bernoullian
in the sense that it may be extended to a two-dimensional
invertible map which is isomorphic to a Bernoulli shift with
an infinite alphabet.} and, in that sense, it is as
random as that of flipping coins (or a roulette wheel).

A map, which describes the ``per bounce" evolution
of $u$ (and not, merely, the ``per major cycle" evolution of the $u$
parameter) is easily found
\footnote{H.H. Rugh and S.E. Rugh (1990),
unpublished. See also B.K. Berger (1991).}
and is known as the ``Farey map":
Make the substitution $ u = 1/x $ in (\ref{BKLuBN})
and write this map in terms of
the parameter $x \in ]0,1]$. This yields the following map on the unit
interval (the ``Farey map''\footnote{cf., e.g.,
M. Feigenbaum (1988) or
Artuso, Aurell and Cvitanovic
(1990), p.378, and references therein.})
\[ x \rightarrow {\cal F}(x) =  \left\{ \begin{array}{ll}
x/(1-x)  & \mbox{if $0 \leq x \leq  1/2$}  \\
(1-x)/x  & \mbox{if $1/2 \leq x \leq 1$}
\end{array}
\right. \]
Whereas the Gauss map has an {\em infinity} of branches
(cf., e.g., pictures in J. Barrow (1982) or R.M. Corless \etal (1990))
the Farey map has only
two branches (the left, $x \leq 1/2$, and the right)
and thus a natural {\em binary} symbolic dynamics (cf.\ sec.3.4)
with a binary alphabet
(which corresponds more directly
to a symmetry-reduced binary symbolic dynamics encoding
of the geodesic motion on the triangular
billiard on the Poincar\'{e} disc).\footnote{The Farey/Gauss
map is closely related to the symbolic description of
geodesic flows on so-called modular surfaces (found by Artin) on the
Poincar\'{e} disc, see e.g. T. Bedford \etal (1991), and
therefore a very nice connection exists (cf.\ also
J. Pullin (1990)) between the Farey/Gauss map encoding of the
mixmaster gravitational collapse and the description of
the mixmaster collapse orbit on the Poincar\'{e} disc.}
The Farey map contains the Gauss map as a
sub-map, since one iteration of the Gauss map corresponds to a transition
from the right branch via oscillations in the left branch back to the right
branch.

\begin{quotation}
\noindent
{\em As regards the evolution of the
Lifshitz-Khalatnikov parameter ``$u$",
the Farey map takes into account both chaotic and ``non-chaotic" segments
of the one-perturbation BKL-combinatorial model for the gravitational
collapse. The watch of the ``Farey map" ticks one step
forward (one iteration of the map) for each bounce against a wall, i.e.
for each oscillation of the scale functions.}
\end{quotation}

\noindent
The Farey map has a {\em marginally stable}
fixed point at the left end ($|{\cal F}'(0)| = 1$).
This has an important influence on the instability
properties of the map. Intuitively, the marginally
stable point of the Farey map at $x= 1/u = 0$ corresponds to
major cycles, containing an ``infinite" number of oscillations
(governed by the $u \rightarrow u-1$ rule), i.e.,
trajectories which penetrate deeply into one of the three corner
channels.

In general, a map ${\cal F}$ with the asymptotic expansions
$ \; {\cal F}(x) \approx x + a x^{\zeta} + ... \; $
(towards the marginally stable point at $x=0$)
and $ \; {\cal F}(x) \approx 1 - b |1 - 2x|^{1/{\alpha} } + ... \;$
(around the top point $x=1/2$) will have an invariant measure
$\; \mu (x) \;$, which has the
asymptotic behavior\footnote{See, e.g.,
Z. Kaufmann and P. Sz\'{e}pfalusy
(1989) or P. Sz\'{e}pfalusy and G. Gy\"{o}rgyi
(1986).}
$$ \mu (x) \sim x^{\eta} \; \; ; \; \; \eta = \alpha + 1 - \zeta $$
near the origin $x=0$.
The situation $\eta > 0$ is then referred to a ``weak intermittency''
(there is a smooth invariant measure and the Lyapunov exponent is
positive, $ h = \lambda = \int \ln |{\cal F}'(x)| \mu (x) dx > 0 $)
while $\eta < 0$ implies ``strong intermittency''
(there is no normalized, smooth measure and $h = \lambda = 0$).
As regards the Farey map we have
$\alpha = 1, \zeta = 2$ ; i.e. $\eta = 1+1-2 = 0 $ which is the borderline
case. However, the Farey map is an example of strong
intermittency.
By direct inspection one may verify that $\mu (x) = 1/x $ is an invariant
measure. This measure is non-normalizable, but is normalized to the
$\delta$-function measure $\delta (0)$ at the marginally stable fixpoint
$x=0$. Hence, all the measure is concentrated at the
{\em stable} fixpoint $x=0$ and the averaged Lyapunov exponent
(the K-entropy) is zero.
Although the overall Lyapunov exponent (Kolmogorov
entropy) of the map is zero,
all periodic orbits (which do not include $x=0$)
are unstable and have $\lambda > 0$.

This strong intermittency means that the
stability properties are completely dominated by the
``infinite number of corner oscillations'' fixpoint at $x=0$.
However, we note that all this is within the
{\em one-perturbation} combinatorial model (\ref{BKLuBN})
for the gravitational collapse, and therefore it is not
valid in the regime $x = 1/u \approx 0 $.
Thus, there is a ``cut off" towards the left end of the
Farey map and the good ergodic properties are regained.
A {\em two-perturbation} analysis has to take over and show
that the trajectory then will leave the corner.\footnote{The avoidance of the
so-called ``dangerous case of small oscillations'' (i.e. dangerous for the
one-perturbation treatment in the {\em BKL}-combinatorial model)
has been discussed in the
work by Belinskii et al., cf.\ Khalatnikov et al. (1985) and references
therein. }

\vspace{0.5 cm}

\noindent
{\bf The binary ``Farey tree''. Self-similarity of
the collapse dynamics.} \\
The near-singularity dynamics of the
gravitational collapse, as regards the parameter $x = 1/u$,
can be symbolically represented by the so-called
``Farey tree", which is a construction associated with the Farey map.
The Farey tree is a ``tree of rational numbers'' (the Farey numbers).
In fig.6 each Farey number has been represented by its
continued fraction.
Originally, the tree is constructed by starting with the endpoints of the
unit interval written as $0/1$ and $1/1$ and
then recursively bisecting intervals
by means of ``Farey mediants"\footnote{See p. 23
in Hardy and Wright (1938) and Cvitanovic and Myrheim (1989).}
$ p/q \oplus p'/q' = (p+p')/(q+q') $.
The Farey tree rationals can be generated by backward iterates of
the number 1/2 by the Farey map, i.e., starting at
$x_0^{(0)} = 1/2$
we get the next layer
$x_0^{(1)} = 1/3$ and $x_1^{(0)} = 2/3$
as the two numbers which are mapped
to $1/2$ under the Farey map ${\cal F}$. Generally,
the $2^{n}$ $n$th inverses of $1/2$ (under the Farey map) are precisely
the $n$th layer of the Farey tree.

\begin{figure}
\vspace{8.0 cm}
\noindent
\caption[abcabc]{{\small A continued fraction representation of
the binary Farey tree (encoding the bounce dynamics
of the gravitational collapse). The dotted lines continue a given branch (and
correspond to repeated bouncing between the same to potential walls)
while the full lines make a ``zig-zag" movement down the tree (which
corresponds to a transition to a third wall).
The Farey tree rationals can be generated by backward iterates of the
number 1/2 by the Farey map, or, what is the same, by
interpolating the rationals downwards by means of
the ``Farey mediants". Figure from Cvitanovic and Myrheim (1989).} }
\end{figure}

Iterating downwards the layers of the Farey tree is to follow the evolution
(backwards) in the parameter $u$ and
corresponds to one bounce
against one of the three walls.
Compare with the displayed trees
of symbols in M.H. Bugalho \etal (1986).

A non-chaotic, segment of the type $u \rightarrow u-1$
(a repeated bounce against two given walls)
corresponds to continue along the given branch of the
Farey tree. For example, drifting along the most
left branch of the Farey tree corresponds to going
backwards through the numbers $x = 1/2 \rightarrow 1/3 \rightarrow 1/4
\rightarrow 1/5 \rightarrow ...$
which in the forward direction translates to the segment of evolution
$ \rightarrow 5 \rightarrow 4
\rightarrow 3 \rightarrow 2 \; $ in the $u$ parameter.

The chaotic part (an inversion $u \rightarrow 1/u$)
is associated with a shift in ``direction" of motion down the
tree, that is when we {\em zig-zag} our way down the tree.
For example, the $[1,1,1,1] = [1]^{\infty}$ orbit,
which only contains one oscillation in every major cycle, would here
be described as an infinite zig-zag route
(each zig and each zag consist of one step, interchanging
without end) and the
associated Farey number then converges to the
golden mean $u = (\sqrt{5}-1)/2$. This is the initial starting value of
the parameter $u$ which in the forward direction generates this
orbit.

Each branch of the Farey tree
is {\em similar} to the entire tree. In the gravitational
collapse dynamics governed by the {\em classical} vacuum
Einstein equations there is also
no scale (the vacuum Einstein equations are scale
invariant\footnote{There is no natural length associated with
general relativity ($G,c$) and no natural length associated
with the quantum principle ($\hbar$), individually. The union
of general relativity and ``the quantum'' $(G,c,\hbar$) does
contain a natural length, $L_{Pl} = \sqrt{(\hbar G)/c^3)}$.
(The Planck length).}).
When the evolution starts to go into the
described sequence of oscillations, it does so forever and does not
distinguish the situations where 10 or 10.000 major cycles
have passed.
According to {\em classical} general relativity nothing prevents this
sequence of (self-similar) oscillations from
going on ``forever". \footnote{However, if we put an
initial scale of the gravitational collapse
in ``by hand" (e.g., for instance,
$ L = c/H \sim 10^{28} \; cm$ which is a characteristic
scale in the present universe)
a restriction concerning the validity of the classical field
equations themselves may reduce the number of such oscillations to
be very small before Planck scales are reached and quantum fluctuations
of the metric become important.}

The gravitational collapse (\ref{metric}) thus
``organizes itself'' (is attracted)
into the described (self-similar) never ending
sequence of oscillations of the scale functions
without any strong fine-tuning of initial-conditions
(as regards the amount of matter\footnote{The perfect
fluid matter contributions (in the case of dust
or a radiation fluid)
in the toy-model gravitational collapse become (in the direction towards
the space-time singularity) unimportant at some point
(the collapse is curvature dominated)
and the model evolves into the
scale-free oscillatory (chaotic) collapse dynamics.}
in the model or the initial ``set up'' of the
gravitational degrees of freedom).

It has been stated
by Belinskii \etal that the
(self-similar) oscillatory character is a
generic ``attractor" (locally, i.e. in the neighborhood of
every space point) even in the larger space of
spatially inhomogeneous gravitational collapses.
(See e.g. Belinskii \etal (1982)
and Zel'dovich and
Novikov (1983) \S 23.3).
This claim is, however, a subject of substantial controversy.

\subsection{Stability properties of the metric (evolved towards
and away from the spacetime singularity). Does it isotropize?}

It was demonstrated some time ago, E.M. Lifshitz (1946), by means of
linear perturbation
analysis, that the FRW cosmological solution is
indeed {\em stable} against
{\em local} perturbations of the metric (perturbations over
small regions of the space, i.e. over
regions whose linear dimensions are small
compared to the FRW scale factor $ R $) in the {\em forward} time
direction (away from the singularity),
but {\em unstable} in the direction towards the space time
singularity!

In the ``class of perturbations'', studied here, which one may call
{\em global} anisotropic $SU(2)$ homogeneous perturbations of the
compact FRW solution, the study shows that in the
direction towards the singularity
a small anisotropy will grow up
\footnote{This  also happens in
the well known ``Taub-solution"
which is an axisymmetric special
case ($ a = b, \; c \neq a \; \; \& C.$)
of the {\em vacuum} mixmaster metric.
This Taub solution is integrable (Taub, 1951) and unstable!}
and the gravitational collapse will go into the described BKL-sequence of
oscillations, which is {\em unpredictable},
even if we knew the initial data with an (almost) infinite amount of
precision!

In the direction
{\em away from the singularity} the matter terms get increasingly
important and the mixmaster
cosmological model {\em isotropizes}\footnote{By ``isotropization''
we here mean that the anisotropic spacetime metric
(\ref{metric}), coupled to perfect fluid matter,
evolves into a stage of (quasi) isotropic expansion rate, that is,
the metric (\ref{metric}) in the forward time direction
evolves into a stage where the expansion is approximately uniform in all
directions and is described by Hubble's law. Note,
however, that the curvature
of the three-dimensional space is very anisotropic and, as
a rule, does not isotropize. See also Zel'dovich and Novikov,
\S 22.7.} (cf., e.g., MTW \S 30.3-30.5,
Ya.B. Zel'dovich and I.D. Novikov (1983), chapt. 22.7,
and V.N. Lukash (1983)) - though not fast enough
to explain the remarkable degree of isotropy we see in the Universe
today!
An idea like, e.g., a Guth/Linde {\em inflationary}
phase of the universe model is needed. However, while at
the one hand inflation occurs in anisotropic cosmological models
under a wide variety of circumstances, there are actually
some difficulties in inflating the compact mixmaster model universe,
if the initial anisotropy is too large, cf., e.g., sec.
6.2 in the review by K.A. Olive (1990).

\section{CHAOTIC ASPECTS OF THE GRAVITATIONAL COLLAPSE (POSITIVE
LYAPUNOV EXPONENTS AND ALL THAT)}

Can we assign an invariant meaning to chaos in the
general relativistic context? It will be clear in the following
that this program of research is still in its infancy and
to pursue this question will require a good deal of technical
apparatus in general relativity.

A standard probe of chaos for dynamical systems with few degrees
of freedom is to look at the spectrum of Lyapunov exponents,
in particular the principal (maximal) Lyapunov exponent
defined in the following way
(J.P. Eckmann and D. Ruelle (1985)): Given a flow
$f^t: {\cal M} \rightarrow {\cal M}$ on a manifold ${\cal M}$
and a metric (a norm) $ || \cdot || $ on the tangent
space $T {\cal M}$, we define for $x \in {\cal M}, \delta x \in
T_x {\cal M}$,
$$ \lambda (x, \delta x) = \lim_{t \rightarrow \infty}
\frac{1}{t} \log || (D_x f^t) \delta x || \; \; .$$
For any ergodic measure $\rho$ on ${\cal M}$ this quantity
for $\rho$-almost all $x \in {\cal M}$ and almost
all $\delta x \in T_x{\cal M}$ defines the principal
(maximal) Lyapunov exponent for the flow $f^t$ w.r.t. the
ergodic measure $\rho$ (and is independent of $x$ and
$\delta x$).

Note, that the maximal Lyapunov exponent, and more generally the spectrum
of Lyapunov exponents, extracted from non-relativistic
{\em Hamiltonian}
flows is {\em invariant} under
non-singular {\em canonical} coordinate transformations
which do not involve transformations of the time coordinate
(see also e.g. H.-D. Meyer (1986)).

In the general relativistic context we may
identify some obstacles for this construction of a
``Lyapunov exponent'' (e.g. for the mixmaster gravitational
collapse, the orbits being three-metrics, ${}^{(3)} g_{ij}$,
evolving in ``time'' towards
the ``final crunch''): \\

{\bf (1)} {\em What should we choose as a
distance measure ``$|| \cdot ||$'' on the solution space? }

{\bf (2)} {\em What should we choose as a time parameter?  }

{\bf (3)} {\em Is there a natural ergodic measure on the solution space?}

\vspace{0.5 cm}

At first, it is natural to try to treat the evolutionary
equations (\ref{numspacespace}) for
the mixmaster metric (\ref{metric}) as a set of ordinary differential
equations on {\em equal footing} with other dynamical systems
governed by some set of ordinary differential equations
and apply the standard probes of ``chaos'' available,
in particular the standard methods for extracting Lyapunov exponents.

\begin{figure}
\vspace{7.0 cm}
\caption[xyzxyz]{{\small
What is a {\em natural} distance measure
$ \; || {}^{(3)} g - {}^{(3)} \tilde{g} || \;  $
between two
nearby space-time metrics ${}^{(3)} g$ and ${}^{(3)} \tilde{g} =
{}^{(3)} g + \delta {}^{(3)} g$,
where $g$ and $\tilde{g}$ are both solutions to the
vacuum Einstein equations but correspond to slightly
different initial conditions, say?
In the case of non-relativistic Hamiltonian dynamical systems,
measures of ``chaos'' such as principal Lyapunov
exponents are calculated by using an {\em Euclidean} distance
measure which is naturally induced from the structure of the kinetic energy
term in the non-relativistic Hamiltonian.
In the general relativistic context, however, it is not obvious
why one should use an
Euclidean distance measure to assign a distance
between two three-metrics.
Do we need such a
distance measure on the space of mixmaster collapses to talk about
chaos? }}
\end{figure}

We decompose the
equations (\ref{numspacespace})
into the form of first order differential equations,
$ \vec{\BFACE{x}}' = \vec{\BFACE{f}} (\vec{\BFACE{x}}) ,\;  \vec{\BFACE{x}} \in
\BFACE{R}^6 $.
Write, e.g., the vacuum
equations (\ref{numspacespace}) as
$ 2 \alpha '' = (e^{2 \beta} - e^{2 \gamma})^2 -  e^{4 \alpha}$
(and cyclic permutations)
and get for the 6-dimensional state vector
\begin{equation}
\vec{\BFACE{x}} = \vec{\BFACE{x}}(\tau) =
(\alpha, \beta, \gamma, \alpha' , \beta' , \gamma')
\end{equation}
the coupled {\em first order} equations
\begin{equation}  \label{decompone}
\frac{d}{d \tau}
\left(
\begin{array}{c}
\alpha \\
\beta \\
\gamma \\
\alpha' \\
\beta' \\
\gamma'
\end{array}
\right)
 =
\left(
\begin{array}{c}
\alpha' \\
\beta'  \\
\gamma'  \\
\frac{1}{2} (e^{2 \beta} - e^{2 \gamma})^2 - \frac{1}{2} e^{4 \alpha} \\
\frac{1}{2} (e^{2 \gamma} - e^{2 \alpha} )^2 - \frac{1}{2} e^{4 \beta} \\
\frac{1}{2} (e^{2 \alpha} - e^{2 \beta})^2 - \frac{1}{2} e^{4 \gamma}
\end{array}
\right)
\end{equation}
which are supplemented with the first integral
constraint $I = 0$ with $I$ given in
(\ref{Ifirstintegral}) as a constraint on the
initial state vector for the gravitational collapse.   \\
Calculations of Lyapunov exponents were in
all previous studies
(cf., e.g., Burd et al. (1990),
Hobill \etal (1991), Berger (1991),
Rugh (1990a), see also Rugh and Jones (1990))
based on the use of an Euclidean distance measure
of the form,
\begin{equation} \label{EuclidianIntroduction}
|| \BFACE{\vec{x}}(\tau)-\BFACE{\vec{x}}^{*}(\tau) || =
\sqrt{ \sum_{i=1..6} (x_{i}(\tau) - x^{*}_{i}(\tau) )^2  }
\end{equation}
assigning a distance between the two state vectors
$\BFACE{\vec{x}}(\tau)$,$\BFACE{\vec{x}}^{*}(\tau) \in \BFACE{R}^6$
in the phase space.\footnote{What is equivalent Lyapunov exponents were
extracted from calculating the
Jacobian matrix from the flow equations (\ref{decompone}) of
the state vector $(\alpha, \beta, \gamma,
\alpha', \beta', \gamma')$ and integrating up along the orbit. Cf. e.g.
appendix A in S.E.Rugh (1990a).}

This choice of Euclidian distance measure is directly inspired
from the study of ordinary (non-relativistic) Hamiltonian
systems.
However, in general relativity there is a priori
no reason why one should use such a distance measure
to give out the distance between ${}^{(3)} g$ and a
``nearby'' three-metric
$ {}^{(3)} g + \delta {}^{(3)} g$.
A distance measure like (\ref{EuclidianIntroduction})
is gauge dependent, i.e.
{\em not invariant} under a change of
coordinates, $g \rightarrow \tilde{g} (g)$.
(Should the distance between two different ``mixmaster collapses''
(at some given time) be an artifact of the
chosen set of coordinates for the description of the
collapse?).\footnote{Being somewhat
acquainted with general relativity and its
Hamiltonian formulation, a more
natural choice may well be a distance measure
which is induced from the structure of the Hamiltonian in general
relativity (just like the Euclidian distance measure
(\ref{EuclidianIntroduction}) is induced from the Hamiltonian
(the kinetic term) in ordinary {\em non-relativistic} Hamiltonian dynamics),
i.e. a distance measure of the type
$
ds^2 = \int d^3 x G_{ijkl} \delta g^{ij} \delta g^{kl}$
where $G_{ijkl}$ is the Wheeler superspace metric on the space
of three metrics. However,
such a distance measure is not positive definite. (Sec.3.8).
}

\vspace{0.5 cm}

If the maximal Lyapunov exponent
is extracted from the flow equations (\ref{decompone})
for the mixmaster collapse with a Euclidian
distance measure
(\ref{EuclidianIntroduction}) on the solution space we
obtain the following result (from S.E. Rugh (1990a)):  \\

\noindent
{\bf In $\tau$-time}:
With respect to the time parameter
$ \tau = \int dt/abc  \; \sim \ln t  $
introduced as a standard time variable (cf.\ Landau and
Lifshitz (1975) \S 118)
for the description, the approach to the
singular point is {\em stretched} out to infinity ($\tau \rightarrow \infty$).
Despite the mentioned aspect of chaotic unpredictability of the model,
as described by the Farey\footnote{We have seen that the
Farey map comes closer to mimic the bounce
dynamics in terms of a simple, non-invertible one-dimensional map and it
has $\lambda = 0$. Moreover, the $\tau = \int dt/abc$ time
between each bounce grows very fast, so it is no surprise to find
$\lambda_{\tau} (\alpha , \beta, \gamma,..) = 0$.
Moreover, since reliable estimates of the time intervals between
bounces may be found in the
literature (cf., e.g., Khalatnikov et al. (1985) and references therein)
it is possible to show this result by upper bound estimates.}
and Gauss map connected to the
combinatorial model (\ref{BKLuBN}) for the axes, this
{\em stretching} is ``so effective" as to
make the Lyapunov characteristic exponent, extracted from the
$(\alpha, \beta, \gamma, \alpha', \beta', \gamma')$
phase flow, zero in $\tau$-time:
$ \lambda_{\tau} = 0 \;\;$ .
(See S.E. Rugh (1990a), S.E. Rugh and B.J.T. Jones (1990), D.
Hobill \etal (1991)\footnote{Note, that Dave Hobill \etal have
extracted the entire spectrum of all six Lyapunov exponents
from the phase flow - and not
only the principal value (the maximal Lyapunov exponent).}
and also A. Burd, N. Buric and R.K. Tavakol (1991)).
This is not standard for a chaotic deterministic model, but
it appears that the zero Lyapunov exponent is simply
a consequence of the time variable chosen for the
description.

\vspace{0.5 cm}

\noindent
{\bf In $t$-time}:
In the original synchronous time
parameter $t = \int abc \; d\tau$ the scale functions of the metric
$$ ds^2 = -dt^2 + \gamma_{ab} (t) \; \BFACE{\omega}^{a}(x)
\BFACE{\omega}^{b}(x) \; \; \;, \; \; \; \gamma_{ab}(t)
= diag(a^2(t),b^2(t),c^2(t)) $$
exhibits an infinite sequence of oscillations
and the successive major cycles condense
infinitely towards the singularity which we, without loss of
generality, may take to be at $t=0$.
(See also the
very illustrative picture in Kip S. Thorne (1985))
of oscillating, fast growing, tidal field
components of the successive oscillations as the spacetime
metric approaches the singular point at $t=0$).

In any finite time interval $ t \in [T , 0] $,
where $T$ denotes an arbitrary small positive number, the model exhibits
an infinite, unpredictable sequence of oscillations.

A Lyapunov characteristic exponent,
although ill defined, is unbounded in $t$ time. \\
(A perturbation will be amplified faster than exponentially
with respect to the metric $t$ time and with respect to the
distance measure (\ref{EuclidianIntroduction})
on approach to the singularity).

\begin{quotation}
\noindent
{\small
The quoted Lyapunov exponents above are for
mixmaster collapses governed by the
{\em vacuum} Einstein equations corresponding to the vanishing
$I = 0$ of the first integral (\ref{Ifirstintegral})
or corresponding to mixmaster collapses coupled to perfect
fluid matter with $p \leq \frac{2}{3} \rho$.
For initial conditions which fail to satisfy the first integral
constraint $I = 0$, the
character of the solutions depends on the sign of the
first integral $I$. We have qualitatively
different behavior for $I < 0$, for $I > 0$ and for $I = 0$.
For $I < 0$ (inclusion of negative ``stiff matter'' energy densities)
the complicated behaviors in fig.1-3 are reflected in positive
Lyapunov exponents, with respect to the $\tau$ time
parameter, extracted from such numerical solutions.
{}From the scale invariance of the
Einstein equations, it follows that if
one ``unwittingly'' put an amount $p = \rho = \rho_0 =
(8 \pi G)^{-1} I < 0 $
of negative mass-density in the model,
the maximal Lyapunov exponent roughly scales as
$$ \lambda \sim \sqrt{- \rho_0} \; \; . $$
This is exactly what
G. Francisco and G.E.A. Matsas (1988) ``unwittingly'' have plotted as
fig.4. in their paper.

For $I > 0$ (i.e. the inclusion of ``stiff matter'' with positive energy
density) all three scale functions will decrease monotonically
after some transient $\tau$ time and the maximal Lyapunov exponent will be
zero.
The reason why even V.A. Belinskii and I.M. Khalatnikov
(1969) had a minor accident in obtaining qualitative accordance between
numerical simulations and their theoretical model
is due to the fact that they had an extremely
small error in choosing the initial data.
They had $I = + 5.5 \times 10^{-2} > 0$, which prevents the
spacetime metric from going into the described
sequence of oscillations according to the {\em BKL}-combinatorial
model.
}
\end{quotation}

For the values of the Lyapunov exponents associated with the
vacuum gravitational collapse one notes that
the fact that Lyapunov exponents, as indicators of
chaos, are not invariant under time reparametrizations
is no surprise:

\begin{quotation}
\noindent
{\em A Lyapunov exponent (which is a ``per time" measure of having \\
exponential separation of nearby trajectories in ``time")
has never been \\
invariant under transformations of the ``time" coordinate!}
\end{quotation}

Moreover, for a given fixed choice of time variable,
transformations of the {\em coordinates for the three-metric}
$g_{ij}$ on the spacelike hypersurfaces $\Sigma_t$
should be accompanied by a transformation of the
distance measure $|| \cdot ||$ on the three-space.
If one does not take proper account of
transforming the distance measure (the metric)
on the solution space when performing a {\em coordinate
transformation} to new variables (e.g. transformation from
the ADM-variables to the Chitre-Misner
variables on the Poincar\'{e} disc),
then some definite value of a Lyapunov exponent
(e.g. $\lambda_{\Omega} = 0$ in the ADM-variables)
may be transformed into any other
value by that coordinate transformation
(and such measures of ``chaos'' which are artifacts of
the coordinate transformations
are not strong candidates for capturing anything interesting about
inherent properties of a dynamical system.\footnote{Thus,
when J. Pullin (1990)
arrived at the conclusion
that the mixmaster toy-model gravitational
collapse is chaotic {\em because} a
positive Lyapunov exponent ``$\lambda = 1$''
can be associated with the
geodesic motion on the Poincar\'{e} disc one should
be careful to project out (separate) what are artifacts of the
coordinate-transformations and what are inherent properties of the
dynamical system. (Invariant under gauge choices).
Certainly the element of chaos and unpredictability of the
mixmaster collapse is not (alone)
due to the negatively curved interior of the
Poincar\'{e} disc. Even the completely integrable Kasner metric may be
mapped to the interior of the negative curved Poincar'{e} disc (by a
set of hyperbolic coordinate transformations identical to those applied
to the mixmaster metric). No chaos arises, since there is no scattering
in any potential boundaries. However, as an artifact of not transforming
the distance measures under the coordinate
transformation, we suddenly appear to have
local exponential instability in all directions
on the Poincar\'{e} disc
- despite that the Kasner metric is completely integrable with linearly
growing separation in the original ADM-variables.}

\subsection{Chaos in the ``coordinates''  or in the
gravitational field?}

The fact that Lyapunov-exponents are found to be strongly gauge
dependent ``touches'' a deeper problem (S.E. Rugh (1990 a,b)
connected with the characterization of ``chaos",
or other dynamical characteristics, in
theories like general relativity
(and to some extent in gauge-theories)
- not present in, e.g., hydrodynamical turbulence studies: How can you be
sure that your measure of ``chaos" is not some artifact of the ``gauge"
variables chosen?
That is: Does the apparent ``chaos" originate from
the gauge-variables chosen (``chaos in the semantics") or is it
present in the ``real world"?

\begin{quotation}
\noindent
{\em To distinguish (in a coordinate invariant way) whether the
``metrical chaos" is due to a ``chaotic choice of coordinates"
(we may call it ``chaos in the semantics")
or reflects
``real chaos and turbulence" in the gravitational field does not seem to be
an easy task, as this separation between coordinate
choices and gravitational fields
is indeed very tricky business in general.

This question
is, in some sense, the
``chaos in general relativity'' (``metrical chaos'') analogue of Eddingtons
worries about the ``spurious gravitational
waves'' as a propagation of ``coordinate changes'' with the
``speed of thought''.   }
\end{quotation}

It is not difficult to
invent (apparently) ``chaotic solutions'' of
``space-time metrics'' where the ``chaotic signal'' is a pure
artifact of gauge (one may consider, e.g., a 2+1 dimensional
study of pure ``gravity'' in a non-linear gauge).
H.B. Nielsen and S.E. Rugh (1992 a).

The problem is rooted in the more general question
(which is itself very interesting):
\begin{quotation}
\noindent
{\em How gauge-invariantly may we capture any dynamics
of the gravitational \\
field (chaotic or not) in the Einsteins theory of general relativity?}
\end{quotation}

It is well known that
it is very difficult to extract gauge-invariant quantities
{\em locally} from general relativity - even in the
weak field limit!\footnote{For example, Roger Penrose
has for several years attempted to extract a measure of ``entropy''
in the metric field itself. But
it is not easy to make a {\em gauge-invariant} and
sensible construction. So far, Roger Penrose has not succeeded.
Pers. comm. with Roger Penrose (at the Niels Bohr Inst.). }

While the collection of the (non-local) Wilson loop variables
$$ \; Tr(P \exp \oint_{\gamma} g A^{a}_{\mu} (x)
\frac{\lambda^{a}}{2} dx^{\mu}) \; $$
in principle offer a tool
to characterize
the dynamics of non-Abelian gauge theories in a complete gauge
invariant manner,\footnote{Thereby loosing locality,
cf.\ also the Aharonov-Bohm effect.}
such a ``projection out'' of ``the gauge-invariant content'' is
troublesome in gravity.
Like in Yang-Mills theories,
we may attempt to characterize the gravitational field
invariantly by loops
(capturing the curvature tensor components
$R^{\alpha}_{\beta \; \gamma \; \delta}$'s,
just as the loops in Yang-Mills theories capture the $F_{\mu \nu}$'s).
But there is a problem of saying (in a diffeomorphism invariant way)
{\em where} the loop is!

One can construct completely gauge-invariant and {\em global}
quantities like for example
\begin{equation}
I(\eta) = \int d^4 x \sqrt{-g}
\; \delta
(C_{\alpha \beta \gamma \delta} C^{\alpha \beta \gamma \delta}
- \eta )
\end{equation}
which measures the invariant 4-volume which has the value
$ C_{\alpha \beta \gamma \delta}
C^{\alpha \beta \gamma \delta} = \eta $.
But it is very hard to capture any dynamics with such a
(space-time global) signal.   \\

\noindent
One may question the very suitability (in more generic cases)
of concepts like
\begin{quotation}
\noindent
$\bullet$ A Lyapunov exponent ``$\lambda$"  (a ``per time''
measure characterizing the ``temporal
chaos'': does the ``distance'' between ``nearby'' orbits grow
up exponentially with time?) \\
$\bullet$ A spatial correlation length ``$ \xi$" (in a spatially
inhomogeneous gravitational field we may have, that
domains separated by distances considerably larger
than $\sim \xi$ cannot ``communicate'' due to spatial disorder
in the field configurations: ``metric turbulence''
indicates a lack of correlations between gravitational signals
in time as well as in space.\footnote{Often (cf.,
e.g., T. Bohr (1990))
temporal and spatial chaos are somewhat connected in a relation of
the type: $ \xi \sim c/\lambda $ where $c$ is the velocity of
propagation (velocity of ``light'' in the case of propagation of
disturbances of spacetime
curvature in general relativity).}
This ``spatially disordered'' aspect of chaos falls away
for the spatially homogeneous toy-models).
\end{quotation}

\noindent
To the extent that we may split the spacetime metric
${}^{(4)} g$ into a three-metric ${}^{(3)}g$ evolving in a
``time''\footnote{Note that we
have been able to study our simple
spatially homogeneous toy-model gravitational collapse in
a {\em (global) synchronous} frame of reference.
One can always choose a synchronous reference frame
{\em locally}, but in generic situations, for strong gravitational
fields, a global construction of a synchronous frame
of reference easily breaks down because of the tendency of these
coordinates to focus to a {\em caustic} (and a coordinate
singularity will develop).}
the very concept of, e.g., a Lyapunov exponent may perhaps be
applicable, though {\em gauge-invariant generalizations}
(i.e. measures of ``metrical chaos'' invariant
under general space-time coordinate transformations
or some {\em more restricted}
class of coordinate transformations)
are (obviously) better to have - if they are possible to construct?

It may very well be that one should head for a ``no go'' theorem
for the most ambitious task of constructing a generalization of a
Lyapunov exponent (extracted from
the continuous evolution equations)
which is meaningful and invariant under the full class
of spacetime diffeomorphisms (H.B. Nielsen and S.E. Rugh).
One should therefore, at first,
try to modify the intentions and construct
(instability) measures which are invariant
under a smaller class of diffeomorphisms
(e.g. coordinate transformations on the spacelike hypersurfaces
$\Sigma_t$ which do not involve transformations of
the time coordinate).

In the mixmaster toy-model context of a spatially homogeneous
spacetime metric this question is particularly simple to address
since all the physical signals are
{\em functions of time} only.

The need for
gauge invariant measures of chaos in this context of general relativity
was noted and discussed
in S.E. Rugh (1990a), emphasized as a major point in
S.E. Rugh (1990b)
and, also,  emphasized in
J. Pullin (1990). See also
discussions in e.g. H.B. Nielsen and S.E. Rugh (1992),
in M. Biesiada and S.E. Rugh (1994) and in S.E. Rugh (1994).

\subsection{Chaos in ever expanding (or ever collapsing) phase spaces}

In the description of the gravitational collapse
involving the $(\alpha, \beta, \gamma,...)$
phase flows (the parametrization
in, e.g., Landau and Lifshitz (1975)) and the $(\Omega,\beta_{+},\beta_{-},..)$
phase flows (the anisotropy-variables of C.W. Misner et al.),
there is no ``recurrence'' of orbits because the phase space
is ever expanding\footnote{If we follow
this metric to the final crunch singularity,
the configuration space variables
$(a,b,c)$ shrink to zero volume $(abc \rightarrow 0)$,
while if we take the logarithmic scale
factors $(\alpha, \beta, \gamma) = (\ln a, \ln b, \ln c)$
the configuration space is not
bounded from below (we have
$\alpha + \beta + \gamma \rightarrow - \infty$).
In terms of anisotropy variables,
also commonly used,
$(\Omega, \beta_{+}, \beta_{-})$ the anisotropy
$ || \vec{\BFACE{\beta}} || $ grows
to infinity and the trajectory of anisotropy $\vec{\BFACE{\beta}}$
does never return to some given
value. }, and it is not straightforward to
define what is meant by a concept like ``ergodicity'' or the
``mixing property'', for example, in
the outward expanding Hamiltonian billiard in the
ADM-Hamiltonian description.
In this sense the gravitational collapse study presents a
dynamical system, which differs from
most other systems treated in the theory of non-linear
dynamical systems.
Thus our gravitational collapse
orbit is also a playing ground for investigating
how to deal with chaos in expanding
(or collapsing) phase spaces.

\begin{quotation}
\noindent
{\em The ideas of chaos apply most naturally to time evolutions with
``eternal return''. These are time evolutions of systems that come
back again and again to near the same situations. In other words,
if the system is in a certain state at a certain time, it will return
arbitrarily near the same state at a later time.
(Cf. D. Ruelle (1991), p. 86)  }
\end{quotation}

\noindent
How do we extract measures of chaos and ergodicity
in an ever expanding phase space?
$\bullet$ As we have already described,
one may project out (extract) lower dimensional
signals from the flow
which may capture (chaotic) ``return properties'' of the dynamics.
An example is the variable ``$x = 1/u \in [0,1]$" which we extracted
from the phase space trajectories in sec.2.4.   \\
$\bullet$ If we  characterize the dynamical evolution of the
gravitational collapse in terms of the infinite bounce sequence
(i.e. encoding the gravitational collapse by
an infinite collision sequence against the three wall boundaries)
we hereby map an ever expanding phase space
(in the $(\alpha, \beta, \gamma,...)$ variables, say)
to an infinite symbolic sequence with return properties (e.g. with
periodic orbits).

\subsection{The many gauges
and different variables describing the mixmaster
gravitational collapse}

The amount of work which has been put in the invention of
different descriptions (different gauges)
and different approaches is substantial: Cf.
the study of the mixmaster gravitational collapse in the ADM
variables (cf.\ e.g. B.K. Berger (1990)),
the Misner-Chitre description (J. Pullin (1990)),
the BKL-variables (D. Hobill \etal (1991), S.E. Rugh (1990 a)),
the Ellis-MacCallum-Wainwright variables
(A.B. Burd, N. Buric and G.F.R. Ellis (1990)), Ashtekhar variables
(Ashtekhar and Pullin (1989)),
the description by O.I. Bogoyavlensky and S.P. Novikov
(e.g. used in
the approach by M. Biesiada, J. Szczesny and M. Szyd{\l}owski).

\begin{table}[ht]
\begin{center}
\begin{tabular}{l}
\topline
The many different gauges and approaches \\
\midline
$\bullet$ BKL variables (Landau and Lifshitz)  \\
$\bullet$ The ADM-Hamiltonian variables (MTW)    \\
$\bullet$ The Misner-Chitre coordinates  \\
$\bullet$ Ellis-MacCallum-Wainwright variables  \\
$\bullet$ Ashtekhar variables  \\
$\bullet$ The approach by Bogoyavlensky and Novikov \\
\bottomline
\end{tabular}
\end{center}
\caption[xxx]{{\small Examples of the many descriptions of the
very same gravitational collapse. What are the ``invariant'' properties
of the dynamics
which can be extracted from all these descriptions?}  }
\end{table}

The ``transformation theory of chaos'' (which
concepts of ``chaos'' are invariant
under the transformations?) between the different
gauges and approaches is not  easy.
As we have seen, in some gauges there is a positive
(finite or infinite, depending on the gauge choice)
value of a ``Lyapunov exponent'' - in other gauges it
is zero.
In many gauges the phase space is {\em ever expanding}
and a notion like that of ``ergodicity'' (relying by definition on
return properties of the flow)
can hardly be assigned any meaning at all.
In other gauges, e.g. the ``Misner-Chitre'' gauge,
the dynamics of our spacetime metric (\ref{metric})
is, it appears, ergodic and, even, a K-flow! (cf., e.g., J. Pullin (1990)).
This rather ``messy situation'' is, of course, an
artifact of not having constructed
gauge invariant measures of chaos (or ergodicity)
in this context of general relativity.

\begin{quotation}
\noindent
{\em
Yet, in the mixmaster gravitational collapse, the ``bounce'' dynamics
\underline{is} there and \underline{that} we must be
able to capture in a gauge
invariant way (i.e. that $\infty$ many bounces take place before
the spacetime singularity is reached).
}
\end{quotation}

Imprints of the bounce structure of the mixmaster dynamics
should be seen in various invariants which can be extracted from the
mixmaster metric, e.g. in the Weyl curvature invariant
$C^2 = C_{\alpha \beta \gamma \delta}
C^{\alpha \beta \gamma \delta} (\tau) $
of the mixmaster metric.
Thus, in order to illustrate how a ``good physical signal''
behaves for a typical
trajectory we have set out to
calculate and display (cf.\ Biesiada and Rugh (1994))
this Weyl curvature invariant.
Premature investigations indicate that
if $C_{\alpha \beta \gamma \delta} C^{\alpha \beta \gamma \delta} $
is plotted with respect to the time variable $\tau$, one observes
(Biesiada and Rugh (1994)) that the bounce structure is also
seen on this curvature invariant.
\footnote{During the workshop
it was discussed (Charles W. Misner, Piotr Chrusciel \etal) that it
is not fully investigated
how the Weyl curvature tensor behaves
near the generic singularity of the mixmaster collapse. For example,
$C^2 = C_{\alpha \beta \gamma \delta}
C^{\alpha \beta \gamma \delta} (\tau) $
will not necessarily increase {\em monotonically} towards
the big crunch singularity.
The near singularity behavior of the Weyl tensor is therefore
interesting in itself and deserves further
examination. }

Also other (higher order) invariants could
be constructed for the mixmaster gravitational collapse
and one should verify whether the
physical degrees of freedom of the mixmaster collapse
in principle may be captured
{\em exhaustively} in terms of the algebraic
invariants constructed from the Weyl curvature tensor field.

\subsection{On symbolic dynamics for the bounce structure of
the gravitational collapse}

Let us now consider a particular ``gauge'', say, the
ADM Hamiltonian description (or the ``Misner Chitre'' gauge)
in which the gravitational collapse is described as a ``ball''
exhibiting an infinite sequence of ``bounces'' against a
potential boundary which is outward expanding\footnote{Cf., e.g.,
MTW \S 30.7 or Ryan and Shepley (1975). A
detailed description of
the ADM-Hamiltonian description of the gravitational
collapse is also given in S.E. Rugh (1990a) (which is available upon
request).} (or stationary in the ``Misner Chitre'' gauge).
Is a ``bounce'' well defined?
Recall, that the scattering walls -
which effectively arise due to the three-curvature
scalar ${}^3 R$ on the
space-like sections of the
metric $ds^2 = -dt^2 + \gamma_{ij} (t)
\BFACE{\omega}^i (x) \BFACE{\omega}^j (x) $ - turn into
``infinitely" hard walls when sufficiently near the space-time singularity
of the metric. Hence, the dynamics resembles that of an ideal
billiard and it gets easier and easier to define when the orbit
``bounces" against the potential wall as the metric approaches the
singular point.

It may be useful to assign a
{\em symbolic dynamics} to the dynamical system, i.e., a
scheme that assigns a unique symbolic string (coding) to each
orbit.
The simplest qualitative way to describe an orbit
is to list the order in which it hits the three boundary walls
$\partial {\cal B}_1, \partial {\cal B}_2, \partial {\cal B}_3$
using the wall labels as symbols (characterizing the orbit by its
infinite collision sequence). This yields a description of the
orbit in terms of ternary symbolic dynamics:
\begin{quotation}
\noindent
``{\bf 1}" for scattering against the boundary
$\partial {\cal B}_1$ (where $c \gg a,b$) \\
``{\bf 2}" for scattering against the boundary
$\partial {\cal B}_2$ (where $a \gg b,c$) \\
``{\bf 3}" for scattering against the boundary
$\partial {\cal B}_3$ (where $b \gg a,c$)
\end{quotation}
As a randomly chosen reference example of a segment of a
``symbolic string" (compare with
Kip. S. Thorne (1985)) we may, for example, have
\begin{equation}  \label{segmentexample}
......\; \; \underbrace{12121212}_{4 \; oscillations} \; \;
\underbrace{131313}_{3 \; oscillations} \; \;
\underbrace{2323232323232323}_{8 \; oscillations} \; \;
\underbrace{121......}_{etc}......
\end{equation}
The sequence of
oscillations between two walls leading to
one of the three corners is denoted as a ``(major) cycle".
Thus, the displayed segment (\ref{segmentexample})
has 4 oscillations in the first
of the displayed (major) cycles
(number ``$i$", say) 3 oscillations in
the next major cycle (number ``$i+1$"), 8 oscillations in cycle ``$i+2$",
etc.    \\

\noindent
{\bf Binary symbolic dynamics suffices} \\
Due to the grammar of the symbolic dynamics, which
forbids the appearance of
$ ...\BFACE{11}...$, $ ...\BFACE{22}...$ or $ ...\BFACE{33}...$ in a
sequence (no orbit can hit the same wall twice without hitting another wall
first) the three-letter alphabet (above) may be reduced to a
{\em two-letter alphabet}. This reduction
can be accomplished in many ways. One
possible way of assigning binary symbolic dynamics to the dynamics is
the following:
Assign a
``$\BFACE{0}$" for scattering against the boundary $\partial {\cal B}_1$ and
a ``$\BFACE{1}$" for scattering against the boundary
$\partial {\cal B}_2$.
Scattering against the last of the three boundaries
$\partial {\cal B}_3$ we denote
``$\BFACE{0}$" if it follows a bounce against the boundary
$\partial {\cal B}_1$ and denote
``$\BFACE{1}$" if following a bounce against the
boundary $\partial {\cal B}_2$.
In this  encoding the motion is considered
as a series of choices of what boundary
to go to from the present boundary (there are always two possibilities,
and binary symbolic dynamics therefore suffices to describe the
chosen route). Except for some arbitrariness in the beginning of a
symbol sequence,
this gives a one-to-one correspondence between the
{\em pruned} \footnote{{\em Pruning}: If some symbolic sequences
have no physical
realizations, the symbolic dynamics is said to be ``pruned".
In that case the symbolic alphabet must be supplemented with a
set of grammatical rules (i.e. a set of pruning rules).
If all possible symbol sequences correspond to physical trajectories,
the symbolic dynamics is said to be ``complete".
In a systematic encoding process,
one will ask first if there is
{\em at most} one orbit to a given symbolic sequence.
If this is the case we
have a {\em covering} symbolic dynamics. If there {\em is} also one
orbit for any symbolic sequence, the symbolic dynamics is said to
be {\em complete}.}
ternary dynamics  and an unrestricted binary dynamics.\footnote{Note,
that in the three disc problem on the hyperbolic Poincar\'{e} disc
we have ``full symbolic dynamics"  while in the
three-disc problem
on an {\em Euclidian} space
(i.e. if the billiard had been in
Euclidian space) there exist
binary symbolic sequences which do not correspond
to physical trajectories.
In that sense, the hyperbolic three-disc dynamics is
simpler to work with than the
Euclidian three-disc problem. See also Giannoni and Ullmo (1990).
} \\
For instance, our reference-segment (\ref{segmentexample}) reads
\begin{equation}
......\; \; \underbrace{01010101}_{4 \; oscillations} \; \;
\underbrace{000000}_{3 \; oscillations} \; \;
\underbrace{1111111111111111}_{8 \; oscillations} \; \;
\underbrace{010......}_{etc}......
\end{equation}

\vspace{0.5 cm}

\noindent
If we were to carry out a systematic
exploration of system characteristics in terms
of the periodic orbit structure (this includes finding periodic orbits
up to some given length,
quantize the system in terms of the periodic orbits,
Selberg zeta functions
and all that) it would be of importance
to use the minimal symbolic dynamics (with no pruning).
We shall, however, just
use the direct symbol assignment
$\BFACE{1}$, $\BFACE{2}$, $\BFACE{3}$
in order to have a fixed notation.

\subsection{Chaos and the topological structure of the
gravitational collapse.}

A simple example
illustrates that
a zero Lyapunov exponent is not contradictory to an intrinsic unpredictable
and chaotic behavior (from S.E. Rugh (1990 a)).

Consider, for example
the famous Lorentz attractor
generated as an asymptotic attractor
by the Lorentz time evolution equations (Lorentz, 1963)
- just like the (BKL) oscillatory behavior is generated as an asymptotic
attractor\footnote{The ``basin of attraction'' are initial conditions
which are not axisymmetric (cf.\ the Taub-solution).}
of the vacuum Einstein equations for the mixmaster
spacetime metric (\ref{metric}).

The ``particle" performs an infinite number of consecutive rotations
about two rotation centers, schematically shown
in fig.8. It rotates, lets say,
$n_1$ times around
the left rotation center (``first major cycle") and then
$n_2$ times around
the right rotation center (``second major cycle")
and return to the left again ...etc...etc...ad infinitum.

One feature of the non-linear Lorentz system is that the number of
consecutive rotations about the two rotation
centers is unpredictable, since these numbers
$(n_1,n_2,...)$
depend sensitively on initial
conditions. (See also Lichtenberg and Lieberman (1983), p.59-62).
In that sense the behavior of the system is
{\em intrinsically} chaotic (stochastic)
and not predictable. With respect to the standard
time parameter of the Lorentz evolution equations, a maximal
Lyapunov characteristic
exponent, extracted from the flow,
mirrors this unpredictability and is greater than zero.

\begin{figure}
\vspace{5.0 cm}
\caption[xxx]{{\small The intrinsic unpredictability of the number of
consecutive rotations about the two rotation centers is
{\em independent}
of the choice of time variable (whereas the spectrum of Lyapunov
exponents is not).
}}
\end{figure}

However, by making some kind of an ``exponential stretching"
of the time parameter axis (thereby making the particle moving
around slower and slower with respect to the new time parameter)
it is easy to make the corresponding Lyapunov exponent, extracted
from the new first order autonomous differential flow, zero.
The unpredictability of the number of consecutive rotations performed
in each ``major cycle" of course still remains!

The gravitational collapse, in the oscillatory BKL-regime
on approach to the singularity, exhibits
consecutive rotations around 3 ``centers" (there are
three ``attraction centers": $\alpha$ and $\beta$ can oscillate
($\gamma$ declines),
$\alpha$ and $\gamma$ can oscillate ($\beta$ declines)
or $\beta$ and $\gamma$ can oscillate
($\alpha$ declines)) corresponding to oscillations in
one of the three channels of the billiard game
in the Hamiltonian description. (See, also, fig.9 of the
triangular symmetric scattering domain in the Poincar\'{e} disc).
If we, by the number of rotations around
a ``center",  mean the number of oscillations in a period (a major cycle)
where  two scale functions oscillate and the third declines,
the BKL-model will be intrinsically unpredictable
as the Lorentz attractor described above
(the number of consecutive rotations around now three
``rotation centers" depends sensitively on initial conditions,
and the ``per cycle" entropy is very big: $h = \pi^2/6 \ln 2 \gg 0$).
But the ``particle" moves too slow with respect to
the standard time parameter (constructed from the  metric $t$ time)
$\tau = \int dt/abc$,
as $\tau \rightarrow \pm \infty$, to mirror this unpredictability.

The property of a dynamical system to behave ``chaotically"
\footnote{I.e. the system displays a complex
evolutionary pattern rather than a simple,
integrable pattern (which is ``just'' quasi-periodic motion in the
action-angle variables).}
should be an intrinsic property of the system and not
rely on the variables chosen for the description.

\subsection{Which concepts should characterize ``chaos'' in the
mixmaster toy-model collapse (and in more general cases)? }

There is nothing mysterious in selecting a particular ``gauge''
and describing physics, e.g. gravitational collapse physics, in that
particular gauge.
Thus, very often, it is not very useful to do the calculations in a
gauge-invariant way (and in general relativity it may in fact
not be possible). So one, usually, ``fixes the gauge''
and performs the calculations in that gauge.

However, in my opinion, it is
of interest to pursue some routes in order to
study how we may eventually deal more invariantly with
the ``chaotic'' mixmaster gravitational collapse:

\begin{quotation}

\noindent
\underline{\bf 1.} How little structure on the solution space is needed to
deduce

information about chaos?

\noindent
\underline{\bf 2.} One may argue that there is a preferred superspace metric
and

a preferred supertime (Wheelers superspace) as a natural structure on

the solution space

\noindent
\underline{\bf 3.} One may attempt a (more) invariant description
(involving

knowledge about extrinsic curvature invariants ``Tr $\BFACE{K}$'' etc.).

\noindent
\underline{\bf 4.} Study our free observer and see what happens with
him (tidal stretches

etc.) in his local time. That is, after all,
a very {\em physical} question!

\end{quotation}

I shall here attempt at preliminary sketches of how to
pursue the first two routes. We have
already, in sec.2.5, sketched the short, yet exciting, story of the free
falling astronomer which reaches the spacetime singularity in
finite proper time
(the tidal stresses exerted on him by
falling freely in the collapsing mixmaster
spacetime metric should be properly
analyzed in terms of the Riemann
tensor $R_{\alpha \beta \gamma \delta}$ for the mixmaster metric
and not in terms of the  Riemann tensor of the Kasner metric).

We shall not pursue route no.3 here but note that in the ``3+1''
splitting of the spacetime into a three-metric ${}^{(3)} g$ evolving
in a ``time'', a complete description of the geometry of the four manifold
also needs, besides the intrinsic 3-geometry (the induced three-metric
${}^{(3)} g_{ij}$ on the spacelike hypersurfaces $\Sigma_t$),
knowledge about how the 3-geometry is embedded in the enveloping
4-geometry, which is characterized by the {\em extrinsic}
curvature $K_{ij}$ (cf., e.g., MTW \S 21.5). One may attempt
a more invariant description (i.e. measures
of local instability of collapse
orbits) by extracting the information from both the
intrinsic geometry and the extrinsic curvature
and its associated invariants (like its trace,``Tr $\BFACE{K}$'').

\subsection{Is it possible to characterize
the complexity of the gravitational collapse without
use of a distance measure $|| \cdot || $ on the solution space?
(and without reference to a particular time variable?)}

Since the standard measure, the Lyapunov exponent $\lambda$,
is a ``per time'' measure, it depends, obviously,
on the choice of time parameter ($\lambda = 0$ in
$\tau=\int dt/abc$ time while
$\lambda=$``$\infty$'' in metric $t$ time).
Besides, the definition of a Lyapunov
exponent requires that we have at hand a metric $|| \cdot ||$
on the solution space, which measures
``distances'' between (neighboring)
trajectories.
In the chain of adding more and more structure to
the solution space,
$$
\left(   \begin{array}{c}
         \mbox{{\small Topological }}  \\
         \mbox{{\small  Space}}
         \end{array}
         \right) \subset ............ \subset
\left(   \begin{array}{c}
         \mbox{{\small Manifold Structure }}  \\
         \mbox{{\small  With a Metric }}
         \end{array}
         \right)
$$
one also hereby gradually introduces more and more arbitrariness.
We need analysis to which extent one may construct measures of chaos
where we do not have to rely on the
(arbitrarily) chosen time-parameter ``$t$'' or
some (arbitrary) chosen distance measure ``$|| \cdot ||$''
on the phase space (cf., also, H.B. Nielsen and S.E. Rugh (1992 a)).

\begin{quotation}
\noindent
{\em How much ``structure'' on the ``solution space'' do we need to
get information about whether a collapse dynamics is of a complex
(chaotic) or regular (integrable) type, i.e.
extract interesting information about ``chaos''?
Do we really need a manifold structure with a metric
$|| \cdot ||$ on it - or can we,
from knowledge of the topological structure alone, say,
(on the space of solutions) tell something about chaos of
interest? }
\end{quotation}
For our gravitational collapse (\ref{metric}) - being spatially
homogeneous - all physics is entirely
given in terms of {\em functions of time}
(e.g. encoded in  ``signals" like
$C_{\alpha \beta \gamma \delta}
C^{\alpha \beta \gamma \delta} (\tau)$ and
higher order algebraic invariants)
and the phase space dynamics is
a low dimensional signal in ``time''.

It is natural to try to
assign a ``good symbolic dynamics''\footnote{This is a very difficult
point. A priori, the
assignment of a symbol-sequence
from phase space dynamics
(coding the trajectory by a discrete ``alphabet'')
is arbitrary
and accomplished by an
arbitrary partitioning of the phase space into different
domains (which are assigned different symbols).
In many systems it is not easy to find a good partition to encode
a generic orbit. }
from the phase space dynamics.
I.e. one may think of the dynamics ``described''
by a string of symbols
\begin{equation} \label{symbolstring}
\left\{ ......, {\cal S}_i, {\cal S}_{i+1}, ... ,
{\cal S}_{i+n}, ...\right\} \; \; \; \; \; , \;  \; \;  {\cal S}_i \in
\left\{ \; 1,2,....n \; \right\} \; \; .
\end{equation}
We have already noticed that there is a naturally
ternary (or binary without pruning) symbolic encoding of
the {\em bounce structure} of the
gravitational collapse.
(Thereby, one maps the ever expanding or collapsing phase
space (without return properties of the collapse orbit), cf.\
sec.3.1,
to symbol sequences {\em with} return properties (e.g.
having periodic orbits)).

We have a {\em gauge-invariant} ``before'' and ``after'' -
but the size of the step in ``time'' between two symbols
in (\ref{symbolstring}) is merely reflecting an arbitrary
gauge choice of a ``time coordinate". \footnote{The continuous
flow of events during
the gravitational collapse is, from the point of
view of ``causality'' (in the sense of which
event is before the other)
invariant under time reparametrizations and diagnostic tools
such as a Fourier transform
(an investigation in this
spirit is also J. Demarett and Y. De Rop (1993))
of a signal -
which can also be performed on
symbol sequences -
do not need a distance measure $|| \cdot ||$ on the solution
space. However, it needs some sort of time variable. Note,
that if we take the infinite string of symbols (\ref{symbolstring})
to mean a symbolic encoding of the infinite {\em bounce}
sequence of the mixmaster gravitational collapse we have,
hereby, not escaped the introduction of a ``time'' variable:
It is a time variable
$i$ (the label of the symbols ${\cal S}_i$) which counts
one unit for each ``bounce'' (a recognizable
``cosmological event''). It is like ``heart beats'' of the mixmaster
gravitational collapse. In terms of these ``heart beats'' the
gravitational collapse is ``infinitely old''
when the ``big crunch'' singularity
is reached. See also C.W. Misner (1969 c) and MTW, \S 30.7, p.
813-814.}

Now, let us say that we are given the set of
{\em all possible sequences} of finite length,
which may
be realized by the physical system (that is usually many!).
Can we then deduce, just by looking at these sequences,
whether the underlying physical system is integrable or
not (i.e. without having any metric $|| \cdot ||$ on the original
solution space)?

Complexity measures for symbol sequences are already at hand.
The {\em topological entropy} $h_{top}$ of
a string of symbols states how the number $N(n)$ of
all the possible sequences of length $n$
grows exponentially with $n$: $N(n) = e^{h_{top} \; n}$.
It is $0$ for an integrable system because of the
quasi- or almost-periodicity\footnote{Quasi-periodicity, or
almost-periodicity, for all realized symbol sequences,
is a sign of integrability. Quasi-periodicity
(multiply periodic motion)
means that the Fourier transform is entirely composed of
$\delta$-functions. The finite sums
(trigonometric polynomials) $s(t) = \sum_{n=1}^{N} \alpha_n e^{\lambda_n t}$
(where the coefficients $\alpha_n$ are arbitrary complex, and the
frequencies $\lambda_n$ are arbitrary real) are quasi-periodic functions.
According to a celebrated theorem by Harald Bohr, the
class of almost periodic functions is identical to the
closure $H \left\{ s(t) \right\} $ of all finite
$s(t)$. (The necessary and sufficient condition for an arbitrary
trigonometric series, $\sum_{n=1}^{\infty} \alpha_n e^{i \lambda_n t}$,
to be the Fourier transform of an almost periodic function is that
$\sum | \alpha_n |^2$ converges, Besicovitch-Bohr).
} (setting severe restrictions on the number of possibilities
for strings of length $n$).
For a random string of symbols the topological entropy is equal
to the logarithm of the number of symbols (e.g. log 2 for the Bernoulli
shift).
For the mixmaster gravitational collapse the number of realized
symbol sequences (characterized by its infinite bounce
sequence)
grows with the length $n$ of the symbol string
like $\sim 2^n$ and the topological entropy is thus
$\log 2 > 0$. That is, for the encoding of the gravitational collapse
with the binary (or ternary) symbolic dynamics, e.g. our segment
$$
......\; \; \underbrace{01010101}_{4 \; oscillations} \; \;
\underbrace{000000}_{3 \; oscillations} \; \;
\underbrace{1111111111111111}_{8 \; oscillations} \; \;
\underbrace{010......}_{etc}......
$$
We have that both ``0'' and ``1'' may be realized at each place
${\cal S}_i$ of the symbol string.
Every refinement (sub-partitioning) of the binary
symbolic dynamics will still lead to an expression
for the topological entropy which is positive.

Somewhat analogously to the concept of a
Lyapunov exponent (which is, roughly, an inverse correlation length in time)
one would like to extract a
``correlation length'' $\xi_{string}$ from the set of all
allowed strings of
symbols (\ref{symbolstring}) measuring, in a quantitative way, that
the correlation between the symbols
(as extracted from the bounce structure captured from
physical signals like
$ C_{\alpha \beta \gamma \delta}
C^{\alpha \beta \gamma \delta} (\tau)$)
decreases with $n$ as we go $n$ steps
along the ``string" (from some $i$ to $i+n$, averaged over $i$).   \\
If all the realized string sequences
(\ref{symbolstring}) are
{\em quasi-or-almost periodic}\footnote{``Quasiperiodicity" is
captured by making a Fourier-transform on the symbol sequence.}
(which we take to be a sign
of integrability), we assign a correlation length of
$\xi_{string} = \infty$.
In a random sequence of symbols,
on the other hand,
there is no correlation between a  symbol and the next, giving
$\xi_{string} = 0$. If a
correlation length can be defined
for a string displaying ``deterministic
chaos" it  should have a correlation length in between these two extremes.

For a given string of symbols the following limit may exist
$$ C(n) = \lim_{N \rightarrow \infty}
\left\{ \frac{1}{2N} \sum_{i = -N}^N {\cal S}_i {\cal S}_{i + n} -
(\frac{1}{2N} \sum_{i = -N}^{N} {\cal S}_i)^2 \right\} $$
and defines a correlation function $C(n)$ for the string ${\cal S}$
(i.e. along the string of symbols
encoding a particular mixmaster gravitational collapse).
If we can find just {\em one}
symbol sequence for which the correlation function
decays exponentially with $n$,
$$ C(n) \sim e^{-n/ \xi} $$
it is not possible (we suspect) for the system to be integrable.

In fact, the mixmaster collapse orbit may very well (within the
validity of the one-perturbation treatment of Belinskii et al.) be strongly
intermittent in the ``per bounce'' time ``n'' and thus the correlation
function will exhibit {\em power law decay}
rather than decay exponentially along
the string of symbols.\footnote{I thank G\'{a}bor Vattay, p.t. at the
Niels Bohr Institute, for fruitful exchanges concerning this point}
Hence we may end up with a power law decaying
correlation function, yet a positive topological entropy, for the
infinite symbol strings of the mixmaster collapse orbits.

We shall not go into further details here.
We remark that, starting out
from our toy-model gravitational collapse (\ref{metric}),
one is apparently naturally driven into speculations of
a general sort in complexity theory. A general analysis
of these problems is
at the ``heart of complexity theory"\footnote{See also
sec.5.4 in P. Grassberger \etal (1991) for references to a number of
different definitions of complexities of time sequences which are
all tied to symbolic dynamics.
Interesting discussions of complexity, in a wider perspective, may also
be found in P. Grassberger (1986) and e.g.
S. Lloyd and H. Pagels (1988).
} and is an extremely difficult subject.

\subsection{Wheeler's superspace metric as distance measure on the space
of three-metrics?  }

If we wish to operate with a distance
measure $|| \cdot ||$ to the
space of three metrics, it is natural to
investigate
whether we have distance measures
in general relativity which are more
{\em natural} than others. In
particular one would like to construct a
distance measure which is invariant
under (a large class of) changes of coordinates for the description.

The original dreams by
Wheeler, DeWitt and others is to consider the dynamics of the three
metrics ${}^{(3)} g$ as geodesics on some manifold
called superspace equipped with the metric tensor
$ G_{ijkl}$  which is named the
supermetric. This supermetric $G_{ijkl}$ induces a norm
on the space of three metrics ${}^{(3)} g $,
\begin{equation} \label{Wheelernorm}
|| \delta g_{ab} ||^2 =
\int d^3 x \sqrt{g} \; G^{ijkl} \delta g_{ij} \delta g_{kl} =
\int d^3 x \sqrt{g} \; G^{AB} \delta g_{A} \delta g_{B}
\end{equation}
and one may measure the distance between two three-metrics
${}^{(3)} g$ and ${}^{(3)} \tilde{g}$ with respect to
the $G_{ijkl}$ tensor.
More explicitly, one
read off the first natural candidate for a metric
on the configuration space (the space of three metrics)
from the structure of the kinetic term in the ADM-Hamiltonian
\begin{equation}
H
=\frac{1}{2} G_{(ij)(kl)}\pi^{ij} \pi^{kl} - \frac{1}{2}
g {}^{(3)}R
=\frac{1}{2} G_{AB}\pi^{A} \pi^{B} - \frac{1}{2}
g \; {}^{(3)}R = 0
\end{equation}
which yields an expression like\footnote{Note that the supermetric,
in this notation (cf.\ C.W. Misner (1972)) differs by a conformal
factor ($\sqrt{g}$) from the expression by
DeWitt (DeWitt (1967)). This is allowed since the
ADM action is invariant with respect to
conformal transformations. I do not want to get entangled in too
much detail concerning this point but refer to
M. Biesiada and S.E. Rugh (1994).
}
\begin{equation}
G_{AB} \equiv
G_{(ij)(kl)} = \frac{1}{2} (g_{ik}g_{jl} + g_{il}g_{jk}
- 2 g_{ij}g_{kl}) \; \; .
\end{equation}
This step corresponds to reading off the ``usual Euclidian metric''
$a^{ij}$  from a non-relativistic Hamiltonian\footnote{
Instead of an analogy with a free particle
in special relativity, with Hamiltonian $H = \frac{1}{2}(\eta^{\mu \nu}
p_{\mu} p_{\nu} + m^2)$ (cf.\ C.W. Misner (1972), p. 451) we
want to point to (cf.\ Biesiada and
Rugh (1994)) the complete one-to-one correspondence between construction of
the superspace and the dynamics of a non-relativistic
particle with the Hamiltonian $H = \frac{1}{2} a^{ij} p_i p_j +
V(\BFACE{q})$ reduced to geodesic flow by virtue of the Maupertuis principle.
}
\begin{equation} \label{nonrelHamiltonian}
H =\frac{1}{2} a^{ij}p_i p_j + V(\BFACE{q}) = E  \; \; .
\end{equation}
Some may stop at this point and say that $G_{ijkl}$ is the superspace
metric. However the dynamics of the
three-metric ${}^{(3)} g$ (the collapse orbit) is not
yet a geodesic flow (cf.\ Biesiada and Rugh (1994)) with respect to
this superspace metric.

By conformal transformation of the metric
$$ \tilde{G}_{AB} = 2 (E - V) G_{AB} = (-2 V) G_{AB}
= {\cal R} G_{AB} = g {}^{(3)} R G_{AB} $$
(where the role of the potential $V$ in the general relativistic context
is played by the quantity $V = -
\frac{1}{2} g {}^{(3)} R \equiv - \frac{1}{2} {\cal R}$)
and rescaling of the parametrization of the collapse
orbit (cf.\ detailed discussion in Biesiada and Rugh (1994))
\begin{equation}
d \tilde{\lambda} = 2(E - V) d \lambda =
(- 2 V) d \lambda = {\cal R} d \lambda =
g {}^{(3)} R \; d \lambda
\end{equation}
one obtains (cf.\ also e.g. C.W.Misner (1972))
that, with respect to this new parameter $\tilde{\lambda}$
and with respect to the conformally rescaled
metric $\tilde{G}_{AB}$, the
three-metric ${}^{(3)} g$ is now an affinely
parametrized geodesic
\begin{equation}  \label{WheelerSupergeodesic}
\frac{d^2 g^A}{d \tilde{\lambda}^{2}} +
\tilde{\Gamma}^{A}_{BC} \frac{dg^B}{d \tilde{\lambda}}
\frac{dg^C}{d \tilde{\lambda}} = 0 \; \; .
\end{equation}
There is no ``magic'' involved (we have ``magic without magic''):
Information about the potential
$V = -\frac{1}{2} g {}^{(3)}R$
has simply been completely encoded in the mathematical definition
of the superspace-metric $\tilde{G}$
which generates the evolution of ${}^{(3)}g$
as geodesic motion (affinely parametrized if
one uses the rescaled parameter
$\tilde{\lambda}$).\footnote{In a completely similar
manner (cf.\ Biesiada and Rugh (1994)) one maps the
non-geodesic motion of the configuration space variable
$q$ governed by the non-relativistic Hamiltonian
(\ref{nonrelHamiltonian}) to affinely parametrized geodesic
motion by conformal transformation $a_{ij} \rightarrow
2(E - V(\BFACE{q})) a_{ij}$ of the metric $a_{ij}$ and rescaling of
the time parameter $dt \rightarrow 2 (E - V(\BFACE{q})) dt$.
}

\vspace{1.0 cm}

$\bullet$
The distance measures induced by the
Wheeler superspace metric $G_{ijkl}$
have the {\em good} property
(which is not shared by artificially introduced Euclidean distance
measures) of being invariant under canonical coordinate transformations.
Thus, if we have a change of coordinates (configuration space variables),
$$g^A \rightarrow g^{*A} = g^{*A} (g^A) $$
the distance measure $ds^2 = G_{AB} \delta g^A \delta g^B $ is invariant
under such transformations, since $G_{AB}$ transforms properly as a
tensor,
\begin{equation} \label{transformation}
G^{*}_{AB}
= \frac{\partial g^C}{\partial g^{*}_A}
\frac{\partial g^D}{\partial g^{*}_B} G_{CD}.
\end{equation}
One sees this by recalling that the coordinate transformation
$g^A \rightarrow g^{*A}$ induces the transformation of momenta
${\pi}_A \to {\pi}_A^{*} = \frac{\partial g^B}{\partial g^{*A}}\;{\pi}_B$.
Therefore the Hamiltonian $\cal H$ reads
$${\cal H} = \frac{1}{2}(G^{AB} {\pi}_A {\pi}_B - {\cal R}) =
\frac{1}{2}(G^{AB}\frac{\partial g^{*C}}{\partial g^A}
\frac{\partial g^{*D}}{\partial g^B}{\pi}^{*}_C {\pi}^{*}_D - {\cal R}) =
\frac{1}{2}(G^{*AB} {\pi}^{*}_A {\pi}^{*}_B - {\cal R}) $$
justifying our formula (\ref{transformation}).

\vspace{0.5 cm}

$\bullet$ A bad property of the distance measure induced by the
Wheeler superspace metric is that such a
distance measure is {\em indefinite}
(not positive definite).
I.e., it appears that one may have situations in which two
spacetime metrics (two mixmaster collapses at some given ``time'')
are at {\em zero} distance but evolve into a distance
different from zero. (This would correspond to a ``Lyapunov
exponent'' of ``$\infty$'' which is of course
an entirely different situation
from having a small finite distance
which in ``time'' evolves into a bigger finite distance).

Note, that it {\em is possible} to
construct a superspace metric $G_{AB}$ which is {\em positive definite}
so that two 3-geometries ${}^{(3)} g$ and ${}^{(3)} \tilde{g}$
are identical if and only if the distance between them is
zero, see e.g. discussions in B.S. DeWitt (1970).
However, such superspace metrics do not make the 3-geometries
evolve as geodesics.  \\

\noindent
{\bf Negativity of the Ricci scalar $R < 0$ as a local
instability criterium?} \\
As a possible criterium for instability of the mixmaster
gravitational collapse,
M. Szyd{\l}owski and {\L}apeta (1990) and
M. Szyd{\l}owski and M. Biesiada (1991)
proposed to look at the
Ricci scalar of the manifold on which the mixmaster dynamics generates
a geodesic flow. (The Maupertius principle was applied to the
Hamiltonian formulation of the mixmaster dynamics
as given in Bogoyavlensky (1985))
and investigate if one could extract
(coordinate invariant) information about instability
properties of the gravitational collapse in this way.

Application of the Maupertuis principle to
the Bogojavlenskii Hamiltonian formulation of the mixmaster dynamics
and the associated
induction of a natural distance measure on the three metrics,
is (cf.\ M. Biesiada and S.E. Rugh (1994))
exactly to implement the
dream by Wheeler (described, e.g., in C.W.Misner (1972) for the mixmaster
metric) to have a superspace metric
``$G_{ijkl}$'' (we call it Wheelers superspace metric) -
with respect to which the three metrics move along geodesics -
and use that metric as a natural distance measure on the space of
mixmaster three-geometries.
Thus, the Ricci scalar calculated
from the Hamiltonian introduced by Bogoyavlenskii,
cf.\ e.g. Szyd{\l}owski and Biesiada (1991),
is exactly to calculate the Ricci scalar of the
Wheelers superspace metric and the criterion $R < 0$ implies that
one has {\em local} instability at least in one direction
(as given by the
{\em geodesic deviation equation} for two
nearby mixmaster metrics
investigated with Wheelers superspace metric as
a distance measure).

An obstacle to this approach (of using the conformally
rescaled Wheeler superspace metric), besides
the obstacle that Wheelers superspace metric is indefinite,
is the accompanying introduction
of a host of singularities of the superspace metric
which makes the original dream of Wheeler
troublesome to achieve
(i.e. the three metrics ${}^{(3)} g$ be geodesics
w.r.t. a superspace metric)
- even in the restricted class of mixmaster
three geometries. Invariants calculated from
the superspace metric, e.g. the Ricci scalar, inherit the singularities
of the Wheeler superspace metric. In particular, the Ricci scalar
quoted e.g. in M. Szyd{\l}owski and M. Biesiada (1991)
has such singularities.
This obstacle has also recently been emphasized
by A. Burd and R. Tavakol (1993) and was previously
discussed in C.W. Misner (1972).

However, as a {\em local}
instability criterium, $R < 0$ indicates - at
nonsingular points where it is defined
- the {\em local} exponential instability in some directions of
the configuration space  w.r.t. the indefinite superspace metric.

\vspace{0.5 cm}

Only insufficient attention, in my opinion,
has been paid towards the
often rather arbitrary distance measures
on the space
of three-metrics which are introduced in
previous studies calculating Lyapunov exponents, etc.

It thus seems worthwhile to
investigate if it is possible
to construct and use good ``gauge invariant'' distance measures
(for example the Wheeler
superspace metric $G_{ijkl}$, with or without
conformal transformation of $G_{ijkl}$, or other distance measures)
on the solution space of three-metrics and if one can
use such distance measures to discuss instability properties
of e.g. the gravitational collapse in a way which
is invariant under some (large) class of gauge transformations.

\subsection{The mixmaster collapse as an interesting
laboratory for testing
ideas about how to apply chaos concepts in the context of general
relativity}

The structure of the mixmaster gravitational collapse
model is so simple\footnote{In the approximation where
the walls are infinitely hard we
have an example of so-called ``algebraic chaos''.
All the interesting dynamics (the chaos)
lies in the transition algebra associated with
the bounces against the wall boundaries.}
that it is not unreasonable to
investigate (cf.\ e.g. Contopoulos \etal (1993))
whether the model should turn
out to be integrable, after all.
I.e., there could be several additional ``hidden''
symmetries in the governing Einstein equations of motion
besides the Hamiltonian.

\begin{quotation}
\noindent
{\em A model like the mixmaster model collapse
is very well understood in terms
of the combinatorial model by Belinskii et al
(controlled by simple maps). However, if we cannot agree on
how to construct indicators of chaos
(from the full phase flow) in this simple example,
how could we then, in principle, dream about constructing
chaos indicators which can deal with more complicated
situations?  }
\end{quotation}

Whether we can assign any invariant
meaning to a concept like ``metrical chaos'' (or
even more ambitious: ``turbulence in spacetime metrics'')
is a theoretical problem of intrinsic interest
in the mathematical study of the dynamics of
the full non-linear Einstein equations.

\section{IS ``CHAOS'' A GENERIC FEATURE OF THE EINSTEIN EQUATIONS?}

\indent
The case study of our mixmaster metric (\ref{metric}) shows that
the ``generic'' expectation of the self-interacting gravitational field
to generate space-time ``chaos and turbulence''
(for strong field strength, e.g. when probing the Einstein equations
near space-time singularities where invariants
like $ R_{\alpha \beta \gamma \delta}R^{\alpha \beta \gamma \delta} $
of the tidal field curvature tensor blow up without upper bound)
is reflected even in such
a {\em highly symmetric} model. This may be considered a very strong property:
\begin{quotation}
\noindent
{\em If such a ``simple" (finite degrees of freedom) deterministic system
shows a chaotic, complex dynamical behavior, then one expects this to
be even more the case in complicated scenarios with a larger number of
interacting degrees of freedom!}
\end{quotation}
In this sense the ``mixmaster" toy model of a gravitational collapse
should be considered as a promising
starting point of subsequent investigations of
more complicated situations (with less symmetry) of the
classical field equations in general relativity.

Compare, e.g,
with the famous Lorentz attractor model (Lorentz, 1963)
for turbulence which is a crude starting
point for an investigation of some properties of
hydrodynamical turbulence. Note, however, that the Lorentz attractor is
a highly truncated model, where only
very few modes of the Fourier expanded equations are
kept in the model; The governing equations for the
mixmaster, $SU(2)$ homogeneous,
metric are {\em the exact Einstein equations}
- involving few degrees of freedom
because of the (a priori) made symmetry-ansatz of the
metric.

\subsection{Aspects of fragility of the gravitational collapse?}

However, also for governing equations of motion
which are modified slightly relative to the Einstein equations
one would expect that ``metrical chaos''
is a generic feature when probing the metric
in regions where the curvature gets strong and non-linear
self-interactions of the involved gravitational fields
get important (e.g. on approach
to curvature singularities).

Let us say that we modify
the Einstein-Hilbert action slightly,
\begin{equation}
S = \int d^4 x \sqrt{-g} \left\{ R + \epsilon \Phi (R, R_{\mu \nu}R^{\mu \nu},
R_{\mu \nu \rho \sigma} R^{\mu \nu \rho \sigma}) \right\}
\end{equation}
where $\Phi$ is a general analytic function and $\epsilon$ is a small
perturbation parameter.
See also discussions in
R.K. Tavakol (1991) and A.A. Coley and R.K. Tavakol (1992).

Chaos in the evolution of the metric (\ref{metric}) should
not merely be a property of the {\em exact} Einstein equations
(i.e. for $\epsilon = 0$), but also for the modified equations,
corresponding to the inclusion of
higher order curvature terms (i.e. for $\epsilon \neq 0$) in the action.
Such higher order curvature terms
- generically arising in underlying theories of gravity like,
for example,
string theories - will {\em blow up} towards the ultraviolet (for example
towards the big crunch singularity of the mixmaster metric).

Thus, when curvature gets big there is no reason why higher
order curvature terms should not appear in the Lagrangian,
$ {\cal L} = \sqrt{-g} \left\{ R + \alpha_1 R^2
+ ......
) \right\}. $ Seen in this perspective, it is puzzlesome
that the chaotic mixmaster oscillations
are unstable and stop at some stage of the collapse
(and proceed in a non-oscillatory manner) if
higher order curvature terms, like $R^2$ terms,
are included in the Lagrangian, according to
J. Barrow and M. Sirousse-Zia (1989).
See, also, J. Barrow and S. Cotsakis (1989) and a recent study
by S. Cotsakis \etal (1993).

A (premature) conclusion arises, that the
mixmaster oscillatory behavior - or the very
fact that the metric (\ref{metric}) is evolving in a
complex (unpredictable) manner towards the big crunch singularity -
is {\em not} structurally robust
towards modifications of the
Einstein-Hilbert action and such modifications
are certainly expected when length scales get small, of the order
of Planck sizes, say.

\subsection{A remark on fragility and structural stability in a wider
context}

In a broader context R. Tavakol (1991) discusses the concept of
``structural {\em fragility}'' and Coley and Tavakol
(1992) advocate that models
which are structurally fragile are
``generic'' among the set of models usually employed
in cosmology.

\begin{quotation}
\noindent
{\small
We have already mentioned  that
isotropization of anisotropic spacetime metrics like the compact mixmaster
metric (\ref{metric})
occurs too slowly within the standard hot big bang model
to explain the remarkable degree of isotropy in the
microwave background radiation observed today.
Thus, isotropy is an unstable property of physically realistic
cosmological initial conditions - yet it is an observed property
of the Universe. Another, famous, example
is the ``flatness'' problem (R.H. Dicke):
Why is the Universe today so near the boundary between open and closed,
i.e. so nearly flat? This is a puzzle because
one may show that $\Omega = \rho/\rho_c = 1$ is an
{\em unstable equilibrium point} of the evolution of the
standard hot big bang theory (i.e. it resembles the situation
of a needle balancing vertically on its tip).
In order for
$\Omega$ to be somewhere in the allowed range today,
$\Omega \sim 1/10 - 10 $, this parameter had to be equal to one
to an accuracy of 49 decimal places if we consider
times around $10^{-35}$ second, say, after the big bang!
This finetuning problem is, however,
no problem at all relative
to the completely vanishing probability for us to have a second law of
thermodynamics which R. Penrose (1989) estimates as one
chance in $\sim e^{10^{123}}$! (The Universe was created in a state
of very low entropy compared to what it might have been).

These finetuning problems are to some extent relaxed within
the inflationary Universe concept.
Inflation
predicts, for example,
that the value of $\Omega$ today should equal one to an accuracy of
about one part in 100.000, cf.\ A.H. Guth (1992).
}
\end{quotation}

\noindent
In view of the variety of simplifying assumptions in the formulation
of any mathematical model of a physical phenomenon in Nature and
the fact that such models have a tremendous power
in describing such phenomena, it appears that such a principle of
``structural fragility'' cannot be implemented in Nature to its
extreme limit.

The extreme opposite principle, the so-called
``random dynamics'' principle,
which concerns the structure of Natural laws themselves,
has been put forward by
Holger Bech Nielsen (at the Niels Bohr Institute).
He contemplates (cf.\ e.g. H.B. Nielsen (1976, 1981))
- and has explored in many contexts
(cf., e.g., C.D. Froggatt and H.B. Nielsen (1991)) -
a principle of the (extreme) structural stability
of the Natural laws (as they appear to us) -
against (almost any) modification of them in the
ultraviolet.

It is a principle which postulates the structure of Natural laws
(at our scales) despite a substantial lack of
structure of laws at a more fundamental level
and the dream, though very ambitious here,
has some analogy to, say, the phenomenon of ``universality in chaos'', i.e.
the realization that many phenomena, like the universality of
Feigenbaum's $\delta$ in a bifurcation scenario,
do not hang severely on the {\em microscopic details} of the underlying
dynamics.\footnote{I.e. there is a huge universality class of
dynamical systems, e.g. all one-dimensional maps with a
quadratic maximum (cf.\ also Cvitanovic (1989)), which generates
the same Feigenbaum $\delta$.} In the context
of our discussion of the mixmaster collapse it also reminds us
of the ``chaotic cosmology'' concept (see also sec.6.3),
developed by C.W. Misner, which
attempts at creating our present Universe from
(almost) arbitrary initial conditions.

If the universality class of fundamental models which leads to the
same infrared phenomenology (the standard model of the electroweak
and strong forces, say) is not too narrow it leaves
(logically) the possibility of having
``{\em chaos in the Natural laws}'' (at the most fundamental ``scales'')
- i.e. the fundamental structure may be selected randomly
if it is just selected from the universality class of models
restricted by the boundaries set by the ``universality class''

However, such boundaries
(if these could be made precise enough to be stated formally,
say, by mathematical formulas) in a certain sense themselves have
status as ``Natural laws'' (regularities) - and so on, ad
infinitum. Therefore, this way, one does not circumvent the
concept of some ``Natural laws'' and ``regularities'' to be implemented in
our Universe.

Preliminary discussions of
boundaries that such a ``random dynamics''
principle may come across
(is it possible to define a concept of
``structural stability for Natural laws''?)
is contained as sec.3 in H.B. Nielsen and S.E. Rugh (1993).
Boundaries that such a project come across are of interest
because hereby one may gradually try to isolate
elements in our description of Nature which can not
(so easily) be modified (and thought differently).
If one contemplates - cf.\ e.g. S. Weinberg (1992) - that the Natural
laws and parameters are the only {\em self-consistent} set of
laws and parameters imaginable, the ``random dynamics'' project
contributes with an analysis of what we could mean by the
word ``self-consistent''.

\subsection{Back to spacetime chaos: ``Chaos'' in other solutions
in classical general relativity?}

It would be a good idea to look for other
toy models of spacetime metrics,
which would exhibit chaotic behavior, e.g.
in connection with solutions obtained in the
(fast advancing) discipline of
{\em numerical relativity} - if one has
good ``codes" available to solve the
Einstein equations without too many symmetry assumptions!

If  possible, it would be very nice if one could
relax the symmetry properties and study the structure of the
cosmological singularities of space-time metrics with only two
spacelike Killing vector fields, say, instead of three. Especially,
it would be wonderful if one could relax the symmetry properties and
generalize the mixmaster gravitational collapse this way.

It is a reasonable expectation that
the severe non-linearity present in the Einstein equations
implies ``chaotic" (non-integrable) solutions
almost ``unavoidably" in scenarios involving high gravitational
field strength (e.g. near space-time singularities).
This ``metrical chaos'' should therefore  be thought of
as a generic feature of the non-linear
Einstein equations in such strong gravitational fields
- and not as a feature connected, merely, to the near-singularity behavior of
some (small) subclass of  metrics, i.e. the spatially homogeneous
toy-models of gravitational collapses.

\begin{quotation}
\noindent
{\small
{\bf On the ``universality class'' for
the intermittent behavior of the ``u'' parameter (discussions
with V.N. Lukash)} \\
Even in the (small) class
of spatially homogeneous metrics
we have, according to Peresetzki and V.N. Lukash, cf.\ Lukash (1983),
as regards the evolution of the
already mentioned ``Lifshitz-Khalatnikov'' parameter ``u",
that their evolution is chaotic (and actually given by the
Farey/Gauss map)
if the invariance groups/algebra's - according to the
standard classification of
three-dimensional Lie-algebras by Bianchi (1897) -
(of the three-spaces) are chosen as
type $VIII$ and $IX$ (type $IX$ is actually the mixmaster space)
in the case of the {\em empty space} equations
$ R_{\mu \nu} = 0 $, but in fact
{\em all the Bianchi metrics} - except for type $I$ (a flat
space) and $V$ (where the curvature is isotropic) - in situations
(a perfect fluid energy momentum tensor is assumed) where
$T_{o \nu} \neq 0$ (but $T_{ij} = 0$), cf.\
Peresetzki and V.N. Lukash.

In fact, one can  show that the spatially homogeneous
toy-model metrics are driven into ``chaos", as regards the evolution of the
parameter ``u", by the presence of {\em curvature anisotropy}
of {\em spiral} character (spiral character:
At least one structure constant is different from zero,
${\cal C}^{i}_{jk} \neq 0$, where all $i,j,k$ are different,
in the Lie-algebra formed by the three spacelike Killing vector
fields); For a space of type $I$ there is no curvature at
all, for a metric of type $V$ it is isotropic and - as a curiosum -
for a space of Kantowski Sachs type,
the anisotropic curvature is not of spiral character!
(V.N. Lukash, private comm., NBI, and V.N. Lukash, Doctorate Thesis
(1983)).
}
\end{quotation}

\noindent
$\bullet$ {\bf Solutions with other ``matter" sources:}   \\
What if other sources than perfect fluid matter are included in the
study? It is known, e.g., that the $SU(2)$ Yang Mills equations in
the limit of infinite wavelength (i.e. the spatially homogeneous mode) admit
``chaotic" solutions\footnote{For more recent discussions of
spatially inhomogeneous chaos in
non-Abelian gauge theories see also e.g. M. Wellner (1992),
B. M\"{u}ller \etal (1992) and T.S. Bir\'{o} \etal.
(1993). A detailed description of chaos in non-Abelian gauge
theories will also be found in S.E. Rugh (1994).}
(cf., e.g., S.G. Matinyan (1985)
and G.K. Savidy (1984))
with a positive metric entropy $ h = \int \sum \lambda^{+} d\mu > 0 $.
(In this limit - the
``extreme infrared" - it is also a famous result of linear stability
analysis that the non-Abelian Yang-Mills configurations are
{\em unstable} under small disturbances, cf.\
N.K. Nielsen and P. Olesen (1978)
and S.J. Chang and N. Weiss (1979)).

If the Einstein and $SU(2)$ Yang Mills equations are coupled, e.g. in the
context of a mixmaster toy model study of a gravitational collapse,
some very complicated and chaotic behavior
must arise!? \footnote{For a study of stability properties of
such coupled $SU(2)$ Einstein-Yang-Mills-Higgs equations
(in the context of monopole and black hole solutions) see
also Straumann and Zhou (1990)
and G.W. Gibbons (1990). } \\

\noindent
$\bullet$ {\bf Spatially inhomogeneous gravitational collapses:}  \\
What is the effect of taking into account spatial degrees of freedom in
the model; i.e., the inclusion of some higher order Fourier modes
(``spin 2" gravitational waves)
superimposed (in a gauge-invariant way?) on
the $SU(2)$ homogeneous (zero-mode expanded)
but anisotropic gravitational collapse
?
Perhaps small amplitude gravitational waves {\em die out} towards the
singularity?
It is, however, far from
obvious that such perturbations
will not do the opposite: Blow up and significantly
alter the evolution of the metric on approach
to the space-time singularity!
(Cf, also, remarks by R. Ove (1990) in a
slightly different context).

Note, that according to
Belinskii, Khalatnikov and Lifshitz (1982),
the simple BKL-combinatorial model for the
alternation of Kasner exponents (derived for the spatially homogeneous
models) remains valid locally
(i.e. in the neighborhood of every space point)
in the general spatially inhomogeneous gravitational
collapse. See also discussion in Ya.B. Zel'dovich and
I.D. Novikov (1983) \S 23.3.  \\

\noindent
$\bullet$ {\bf Chaos in colliding gravitational waves?} \\
One may speculate whether yet another example of chaos in the
classical Einstein equations occurs in the
dynamics of the metric field (and extracted gauge invariant
quantities) when two non-linearly interacting plane gravitational waves
extending to infinity
collide and generate black hole singularities -
via {\em nonlinear focusing} effects - of the ``spacetime" structure
\footnote{Cf., e.g., S. Chandrasekhar, talk at the
``300 years of Gravity" celebrations at D.A.M.T.P.
(1987).
However, it should be realized that the
self-focusing singularities most likely arise
because they are colliding plane waves which extend to
{\em infinity} -
and such waves do not exist!
(M. Demianski and D. Christodoulou, pers. comm.).}.
The detailed nature of the evolutionary tracks of the gravitational
field ``$ \left\{ g_{\mu \nu} (x) \right\} $ `` - when
``bouncing off" parts of the spacetime manifold in the form of black hole
singularities - must almost unavoidably (?)
be {\em chaotically unpredictable!}
Highly nonlinear effects (when the curvature gets strong)
determines the detailed nature of the dynamics.

\section{IMPLICATIONS OF ``METRICAL CHAOS'' ON THE QUANTUM
LEVEL?}

Since the classical Einstein equations are scale-independent,
there is - a priori - no scale
built in our toy-model study (\ref{metric})
of the oscillatory, chaotic collapse - when evolved according to the
classical Einstein equations.
However, if we - by hand -
put in a length-scale (lets say ``the size" of the Universe
$ \sim c/H \sim 10^{28} cm $, or somewhat less)
as a starting condition on the scale factors
of the gravitational collapse,
one notes (cf.\ also the displayed fig.5)
that already after few major cycles of bounces
some of the length scales of the gravitational collapse get
very small and, in fact, very fast reach Planck-scales
$$ l_{Pl} = \sqrt{\frac{\hbar G}{c^3}} \sim 10^{-33} cm $$
and, correspondingly, the space-time curvature blows up
$$
\sqrt{R_{\alpha \beta \gamma \delta} R^{\alpha \beta \gamma \delta} }
\sim l_{Pl}^{-2} = \frac{c^3}{\hbar G}
\sim 10^{66} cm^{-2} \; \; .
$$
Such scales may set characteristic
scales when quantum effects become important
and the gravitational degrees of freedom should be quantized.
\begin{quotation}
\noindent
{\em Thus, the mixmaster gravitational collapse turns very
fast into a quantum problem! (and is chaotic in a
domain where it should be quantized)}
\end{quotation}
The fact that the scale functions for the mixmaster
spacetime metric change so fast can be understood
from the observation that the Einstein equations for the metric
(\ref{metric}) involve exponential functions
(cf.\ (\ref{numspacespace}) or (\ref{decompone}))
and such exponential functions are well known to give out
``large numbers'' very fast!  \\
We are lead to consider our toy-model study of a chaotic gravitational
collapse as:
\begin{quotation}
\noindent
{\bf 1.} Merely a look into a ``chaotic sector"
of the solution space to the {\em classical} Einstein equations.
One notes
that the classical Einstein equations do not internally lead to
contradictions at any scale
(see also Landau and Lifshitz, Vol.II, \S 119)
and considered as a {\em mathematical study} of the
non-linear Einstein equations,
``metrical chaos'' is a concept (not yet properly defined)
of interest in itself! \\
{\bf 2.} Presenting some evidence in favor of
speculations that
the process of quantization of the gravitational field - occurring at
small lengthscales and correspondingly
high curvatures ($\sqrt{R_{\alpha \beta \gamma \delta}
R^{\alpha \beta \gamma \delta}} > l_{Pl}^{-2} $) -
in generic situations rather
will occur ``around" field configurations which are
``turbulent" (space-time chaos) than around well behaved (integrable)
solutions\footnote{Thus, our ``intuition" (on gravitational
collapses) based on the well known textbook example of the
FRW-type ``big bang"/``big crunch" collapse - with simple dynamical behavior -
is a non-typical classical field configuration to have in mind
when ``quantizing gravity".
}. In the semiclassical limit, which
is built around the classical configurations,
this may be of importance.
\end{quotation}

Does classical chaos obstruct the
quantization of the theory?
Lee Smolin (1990) makes
the point that one will - very likely -
only make progress on the construction
of quantum general relativity to the extent that one uses
{\em non-trivial
information} about the {\em dynamics} of general relativity.

In perturbation theory and Monte-Carlo simulations
(methods of
quantization in which no special information about the dynamics of the
theory is used) chaos usually does not obstruct the quantization.
Ashtekhar and Pullin (1989), for example, emphasize that chaotic
behavior of Yang-Mills fields has not obstructed the quantization
of the theory. Such questions, being far from settled, are
central topics in the
theory of ``Quantum Chaos'' and
will also be discussed (and references may be found)
in S.E. Rugh (1994).

Spatially homogeneous cosmologies - having a finite number of
(anisotropy) degrees of
freedom - are often considered a laboratory for testing ideas
about the application
of quantum principles to gravitational degrees of freedom.
Examples could be the Robertson Walker, the
Kasner or the mixmaster spaces.
Now, the Kasner or FRW solutions (being integrable) do not display any
``chaos" at the classical level, while the evolution of the mixmaster
metric is chaotically complicated (and in metric $t$ time
the dynamical evolution condenses, as regards the
oscillations of the anisotropy degrees of freedom, infinitely towards the
two spacetime singularities).

\begin{quotation}
\noindent
{\em Does this striking difference in
the dynamical behavior at the classical level
mirror itself in quantum effects in
the ``quantized" model cosmologies? I.e. what
is the effect of applying quantum
principles to a
chaotic, dynamically complicated gravitational collapse -
the mixmaster, say - compared to more symmetric
gravitational collapses (of the FRW- or Kasner-types, say)
which are integrable all the way to the ``final crunch" singularity? }
\end{quotation}

In a {\em semiclassical} treatment (i.e. when the collapse has not yet
reached Planck lengths) the quantum solution builds
around the classical dynamics. So in the semi-classical limit
it of course, for that reason,
matters what the classical solution looks like!

What about the {\em fully quantized} regimes, when the length scales of the
gravitational metric have collapsed down to Planck sizes, or less?
It is hard to know what happens in the fully quantized regime of our
gravitational collapse (despite that it reaches these small scales very fast)
let alone that no good
theory of quantum gravity is available.

\subsection{The Wheeler-DeWitt quantization
of the gravitational collapse}

\indent Since the pioneering works of P.A.M. Dirac in the fifties
(cf., e.g., P.A.M. Dirac (1959) and references therein) and
R. Arnowitt, S. Deser and C.W. Misner (1962),
it is well known that a
Hamiltonian formalism (ADM 3+1) may be set up for the Einstein
equations and this Hamiltonian formalism is often
used in a ``toy-application'' of quantum principles (the
Wheeler-DeWitt equation)
to such a model of an anisotropic
gravitational collapse.
See, e.g., the recent bibliographies on these topics
by C. Teitelboim and J.J. Halliwell in
S. Coleman \etal (1992).

More explicitly, the Wheeler-DeWitt equation
(J.A. Wheeler (1968), B.S. DeWitt (1967))
arises after appropriate translation of the coordinate and momentum
variables into operators
by imposing the Hamiltonian constraint \footnote{Such
constraints always arise in field theories where the field
variables have a ``gauge arbitrariness''
(e.g. also in the Hamiltonian formulation of Yang-Mills theory)
and the vanishing of a
Hamiltonian, like in equation (\ref{WheelerDeWitt}),
is a characteristic feature of theories which are invariant
under reparametrizations of time (cf., e.g. R.M. Wald
(1984), appendix E).
There is, however, a crucial difference between gauge
theories and {\em parametrized} theories like
general relativity (also invariant under reparametrizations
which involve time). In a parametrized theory the
{\em constraints} give all the {\em dynamics}. See also e.g.
J.B. Hartle and K.V. Kuchar (1984).}
as an operator constraint on the ``quantum wave function $\Psi$''
of the gravitational collapse.

For the sake
of notational simplification,
we put $16 \pi G = 1$, $\hbar = 1$, etc.,
and the Wheeler-DeWitt equation (the operator form of the Hamiltonian
constraint) has the form
\begin{equation} \label{WheelerDeWitt}
\hat{\cal H} \; \Psi = (G_{ijkl}
\frac{\delta^2}{\delta {g}_{ij} \delta {g}_{kl} }
+ g^{1/2} \; {}^{(3)}R -
{}^{(3)}{\cal L}_m ) \; \Psi = 0
\end{equation}
where $\Psi = \Psi({}^{(3)} g) = \Psi (g_{ij})$
is a quantum mechanical wave-function
(for our gravitational collapse), ${}^{(3)} R $ is the curvature
scalar of the three-metric in question, ${}^{(3)} {\cal L}_m $
denotes the three-dimensional
Lagrangian density of the (non-gravitational) matter
fields and
$$G_{ijkl} \equiv \frac{1}{2} g^{-1/2} (g_{ik} g_{jl}
+ g_{il} g_{jk} - g_{ij} g_{kl} ) $$
is the Wheeler-DeWitt superspace metric (cf., e.g, discussion
in B.S. DeWitt (1967)) on the space of the
three-geometries.

A substantial amount of uncertainty could be expressed as to whether a
toy-application of quantum principles along such lines
makes good sense when implemented in a context like
the mixmaster gravitational collapse (\ref{metric}). Some points of
uncertainty:

\begin{quotation}
{\small

{\bf 1.} The ``factor ordering'' ambiguity
in translating a classical equation like the classical Hamiltonian constraint
into an operator identity (involving non-commutating quantum operators)
is a severe problem.
For example, one may select a factor ordering
- in the case of our
mixmaster toy-model gravitational collapse -
that makes {\em any given function}
$\Psi$ (of the
mixmaster three-metric
variables
$\Omega, \beta_{\pm}$)
a solution of the Wheeler-DeWitt equation for this factor ordering!
(cf.\ investigations by Kuchar quoted in Moncrief and Ryan
(1991), p.2377).

{\bf 2.} It is not clear that quantum solutions
of the highest-symmetry models are approximations to quantum
solutions of models with less
symmetry (e.g. that ``quantizing'' the spatially homogeneous
mixmaster collapse is an approximation to some ``true quantum
gravity solution'', which includes space-dependent modes).
Cf. Kuchar and Ryan (1989).

{\bf 3.} If space-dependent gravitational waves are superimposed
on the mixmaster metric and
taken into account perturbatively in
the Wheeler-DeWitt equation one will most likely end up
with non-renormalizable divergencies?

{\bf 4.} The symbol ``$\Psi$" (the ``wave-function'' of
the quantized gravitational collapse including matter
degrees of freedom etc.) appearing in equation (\ref{WheelerDeWitt})
does not have a straightforward interpretation if it also includes
the observer, for instance in an application of quantum principles to
an entire model-cosmology.
One of the difficulties one encounters is how to
allow for an experimentalist with a ``free will'' to
perform experiments, if {\em everything} (including the
experimentalist) is described by a big
(completely deterministic) ``Schr\"{o}dinger equation''?
The very notion of an experiment seems meaningless, since the
measurement was determined in advance (at the ``big bang'')?
\footnote{See also, e.g., chapt. 15
``Quantum Mechanics for Cosmologists" in J.S. Bell (1987).}

}
\end{quotation}

In lack of a truly renormalizable underlying theory of
quantum gravity, one may choose to ignore
such obstacles and consider a toy-application of the Wheeler-DeWitt
equation (\ref{WheelerDeWitt})
implemented in the case of the mixmaster
gravitational collapse and hope that it retains aspects of the
full theory.\footnote{According to R. Graham (1993):
``Given a dynamical system of Hamiltonian form, the temptation
to quantize seems to be irresistible''.}

What do we obtain? We yield
an operator constraint equation which
roughly  has a  mathematical form as given below
if we consider the simplest possible choice of factor ordering
of the operators
(we neglect non-gravitational matter fields)
\begin{equation} \label{WheelerDeWittBIX}
\left\{
\;- \frac{\partial^2}{\partial \Omega^2} +
\frac{\partial^2}{\partial \beta_{+}^2} +
\frac{\partial^2}{\partial \beta_{-}^2} +
\; e^{-6 \Omega} \; \; {}^{(3)} R \; (\Omega, \beta_{\pm}) \;
\right\}  \;
\Psi (\Omega, \beta_{\pm}) = 0  \; \; .
\end{equation}
It looks like a
{\em zero energy} ``Klein-Gordon-Schr\"{o}dinger'' wave
equation.
$\Omega, \beta_{+}, \beta_{-}$
are the degrees of freedom in the
parameterization (\ref{metricADMvariables})
of the metric (\ref{metric}) and $\Omega, \beta_{+}, \beta_{-}$
are
one-to-one related to the scale functions
$a,b,c$ in (\ref{metric}) via
\begin{eqnarray}
a &=& \exp (-\Omega + \beta_{+} + \sqrt{3}\beta_{-})
\nonumber \\
b &=& \exp (-\Omega + \beta_{+} - \sqrt{3}\beta_{-})
\nonumber \\
c &=& \exp (-\Omega - 2\beta_{+})
\nonumber
\end{eqnarray}
$ \Omega = -\frac{1}{3} \ln (abc)  \propto - \log (Volume) \;$
may function as a ``time'' parameter of the gravitational collapse
since, as we have already noted, the
three-volume $V = 16 \pi^2 abc$  of the metric (\ref{metric})
is monotonically decreasing
on approach to the space-time singularity and the variables
$ \beta_{+} = \frac{1}{6} \ln (ab/c^2), \; \;  \;
\beta_{-} = \frac{1}{2\sqrt{3}} \ln (a/b) $
denote the ``state of anisotropy''
of the gravitational collapse.
The three-curvature scalar on the three-space,
$ \; {}^{(3)} R \;$, can be calculated
from the metric (\ref{metric}) and has the  following form
\begin{eqnarray}  \label{threeR}
{}^{(3)} R & = &
- e^{2\Omega} \left\{
e^{4\beta_{+}}( \cosh(4 \sqrt{3} \beta_{-}) - 1) +
\frac{1}{2} e^{-8 \beta_{+}} -
2 e^{-2 \beta_{-}}( \cosh(2 \sqrt{3} \beta_{-}) \right\}  \nonumber \\
& = &
- \frac{1}{2 a^2 b^2 c^2} \left\{
(a^2 - (b+c)^2 )(a^2 - (b-c)^2 )\right\} \; \; .
\end{eqnarray}
The structure of ${}^{(3)} R$  gives rise to a scattering
potential (in the initial stages of the collapse describing
soft scattering walls which
fast develop, however, into ``infinitely hard'' walls
when approaching the space-time singularity of the metric)
which makes the classical Hamiltonian dynamical problem resemble
that of a billiard ball being played on a billiard table
(the anisotropy plane
$(\beta_{+},\beta_{-}) \in $ {\bf R}${}^2$) with this
potential as the boundary walls.
Of course, in the isotropic case  $a = b = c = R/2$
we reobtain the expression for the three-curvature scalar
$ {}^{(3)} R = + 3 / 2 a^2 = + 6 / R^2 $
corresponding to the compact ($k=+1$) $FRW$ space.

The Wheeler-DeWitt quantized mixmaster collapse has been studied
by various authors, for example C.W. Misner (1972)
(see also, e.g., S.W. Hawking and J.C. Luttrell
(1984), R. Graham and P. Sz\'{e}pfalusy (1990) and
V. Moncrief and M.P. Ryan (1991), and references
therein, for a toy ``quantization" of the mixmaster metric).

Recently, exact solutions have been found by R. Graham
if an additional supersymmetry is introduced in the study,
R. Graham (1991, 1992).
(See, also, P.D. Eath, S.W. Hawking and O. Obreg\'{o}n (1993)).
These solutions describe virtual quantum wormholes, see also
R.Graham (1993).

In the context of the Wheeler-DeWitt equation, applied
to spatially homogeneous gravitational collapses,
we could hope (preliminarily) to
address the question: What is the effect of having
a chaotic gravitational collapse (as opposed to an
integrable solution) when quantizing it?
Since in the Wheeler-DeWitt equation
we are merely looking at the zero
energy solution $\hat{\cal H} \Psi = 0$,
it is at first sight not obvious how we can relate to, say,
an effect (cf.,e.g., M.V. Berry (1981))
like that  of nearest neighbour energy repulsions.
Prof. R. Graham, however,  pointed out that
C.W. Misner already twenty
years ago offered
a separation of the Wheeler-DeWitt equation in the region
sufficiently near the space-time singularity where the
scattering potential $ \; g \; {}^{(3)}R \; $
is approximately infinitely hard and where
the solutions to the Wheeler-DeWitt equation are related to
the spectral properties of the Laplace operator on the
Poincar\'{e} disc.

Before I sketch this last point
I will digress shortly
into a description of the mixmaster gravitational collapse
of remarkable beauty.

\subsection{The Poincar\'{e} disc description of
the gravitational collapse}

\indent The dynamics may be transformed
\footnote{Note, that the Poincar\'{e}
disc description of the mixmaster collapse
has gone almost unnoticed for twenty years. See, however,
J. Barrow (1982). Suddenly
several people looked at it again, J.Pullin (1990),
Graham and Sz\'{e}pfalusy (1990), (S.E. Rugh (1991)).}
into that
of piecewise geodesic motion in a
non-compact domain with triangular symmetry
(corresponding to the symmetry under the interchange
$a \leftrightarrow b \leftrightarrow c$ of the scale factors in the
metric (\ref{metric}))
on the two dimensional Poincar\'{e} disc
(the Lobachevsky space) of constant negative curvature!
As is well known, geodesic motion on surfaces of constantly negative
curvature
is a standard laboratory for testing ideas in classical and quantum
chaos, cf.\ e.g. Balazs and Voros (1986).
However, the mixmaster gravitational
collapse may very well be one of the
only physically motivated\footnote{If this system is not a ``physical
system'' in an {\em empirical} sense, it is certainly a physically motivated
system, being a (globally) perturbed FRW metric, evolved by the
full, non-linear Einstein equations.}
dynamical systems known, which realizes such
geodesic motion on surfaces of constantly negative curvature!
It is thus
clearly an interesting and beautiful aspect of the
mixmaster collapse which deserves further investigations.
(The Poincar\'{e} disc description has been emphasized
also by J. Pullin, recently, cf.\ J. Pullin (1990) and
R. Graham and P. Sz\'{e}pfalusy (1990)).

A very brief discussion
of the set of transformations which brings the mixmaster
gravitational collapse into that of geodesic motion on the Poincar\'{e} disc
is offered in Misner, Thorne and Wheeler (1973) \S 30.7.
I will not drift into details here - since that is far beyond the
scope of this presentation (a detailed description will be available in
S.E. Rugh (1994), where the ``Poincar\'{e} disc''
description is worked out and discussed in exhaustive detail).

The crucial observation
(which is originally due to C.W. Misner and D.M. Chitre) is
that while the original Hamiltonian
\begin{equation} \label{HamiltonianOmegaBeta}
{\cal H} =  \; - p_{\Omega}^2 + p_{+}^2
+ p_{-}^2 + g \; {}^{(3)} R (\Omega, \beta_+, \beta_-)
\end{equation}
has outward, time dependent expanding
potential boundaries
in the $\Omega, \beta_{\pm}$ variables,
a set of transformations
$$(\Omega, \beta_{+},\beta_{-}) \leftrightarrow (t,\xi,\phi)
\leftrightarrow (t,x,y)$$
may be devised
\begin{equation} \label{hyperbolicone}
\left(
\begin{array}{c}
\Omega - \Omega_0 \\
\beta_{+} \\
\beta_{-}
\end{array}
\right)
 =
e^{t} \;
\left(
\begin{array}{c}
\cosh \xi  \\
\sinh \xi \cos \phi \\
\sinh \xi \sin \phi
\end{array}
\right)
=
e^t \; \frac{1}{1-(x^2 + y^2)}
\left(
\begin{array}{c}
1 + (x^2 + y^2) \\
2 x \\
2 y
\end{array}
\right)
\end{equation}
which makes  the location of the potential boundaries
{\em time-independent} with respect to the new ``$t$'' time
parameter when we are in a region sufficiently near the space-time
singularity of the metric.
(The second part of the transformation (\ref{hyperbolicone})
is a mapping\footnote{There are some
discrepancies with the expressions in
J. Pullin (dec. 1990). I, however, get these set of transformations.}
of $(\xi,\phi)$ into coordinates $(x,y)$
inside the Poincar\'{e} unit disc ${\cal D}$ in the complex plane, see
below).

The transformations of the momenta $p_{\Omega}$, $p_+$, $p_-$
which appear in the Hamiltonian
(\ref{HamiltonianOmegaBeta})
are constructed to make the
transformations of coordinates and momenta become
{\em canonical} transformations,
$ p_{i'} = (\partial q^i/ \partial q^{i'}) p_i $,
and the Hamiltonian is (after proper rescalings)
transformed into a  Hamiltonian which,
in $(t, \xi, \phi)$ coordinates, has the  form
\begin{equation}  \label{Hamiltonxi}
\tilde{\cal H} =  -p_{t}^2 + p_{\xi}^2 +
\frac{p_{\phi}^2}{\sinh^2 \xi}
\; + \;
\tilde{V} (t, \xi, \phi) \; \; .
\end{equation}
The scattering potential
$\tilde{V} (t, \xi, \phi)$
has (in the asymptotic region
sufficiently near the singularity of the metric)
infinitely steep potential
boundaries and the location of these potential boundaries
(which show up to form a triangular domain ${\cal B} \subset {\cal D}$
inside the Poincar\'{e} disc, see below)
are {\em independent} of the new ``time'' variable $t$.\footnote{Note,
according to (\ref{hyperbolicone}),
that the new ``time'' parameter $t$ is a somewhat ``strange''
new time variable,
$$ t = \frac{1}{2} \ln ( (\Omega - \Omega_0)^2 -
(\beta_+^2 + \beta_-^2) ) = \frac{1}{2} \ln ( (\Omega - \Omega_0)^2 -
|| \vec{\BFACE{\beta}} ||^2 ) $$
which mixes the state of anisotropy $\vec{\BFACE{\beta}}$ (the position
of the ``billiard ball'') with the original $\Omega$-time variable,
$\Omega = -\frac{1}{3} \ln (abc)$, in the ADM-Hamiltonian variables.}.
The location of the potential boundaries are
in the $(\xi,\phi)$ space
determined by the equations, cf.\ MTW (1973) \S 30.7,
\begin{equation}  \label{wallstationary}
2 \tanh \xi = -  \frac{1}{\cos (\phi + m \frac{2 \pi}{3}) }
\; \; \; \; , \; \; \;
m = -1, 0, 1 \; .
\end{equation}
Having obtained in this way the stationary (in the $(\xi, \phi)$ plane)
scattering potential $ \tilde{V}(\xi, \phi)$ which is
zero inside the domain (we call this domain ${\cal B}$)
bounded by (\ref{wallstationary}) and $ \infty$
in the region outside this domain
(an approximation which is extraordinarily
good when sufficiently near the ``big crunch" singularity of
the mixmaster gravitational collapse)

\begin{equation} \label{wallstatichard}
\tilde{V} (t, \xi,\phi) \rightarrow
\tilde{V} (\xi,\phi)
= \left\{ \begin{array}{ll}
0 & \mbox{inside domain ${\cal B}$ }  \\
+ \infty & \mbox{outside domain ${\cal B}$ }
\end{array}
\right.
\end{equation}
\noindent
we have obtained a Hamiltonian (\ref{Hamiltonxi}) which,
in the interior of the scattering domain ${\cal B}$,
resembles the Hamiltonian for a free particle\footnote{Since
the Hamiltonian (\ref{Hamiltonxi}) ceases to be time-dependent
in the asymptotic region near the spacetime singularity,
we have $\dot{p}_t = - \partial \tilde{H}/ \partial t = 0$.
Hence, $p_t$ is a constant of motion and the bouncing of the
$(\xi, \phi)$ variables (or the $(x,y)$ variables
(\ref{transfChitre2})) within the scattering domain
${\cal B}$ will take place at {\em constant speed} with respect
to the $t$ time coordinate.}
which moves along the geodesics on a curved manifold with constant
negative curvature.

To see this, note, that in general
the Hamiltonian for a particle that moves along the
geodesics ($ \delta \int ds = 0 $) of a manifold with the metric
$g_{\mu \nu}$,
\begin{equation}
ds^2 = g_{\mu \nu} \; dx^{\mu} \; dx^{\nu}
\end{equation}
has the form
\begin{equation}
{\cal H} = \frac{1}{2m} \; g^{\mu \nu} \; p_{\mu} \; p_{\nu}
\end{equation}
where the metric tensor $g^{\mu \nu}$ is the inverse of
$g_{\mu \nu}$, i.e.
$ g_{\mu \nu} g^{\nu \rho} \equiv
\sum_{\nu} g_{\mu \nu} g^{\nu \rho} = \delta_{\mu}^{\rho} \; $.

Inside the (stationary) potential walls
(\ref{wallstatichard}) we have arrived at the stationary
Hamiltonian (\ref{Hamiltonxi}).
As regards the {\em projected}
motion on the two dimensional plane $(\xi, \phi)$,
we may therefore conclude from the form
of the Hamiltonian
(\ref{Hamiltonxi}) that the
associated flow is a geodesic flow on a Riemannian manifold with the
metric
\begin{equation}  \label{geodesicfundamental}
ds^2 = d \xi^2 + \sinh^2 \xi \; d\phi^2
\end{equation}
inside the
(stationary) region (the ``billiard table" domain ${\cal B}$)
bounded by infinitely steep potential walls (\ref{wallstationary}).
This metric (\ref{geodesicfundamental})
is the metric of the Poincar\'{e} disc of constant negative
curvature! (cf.\ also, e.g., Balazs and Voros (1986)).
Explicit computation of the Gaussian curvature
$K$ at the point $(\xi,\phi)$  gives
$$ K = - \frac{1}{\sinh \xi} \; \frac{\partial^2}{\partial \xi^2}
\; \sinh \xi = - 1  $$

As is well known, by suitable transformations of the coordinates,
the geodesic flow on the Riemannian manifold
(\ref{geodesicfundamental})
may
be represented in various (equivalent) ways. Among these equivalent
models (representations) we will consider the
{\em ``Poincar\'{e} unit-disc model"} for the hyperbolic geometry.
By identifying a point $(\xi,\phi)$ with the point
$z = (x,y) = r(\cos \phi, sin \phi) \equiv r e^{i \phi} \in
\BFACE{C}$ (in the complex plane) via the
coordinate transformation
\begin{equation}  \label{transfChitre1}
z =  x + i y = r e^{i \phi} =  \tanh (\xi / 2) \;
e^{i \phi} \; \; , \; i.e.
\end{equation}
\begin{equation}  \label{transfChitre2}
(
\begin{array}{c}
x  \\
y
\end{array}
)
=
(
\begin{array}{c}
r \cos \phi   \\
r \sin \phi
\end{array}
)
=
(
\begin{array}{c}
\tanh \xi / 2 \; \cos \phi  \\
\tanh \xi / 2 \; \sin \phi
\end{array}
)
\end{equation}
we obtain a hyperbolic flow (of constant
negative curvature $K=-1$) on the unit disc
\begin{equation} \label{unitdisc}
{\cal D} = \left\{ z = x + i y \in \BFACE{C} \; |
\; |z| = \sqrt{x^2 + y^2} \leq 1 \right\}  \; \; .
\end{equation}
Any point of the
$(\xi,\phi)$ space is transformed into the unit
disc (\ref{unitdisc}). The boundary
(i.e. points at $r=1$) of the disc ${\cal D}$
corresponds to the points at infinity $ \xi = \pm \infty$.
By this coordinate transformation (cf., also,
Balazs and Voros (1986),
p. 117-118), the metric (\ref{geodesicfundamental}) is
transformed into
\begin{equation}   \label{metricunitdisc}
ds^2 =
\frac{4 (dr^2 + r^2 d\phi^2) }{(1 - r^2)^2 } =
\frac{4 (dx^2 + dy^2) }{(1 - x^2 - y^2)^2 }  \equiv
\frac{4 | d z |^2 }{(1 - |z|^2)^2 }
\end{equation}
in the interior of the unit disc.

To summarize shortly: The coordinate transformations
in (\ref{hyperbolicone}) have, in the region sufficiently near
the space-time singularity,
accomplished to transform the Hamiltonian dynamics (originally
formulated in the $(\Omega, \beta_{+},\beta_{-})$ variables and
their canonical conjugate momenta)
into  that of piecewise ``geodesic motion" (free motion)
which takes place on the Poincar\'{e} unit disc (with constant
negative curvature) inside a scattering
domain ${\cal B} \subset {\cal D}$ (bounded by three
walls) in the interior of the unit disc ${\cal D}$.

If one translates - via the defining
coordinate transformations (\ref{transfChitre2}) -
the equations (\ref{wallstationary})
of the boundary of the scattering domain ${\cal B}$,
one finds that they describe an equilateral triangle of geodesics
with the three corners at ``infinity'' (at the
boundary circle of the Poincar\'{e} disc). ${\cal B} \subset {\cal D}$
is thus a zero angle
triangle on the Poincar\'{e} disc with finite
hyperbolic area.
\begin{quotation}
\noindent
{\small
Using the identity
$\; \tanh \xi \equiv 2 \tanh (\frac{\xi}{2}) /
(1 + \tanh^2 (\frac{\xi}{2})) $, we
express the formula for the left wall in the $(x,y)$ coordinates. We get
$ 4 \; x/(1 + (x^2 + y^2)) = -1 $, which is a circle
$ (x + 2)^2 + y^2 = 3 $
with center $(x,y) = (-2,0)$ and radius $r = \sqrt{3}$.
The part of this circle which overlaps with the Poincar\'{e} disc ${\cal D}$
governs the equation
of the {\em left} wall  $\partial {\cal B}_1$
($ 2 \pi/3 \leq \; \phi  \; \leq 4 \pi/3 $).
It is easily verified that
$\partial {\cal B}_1$ cuts the boundary
$\partial {\cal D} = \left\{ (x,y) | x^2 + y^2 = 1 \right\} $ of the
Poincar\'{e} disc at the points $z = -\frac{1}{2} \;
\pm  \; i \frac{\sqrt{3}}{2} ) =
\left\{ e^{i \frac{2 \pi}{3} } \; , \; e^{i \frac{4 \pi}{3} } \right\}$
and at a right angle
$\varphi = \pi/2$. Hence, we conclude that the left
boundary wall $\partial {\cal B}_{1}$ is a
geodesic curve on the Poincar\'{e} unit disc ${\cal D}$.
The triangular symmetry of the dynamical system gives us
two similar ($2 \pi/3$ rotated) wall boundaries consisting of circular
arcs, the radii being $\sqrt{3}$ and
centres located at $z = -2\; e^{\pm i 2 \pi/3} = 1 \pm i\sqrt{3}$.
A {\em polygon} in the Poincar\'{e} disc ${\cal D}$ is
(as in Euclidian geometry) defined as
a closed, connected area whose boundary consists
in parts of geodesics, called its sides.
Thus, the scattering domain ${\cal B}$ is a regular
{\em triangular polygon}; a regular 3-sided billiard (the three sides
which we denote $\partial {\cal B}_{1}$, $\partial {\cal B}_{2}$ and
$\partial {\cal B}_{3}$) on the Poincar\'{e} disc.
(Cf. fig.9).
}
\end{quotation}

\begin{figure}
\vspace{8.0 cm}
\noindent
\caption[abcxxxx]{ {\small The triangular
billiard ${\cal B} \subset {\cal D}$
forming the scattering domain
inside the Poincar\'{e} disc $\partial {\cal D}$. The
``billiard table" is bounded by the ``infinitely hard" wall boundaries
$\partial {\cal B} = \partial {\cal B}_{1} \cup
\partial {\cal B}_{2} \cup \partial {\cal B}_{3} $.
(Each side of the 3-sided polygon,
$\partial {\cal B}_{1}$, $\partial {\cal B}_{2}$ and $\partial {\cal B}_{3}$,
are geodesics on the Poincar\'{e} disc, being circular arcs
orthogonal to the boundary circle $\partial {\cal D}$ (i.e.
cutting the Poincar\'{e} disc boundary $\partial {\cal D}$ at right angles
$\varphi = \pi/2$). Hence, ${\cal B}$ is a
triangular polygon, where all the interior angles
between two sides at each of the three vertices
${\cal C}_1, {\cal C}_2, {\cal C}_3$
are zero.
The {\em approximations} involved in arriving at the description
of the gravitational toy-model collapse are
{\em very accurate} when sufficiently near the space-time
``big crunch" singularity of the metric;
The scattering potential $\tilde{V}(x,y)$
then (to a very high accuracy) vanishes
identically in the interior of the domain ${\cal B}$
(inside the wall boundaries $\partial {\cal B}$)
and equals $+ \infty$ outside ${\cal B}$.  }  }
\end{figure}

The periods of geodesic motion (free motion)
which occur between the bounces against one of the
three walls of the triangle
correspond to the periods of straight line
behavior (``Kasner epochs'') in fig.5.
The source of the chaos of the model is to be thought of
as due to the bounces
in the potential boundaries $\partial {\cal B}$
in conjunction with the negative curved interior
(which gives sensitive
dependence on initial conditions.
To the bounces are associated algebraic transition rules
related to continued fraction expansions, cf.\ table 1.
In fact, due to the symmetry properties of the domain (tiling the
Poincar\'{e} disc by the action of an arithmetic group
(E.B. Bogomolny \etal (1992), J. Bolte \etal (1992)),
the chaos in the mixmaster collapse
(in the region sufficiently near the singularity) is an example of
arithmetic (algebraic) chaos. We thus appear to have
an example of chaos of remarkable beauty.

\begin{quotation}
\noindent
{\small
{\bf Are the exact numerical values of the
Lyapunov exponents $\lambda$
an ``artifact'' of the considered set of transformations?} \\
One observes, cf.\ J. Pullin (1990),
that the constant negative curvature $K < 0$ of the
interior of the Poincar\'{e}
disc may be translated into a statement about
a positive Lyapunov exponent. The
geodesic motion takes place at constant velocity
and we have for the Lyapunov exponent
$\lambda = \left\{ velocity \right\} \times \sqrt{-K} > 0$.
(cf.\ Balazs and Voros (1986), p. 147).
This expression scales with $\sqrt{-K}$ and the precise value of
the negative curvature $K$ of the Poincar\'{e} disc
appears to be an
artifact\footnote{One could transform the collapse orbit
to a Poincar\'{e}
disc with any negative curvature $K < 0$ (One should, however,
verify how the
velocity of the ``billiard ball'' scales in that case).}
of the selected set of coordinate transformations
(\ref{hyperbolicone}).
However, the instability of a closed orbit $\gamma$ in the
triangular billiard on the Poincar\'{e} disc is given by multiplying
the apparent ``Lyapunov exponent'' $\lambda \propto \sqrt{-K}$
with the hyperbolic length of the closed orbit.
Since the length of a closed orbit $\oint_{\gamma} ds$
scales as $|K|^{-1/2}$ the expression for the instability exponents
of the closed orbits is {\em invariant} under rescalings of $K$:
$$ \left\{ \; instability \; \; of \; \; closed \; \; orbit \; \right\}
\propto  \sqrt{-K} \times
\left\{ \; length \; \; of \; \; closed \; \; orbit \; \right\} = $$
$$ = \sqrt{-K} \oint_{\gamma} ds =
\sqrt{-K} \oint_{\gamma} \frac{2 |dz|}{\sqrt{-K} (1 - |z|^2)}
=  \oint_{\gamma} \frac{2 |dz|}{ (1 - |z|^2)} \; \; .  $$
Thus the instability properties of the
periodic orbit structure
seem to be a (slightly) more invariant characterization.

}
\end{quotation}

\vspace{0.5 cm}


As concerns  the Wheeler-DeWitt toy-quantization of our gravitational
collapse in the $(t, \xi, \phi)$ variables, say,
it will actually have a complicated form (cf.\ equation (74) in C.W. Misner
(1972)). Being sufficiently near the singularity of the metric
where the walls
are approximately infinitely hard we have an operator
constraint on the wave function $\Psi = \Psi ({}^{(3)} g) =
\Psi (t, \xi, \psi)$ of the form
\begin{equation}
\left\{ e^{-t} \frac{\partial}{\partial t} (-e^{t}
\frac{\partial}{\partial t}) +
\frac{1}{\sinh \xi}
\frac{\partial}{\partial \xi}(\sinh \xi \frac{\partial}{\partial \xi})
+ \frac{1}{\sinh^2 \xi}\frac{\partial^2}{\partial \phi^2} \right\}
\Psi(t, \xi, \phi) = 0
\end{equation}
inside the domain ${\cal B} \subset {\cal D}$
with the boundary condition that $\Psi$
vanishes at the boundary $\partial {\cal B}$.

The substitution
\begin{equation}
\Psi = \Psi (t, \xi, \phi) =
e^{-(1/2 + i \omega) t} \; \psi (\xi, \phi)
\end{equation}
yields for the $\psi$ component
\begin{equation}
\left\{ \frac{1}{\sinh \xi}
\frac{\partial}{\partial \xi}(\sinh \xi \frac{\partial}{\partial \xi})
+ \frac{1}{\sinh^2 \xi}\frac{\partial^2}{\partial \phi^2} \right\}
\psi (\xi, \phi) = (\omega^2 + \frac{1}{4}) \;
\psi (\xi, \phi) \; \; .
\end{equation}
The left hand side is recognized as
the Laplace operator on the
two-dimensional Poincar\'{e} disc of constant negative curvature
with metric $ds^2 = d\xi^2 + \sinh^2 d \phi^2 $.
We conclude that
properties of solutions to the Wheeler DeWitt equation for the mixmaster
collapse  are one-to-one related to
spectral properties of the Laplace operator
on the Poincar\'{e} disc, that is the eigenvalue equation
which in the more familiar Poincar\'{e} disc variables $(x,y)$ reads
\begin{equation}
\bigtriangleup \psi  =  \frac{1}{4} (1 - x^2 - y^2)^2
(\frac{\partial^2}{\partial x^2} +
\frac{\partial^2}{\partial y^2} ) \;
\psi = (\omega^2 + \frac{1}{4}) \; \psi  \; \; .
\end{equation}
with the boundary condition that
$\psi$ vanishes at the boundary
$\partial {\cal B}$ of the triangular domain.

Moreover, spectral properties of the Laplace operator
are related to the Gutzwiller trace formula
involving a sum over the periodic
orbits inside the triangular domain ${\cal B} \subset {\cal D}$.
(Cf. Gutzwiller (1990) and also Cvitanovic (1990) and
references therein).

It is amusing that the spectrum of the Laplace operator
exhibits ``{\em non-generic}'' energy-level statistics
in the sense that
the distributions of nearest neighbor level spacings
displays Poisson rather
than Wigner statistics. This
is due to the symmetry
properties of our triangular domain (the domain
tiles the disc under the action of a so-called
``{\em arithmetic group}'')
with a resulting exponentially large {\em degeneracy} of lengths of the
periodic orbits.
Note, also, that the (approximate) semiclassical Gutzwiller trace formula,
cf.\ Gutzwiller (1971, 1990), is exact in this case, since it coincides
with the exact Selberg trace formula (giving an exact relationship
between the quantum spectrum and the classical
periodic orbits).\footnote{For a discussion of the most recent results
as concerns the the Wheeler-DeWitt quantization of the mixmaster gravitational
collapse, see R. Graham (1993).}

In the context of the toy-model mixmaster gravitational collapse, a contact
is thus made between the disciplines of ``Quantum Cosmology''
and ``Quantum Chaos'', each dating back twenty years to pioneering
works of C.W. Misner and M.C. Gutzwiller!

\section{DISCUSSION, OPEN PROBLEMS AND DREAMS (SOME FINAL REMARKS)}

The toy-model sketched (the mixmaster gravitational collapse) is
a very simple model.
Nevertheless, it has enough degrees of freedom to behave
in a dynamically complicated  and
non-predictable way (with positive Kolmogorov entropy)
on approach to the singular space-time point (the ``big crunch"
point) and in this way
the toy-model space-time metric captures an interesting
non-linear aspect of the classical Einstein equations
(probing the Einstein equations
in a domain where the field strength gets strong
(as captured by an invariant like
$R_{\alpha \beta \gamma \delta} R^{\alpha \beta \gamma \delta}$,
which grows without limit) and non-linearities of the Einstein
equations are important).

In fact, if one were to imagine a
{\em generic} gravitational collapse (e.g. for an entire
``model-universe") to a singular point, then it would
misguide our intuition to imagine a
Friedmann Robertson Walker type of collapse solution
(the ``standard" big crunch/big bang). This (non-generic)
solution is completely integrable (and behaves in a
dynamically trivial
way) as an artifact of
the enormous amount of symmetry imposed on that solution.

Rather one should try to imagine a sort of
``metrical chaos'' (chaos in the
space-time metric) being developed on approach to
the curvature singularity - as the gravitational
field strength gets stronger and non-linear
aspects of the Einstein equations are increasingly important.

\subsection{How to build up concepts which characterize such
``metrical chaos''? }

The general problem of transferring standard indicators of chaos to the
situation of general relativity (pointed out in S.E.Rugh (1990a),
emphasized in S.E. Rugh (1990b) and by J. Pullin (1990)) is
well illustrated by the simple spatially homogeneous mixmaster
collapse.

We have noticed (S.E. Rugh (1990 a,b)) that a naive application of a
``standard'' Lyapunov exponent will be a highly
gauge-dependent measure, even
under the small subclass of gauge-transformations, which only
involve (the perfectly allowed) reparametrizations of the
time-coordinate.
To construct a gauge invariant generalization of a
``Lyapunov exponent''
is a yet unsolved problem.

Perhaps, there is a ``no go'' theorem to be constructed
towards a completely gauge invariant construction?
(H.B. Nielsen and S.E. Rugh)

We have noticed that the ergodicity properties of the toy-model
mixmaster collapse  are interesting (and subtle):
In some ``gauges'' the phase space is ever expanding (or ever
collapsing) and there are no return properties of orbits
and a concept like ``ergodicity'', for example, is very
ill defined. By a suitable set of coordinate
transformations, however, and some
very reliable approximations, the Hamiltonian
can be brought
into that of a geodesic flow on a two-dimensional
Riemann manifold of constant negative curvature (the Poincar\'{e} disc)
bounded by some stationary walls forming an equilateral triangle with
three corner infinities on the Poincar\'{e} disc.
In this picture a number of stochastic
properties of the gravitational collapse may be stated, e.g., the property
of ergodicity. It is even a K-flow!

\begin{quotation}
{\em
\noindent
However, the chaotic properties of
a toy-model like the
mixmaster gravitational collapse
is of course - inherently - the same, irrespective of
the choice of ``gauge'' (choice of description).
If we get apparent ``chaotic behavior'' in one description (in one gauge)
but absence of chaotic behavior in another description
- it is because one has posed the wrong question! }
\end{quotation}
I thus disagree with the conclusion arrived at by Pullin
(cf.\ also J. Pullin (1990))
about the superiority of the
``Poincar\'{e} disc'' gauge-choice over other
gauge-choices.

It appears to me that
no ``gauge" is better than others. Rather, one
ought to seek measures (of chaos) which are invariant under
(some large class of) gauge transformations (to the extent they are
possible to construct).

\subsection{Is chaos in spacetime metrics of
importance for observable phenomena in Nature?}

The extreme conditions for the gravitational field
referred to in this manuscript
(probing the Einstein equations near space-time curvature singularities)
are of course far removed from ``daily life'' gravity.
It is hard to find observable {\em astrophysical} phenomena
which involve very strong gravitational fields and
thus probe the {\em non-linearities}
of the Einstein equations at a really deep level.
(For instance, the rather large perihelion
precession
in binary pulsar systems, which
are indeed marvellous relativistic laboratories, do
not qualify as an observational test of
non-linearities of the Einstein equations at a
deep level).
In the description of many gravitational phenomena, even the
(quasi) {\em linearized} equations suffice.

The relevant laboratory for probing deep non-linearities of the
Einstein equations seems to be (1) violent astrophysical phenomena
\footnote{A gravitational phenomenon like the generation of
gravitational waves (which will be observable in the next century)
emitted from {\em strong} astrophysical sources needs that we take full
account of the nonlinearity of the
Einstein equations, cf.\ e.g. D. Christodoulou (1991). }
(2) early moments of the Universe (could the Universe itself - and
the large scale structure of the spacetime metric of the Universe -
have had a violent past?
Would there be observable fingerprints of this today?
Cf. also next subsection 6.3)

Turning to our toy-model metric (\ref{metric})
the Einstein equations for the metric
involve exponential functions.
Such exponential functions are well known to give out
``large numbers'' very fast! In fact, it is so that if we
- by hand - put some ``reasonable
scale'' (like $L \sim c/H \sim
10^{28} \; cm \sim$ the characteristic size of the present Universe)
for  the initial length scales,
$a \sim b \sim c \sim 10^{28} \; cm$ of the gravitational
collapse, then only few (chaotic) cycles of the
the scale-functions $a,b,c$ occur before they reach and pass
the Planck regime,
$R_{\alpha \beta \gamma \delta} R^{\alpha \beta \gamma \delta}
\sim l_{Pl}^{-4} $. Thus, the gravitational collapse is chaotic
(``metrical chaos'') in the very same region where
it should be treated quantum mechanically.
In that sense this gravitational collapse is
very much addressing a topic on
the interface of quantum physics, cosmology and chaos.

If the large scale structure of spacetime (our Universe)
started out (near Planck time, say)
in a chaotic, oscillatory behavior given by
the mixmaster metric (\ref{metric}) say,
would we then be able to detect signals
of this behavior today?
Most likely, we would observe a
substantially larger anisotropy of the background microwave
radiation temperature than the high degree of isotropy observed
today, since the Universe dissipates away the initial anisotropies
too slowly. (See also next subsection).
In a Universe with an {\em inflationary}
phase, the existence of anisotropies and
inhomogeneities before the onset of inflation is not excluded,
but it appears that there is not much time for spacetime dynamics
to perform ``chaotic oscillations'' in the short interval from
Planck times $t_{Pl} = \sqrt{(\hbar G/c^5)} \sim 10^{-43}$
seconds after the ``big bang''
to the onset of an inflationary phase at the
G.U.T. scales $\sim 10^{14}$
GeV occurring $\sim 10^{-34}$ seconds after
the ``big bang''.\footnote{Using $R \propto T^{-1}$ (valid after inflation
has occurred), it is fascinating to imagine that
our current Hubble volume (size of the Universe)
had a size of about $\sim 10$ cm at those early times!
(i.e. when the temperature in the
Universe was $\sim 10^{14}$ GeV).}
Moreover, inflation tends to smooth things out so
pre-inflationary history (e.g. dynamics of the spacetime metric)
is rendered more irrelevant.

\newpage

\begin{quotation}
\noindent
{\small
{\bf A remark concerning dependence of the ``present'' on the
``past''. I.} \\
A more serious problem than whether we will be able to detect
evidence of ``chaotic spacetimes'' in the early moments of the
Universe is whether
detectable signals (now) in cosmology contain
information which will prove useful
in probing the structure of the Natural laws
at higher energies than those which can be created
here on Earth.
In view of the forseeable limitations
in {\em our} abilities of putting
substantially more than present energies $\sim 100 GeV$ on a
single elementary particle (experiments here on Earth),
one would like to ``resort to'' our Universe - in its very early
moments - as a laboratory to
probe the structure of fundamental laws at energies
beyond $\sim 100 GeV$. In cosmology, however, we really see the
big bang physics through an extremely cloudy
and {\em little informative filter!} See also, e.g.,
H.B. Nielsen, S.E. Rugh and C. Surlykke (1993).
Every time {\em thermal equilibrium phases} are reached at certain
stages in the evolution of the Universe
then only very little information
of the physics that went on {\em before} that phase can reach us today.
It is  difficult for a signal to
survive through an equilibrium phase!
Basically, only {\em conserved}
quantum numbers {\em survive}, like baryon and lepton
numbers and energy. However, if the thermodynamic equilibrium is
not reached globally, there may survive some information in the
correlations or, rather, in the {\em spatial variations} in the densities of
these conserved quantities. Gravitational signals, i.e. spacetime
dynamics, may contribute to
establish correlations over big distances, since gravity is a
long range force.
}
\end{quotation}

\subsection{A chaotic initial ``big bang''? }

Could our Universe have started out in a chaotic state? Cf.
the ``chaotic cosmology'' concept, developed by C.W. Misner \etal,
which attempts at creating our present Universe from (almost)
arbitrary irregular initial conditions, e.g. from a
spacetime metric with large initial anisotropies
and inhomogeneities.\footnote{See also, e.g.,
discussions in B.L. Hu \etal (1993). }

\begin{quotation}
\noindent
{\small
{\bf A remark concerning dependence of the ``present'' on the
``past''. II.} \\
Whereas one would (in cosmology) be tempted to
explain the present Universe
without having to resort to considerable
fine tunings of the initial data
of the Universe (at Planck scales, say)
a ``high energy physicist'' would
prefer that the present
depends very sensitively on the past!\footnote{Inflation, which only lessens
the dependence of the present from the initial data of the past,
is in that perspective not good news. (Even within the
inflationary Universe, however,
there will be an arbitrary large number of initial conditions which are
inconsistent with observations today).}
- in order to be able to deduce interesting {\em information
about the past} (in particular, to get a detailed look into the Natural
laws beyond $\sim 100 GeV$, via signals
from cosmology).

The viewpoint depends on the
question(s) which drives our study: Do we ``want'' to give
a ``{\em probabilistic}'' explanation of the
present observed Universe (I.e.: if the Universe we see today
results from a larger basin
of initial conditions then it is more ``likely'' to have it?) or do we
``want'' signals coming from cosmology
to be a very sharp
probe exploring into the structure of Natural laws at very high energies
(corresponding to hot dense phases of the Universe)?

If one contemplates, cf.\ e.g. S. Weinberg (1992), that
the actually implemented Natural laws are
the only {\em self-consistent}
set of laws and parameters imaginable, and
also consider initial conditions of the Universe as resulting
from such implemented regularities (the division line between Natural
laws and initial conditions is rather arbitrary anyway)
then I must admit that I
have no problem with reconciling myself to the fact
that {\em initial conditions} of the Universe may be as
fine tuned as
{\em Natural laws} appear to be (so far explored).
}
\end{quotation}

\noindent
If the Universe had very large irregularities initially, we have -
in order to obtain the {\em present} Universe with large scale
smoothness -
to introduce some physical mechanisms to ``dissipate away'' all
these initial irregularities (anisotropies and inhomogeneities).

A very effective dissipative process, suggested
by Zel'dovich, is that of vacuum particle creation from the
changing gravitational field in anisotropic spacetimes. (Gravitational
fields can create particles somewhat similar to the way magnetic
fields can create electron pairs). In a test field approximation where
the spacetime anisotropies are not too big, one may estimate
the probability ${\cal P}$
of producing a pair of conformally invariant scalar particles
in a world tube $T$ of constant, co-moving cross section extending
from the singularity to the present and it turns out to
be proportional
to the integrated Weyl-curvature invariant over the world
tube (cf., e.g., J.B. Hartle (1981)),
$$ {\cal P} \sim \int_T d^4 x \sqrt{-g} \;
C_{\alpha \beta \gamma \delta}
C^{\alpha \beta \gamma \delta} \; \; . $$
If the initial Weyl tensor invariant is too large and
if the Universe has managed to dissipate away all this initial
anisotropy, the heat entropy generated would be very large,
and the Universe would
{\em already} have reached ``heat death'' which
is not the case. In fact, the ``small'' entropy per baryon
now severely limits the amount of
dissipation that has taken place in the initial phases of the
Universe\footnote{In order not to over-produce entropy. See also Barrow and
Matzner (1977) and Barrow (1978)).}

In order not to reach ``heat death'' too fast, the Universe
has to start out in a state of low entropy\footnote{
This fact, and the enormous amount of fine tuning and
precision in the organization of the starting conditions of the
Universe this implies, has
been emphasized very strongly in e.g. R. Penrose (1989).
Perhaps, the Guth/Linde inflationary Universe
provides us, more naturally, with
a ``low entropy'' initial condition? (see, e.g.,
D. Goldwith and T. Piran (1991) and references therein). }.
According to the Weyl tensor hypothesis of R. Penrose (1979),
gravitational entropy should, somehow, be measured by the
Weyl curvature
tensor, and the constraint imposed on the Universe to start out in
a state of low entropy translates, e.g., into a requirement
of creating a Universe with a small Weyl curvature (whereas
a final ``big crunch'', if the Universe is closed, may -
and is expected to - have a large
Weyl curvature and a large degree of disorder).

The mixmaster space-time metric appears to be a
well suited toy-model for
implementation of the Weyl tensor hypothesis, having a very large
Weyl curvature near the space-time singularity.\footnote{M.
Biesiada and S.E. Rugh, in preparation.}
Thus, according to the Weyl-curvature hypothesis, the chaotic collapse
should be considered a (simplified) candidate for the
large scale metric at a final crunch\footnote{Cf. also
the title ``Gravity's chaotic future'' in a recent issue of
Science News, Vol. 144, p. 369-384. Dec.4 (1993).}
rather than at the initial bang.

Turning back to our question about observational consequences
of having chaos in spacetime metrics:
If it is a chaotic collapse at a ``big crunch'' point -
it will probably be too hot
for us to be there and observe it!

\section*{ACKNOWLEDGMENTS}

I would like to thank Holger Bech Nielsen for all sorts of fruitful
exchanges, from chaos in dynamical systems to ``chaos in
Natural laws'' to viewpoints concerning what our Creator should
or should not do.

I am indebted to the ``Chaos Group'' at the Niels Bohr Institute
and NORDITA
and John Barrow, Freddy Christiansen,
Marek Demianski, Robert Graham, Bernard Jones,
V.N. Lukash and Igor Novikov and, especially, I am indebted
to Holger Bech Nielsen and Hans Henrik Rugh
for discussions and collaboration on various aspects of the
gravitational collapse.

Special thanks are also due to Marek Biesiada for
discussions and collaboration
in Copenhagen, Poland and Japan
and hospitality at the Copernicus Astronomical Center at Warsaw, Poland.

Support from
the Danish Natural Science Research Council Grant No. 11-8705-1
is gratefully acknowledged and I thank Dave Hobill for the
invitation to participate in this interesting workshop.


\end{document}